\documentclass[12pt,draftcls,journal,onecolumn,letterpaper]{IEEEtran}
\usepackage[english]{babel}
\usepackage{amsmath,amssymb,amscd,latexsym,dsfont,bbm,amsbsy}
\usepackage{subfigure,color,graphicx,enumerate,stfloats,psfrag}
\usepackage{cite,multirow}
\newcommand{\ie}{i.e.,~}
\newcommand{\eg}{e.g.,~}
\newcommand{\cf}{cf.~}
\newcommand{\argmax}{\mathop{\mathrm{argmax}}}
\newcommand{\tr}[1]{\mathrm{#1}}

\newcommand{\mbbmss}[1]{\mathbbmss{#1}}
\newcommand{\mb}[1]{\boldsymbol{#1}}
\newcommand{\mc}[1]{\mathcal{#1}}
\newcommand{\mf}[1]{\mathsf{#1}}
\newcommand{\ms}[1]{\mathds{#1}}

\newcommand{\un}[1]{\underline{#1}}
\newcommand{\set}[1]{\{#1\}}
\newcommand{\cd}{\cdot}

\newcommand{\ld}{\ldots}
\newcommand{\ds}[1]{\displaystyle{#1}}
\newcommand{\mat}[1]{\mathbb{#1}}
\newcommand{\eqlab}[2]{\begin{align} \label{#1} #2 \end{align}}
\newcommand{\eq}[1]{\begin{align*} #1 \end{align*}}

\newcommand{\PP}{P}
\newcommand{\tx}{\tilde{\mb{x}}}
\newcommand{\trace}[1]{\mathop{\mathrm{trace}\left(#1\right)}\nolimits}
\newcommand{\diag}[1]{\mathop{\mathrm{diag}\left(#1\right)}\nolimits}
\newcommand{\cov}[1]{\mathop{\mathrm{cov}\left(#1\right)}\nolimits}

\newcommand{\bv}{\mb{v}}
\newcommand{\bx}{\mb{x}}
\newcommand{\XX}{\mat{X}}
\newcommand{\LL}{\mat{L}}
\newcommand{\NN}{\mat{N}}

\newcommand{\Eb}{E_\tr{b}}
\newcommand{\Ns}{N_\tr{s}}
\newcommand{\Es}{E_\tr{s}}
\newcommand{\Rc}{R_\tr{c}}
\newcommand{\Imax}{\mf{C}}
\newcommand{\SNR}[0]{\mathsf{SNR}}
\newtheorem{theorem}{Theorem}
\newtheorem{example}{Example}
\newtheorem{corollary}[theorem]{Corollary}

\newtheorem{lemma}[theorem]{Lemma}

\newcommand{\bX}{\boldsymbol{X}}

\newcommand{\bH}{\boldsymbol{H}}
\newcommand{\bY}{\boldsymbol{Y}}

\newcommand{\bh}{\boldsymbol{h}}
\newcommand{\by}{\boldsymbol{y}}

\title{On the BICM Capacity}
\author{%
\IEEEauthorblockN{Erik Agrell and Alex Alvarado\\}
\IEEEauthorblockA{Department of Signals and Systems, Communication Systems Group\\ Chalmers University of Technology, Gothenburg, Sweden\\}
\emph{\set{agrell,alex.alvarado}@chalmers.se}
\thanks{Research supported by the Swedish Research Council, Sweden under grant \#2006-5599. This work has been presented in part at the International Wireless Communications and Mobile Computing Conference IWCMC 2009, Leipzig, Germany, June 2009, and the IEEE Information Theory Workshop ITW 2009, Taormina, Italy, Oct.~2009.}
}%

\begin{document}
\maketitle
\begin{abstract}
Optimal binary labelings, input distributions, and input alphabets are analyzed for the so-called bit-interleaved coded modulation (BICM) capacity, paying special attention to the low signal-to-noise ratio (SNR) regime. For 8-ary pulse amplitude modulation (PAM) and for 0.75~bit/symbol, the folded binary code results in a higher capacity than the binary reflected gray code (BRGC) and the natural binary code (NBC). The 1~dB gap between the additive white Gaussian noise (AWGN) capacity and the BICM capacity with the BRGC can be almost completely removed if the input symbol distribution is properly selected. First-order asymptotics of the BICM capacity for arbitrary input alphabets and distributions, dimensions, mean, variance, and binary labeling are developed. These asymptotics are used to define first-order optimal (FOO) constellations for BICM, \ie constellations that make BICM achieve the Shannon limit $-1.59~\tr{dB}$. It is shown that the $\Eb/N_0$ required for reliable transmission at asymptotically low rates in BICM can be as high as infinity, that for uniform input distributions and 8-PAM there are only 72 classes of binary labelings with a different first-order asymptotic behavior, and that this number is reduced to only 26 for 8-ary phase shift keying (PSK). A general answer to the question of FOO constellations for BICM is also given: using the Hadamard transform, it is found that for uniform input distributions, a constellation for BICM is FOO if and only if it is a linear projection of a hypercube. A constellation based on PAM or quadrature amplitude modulation input alphabets is FOO if and only if they are labeled by the NBC; if the constellation is based on PSK input alphabets instead, it can never be FOO if the input alphabet has more than four points, regardless of the labeling.
\end{abstract}


\section{Introduction}

The problem of reliable transmission of digital information through a noisy channel dates back to the works of Nyquist \cite{Nyquist24,Nyquist28} and Hartley \cite{Hartley28} almost 90 years ago. Their efforts were capitalized by C.~E.~Shannon who formulated a unified mathematical theory of communication in 1948 \cite{Shannon48,Shannon49}\footnote{An excellent summary of the contributions that influenced Shannon's work can be found in \cite[Sec.~I]{Pierce73}.}. After he introduced the famous capacity formula for the additive white Gaussian noise (AWGN) channel, the problem of designing a system that operates close to that limit has been one of the most important and challenging problems in information/communication theory. While low spectral efficiencies can be obtained by combining binary signaling and a channel encoder, high spectral efficiencies are usually obtained by using a coded modulation (CM) scheme based on a multilevel modulator.

In 1974, Massey proposed the idea of jointly designing the channel encoder and modulator \cite{Massey74}, which inspired Ungerboeck's trellis-coded modulation (TCM) \cite{Ungerboeck76,Ungerboeck82} and Imai and Hirakawa's multilevel coding (MLC) \cite{Imai77,Imai77b}. Since both TCM and MLC aim to maximize a Euclidean distance measure, they perform very well over the AWGN channel. However, their performance over fading channels is rather poor. The next breakthrough came in 1992, when Zehavi introduced the so-called bit-interleaved coded modulation (BICM) \cite{Zehavi92} (later analyzed in \cite{Caire98}), which is a serial concatenation of a binary channel encoder, a bit-level interleaver, and a memoryless mapper. BICM aims to increase the code diversity---the key performance measure in fading channels---and therefore outperforms TCM in this scenario \cite[Table~III]{Caire98}. BICM is very attractive from an implementation point view because of its flexibility, \ie the channel encoder and the modulator can be selected independently, somehow breaking Massey's joint design paradigm. BICM is nowadays a \emph{de facto} standard, and it is used in most of the existing wireless systems, \eg HSPA (HSDPA and HSUPA) \cite{3gpp25212_2009}\cite[Ch.~12]{Dahlman08_Book}, IEEE 802.11a/g \cite{IEEE80211_July99} IEEE 802.11n \cite[Sec.~20.3.3]{IEEE80211n_October09}, and the latest DVB standards (DVB-T2 \cite{ETSI_EN_302_755_v111}, DVB-S2 \cite{ETSI_EN_302_307_v121}, and DVB-C2 \cite{ETSI_EN_302_769_v111}).

Plots of the BICM capacity vs. $\Eb/N_0$ reveal that BICM does not always achieve the Shannon limit (SL) $-1.59~\tr{dB}$. This can be explained based on first-order asymptotics of the BICM capacity, which were recently developed by Martinez \emph{et al.} for uniform input distributions and one- and two-dimensional input alphabets \cite{Martinez08b,Alvarado10c}. It was shown that there is a bounded loss between the BICM capacity and the SL when pulse amplitude modulation (PAM) input alphabets labeled by the binary reflected gray code (BRGC) is used. Recently, Stierstorfer and Fischer showed in \cite{Stierstorfer09a} that this is caused by the selection of the binary labeling and that equally spaced PAM and quadrature amplitude modulation (QAM) input alphabets with uniform input distributions labeled by the natural binary code (NBC) achieve the SL. Moreover, the same authors showed in \cite{Stierstorfer07a} that for low to medium signal-to-noise ratios (SNR), the NBC results in a higher capacity than the BRGC for PAM and QAM input alphabets and uniform input distributions. 

The fact that the BICM capacity does not always achieve the SL raises the fundamental question about first-order optimal (FOO) constellations for BICM, \ie constellations that make the BICM achieve the SL. In this paper, we generalize the first-order asymptotics of the BICM capacity presented in \cite{Martinez08b} to input alphabets with arbitrary dimensions, input distributions, mean, variance, and binary labelings. Based on this model, we present asymptotic results for PAM and phase shift keying (PSK) input alphabets with uniform input distribution and different binary labelings. Our analysis is based on the so-called Hadamard transform \cite[pp.~53--54]{MacWilliams77}, which allows us to fully characterize FOO constellations for BICM with uniform input distributions for fading and nonfading channels. A complete answer to the question about FOO constellations for BICM with uniform input distributions is given: a constellation is FOO if and only if it is a linear projection of a hypercube. Furthermore, binary labelings for the traditional input alphabets PAM, QAM, and PSK are studied. In particular, it is proven that for PAM and QAM input alphabets, the NBC is the only binary labeling that results in an FOO constellation. It is also proven that PSK input alphabets with more than four points can never yield an FOO constellation, regardless of the binary labeling. When 8-PAM with a uniform input distribution is considered, the folded binary code (FBC) results in a higher capacity than the BRGC and the NBC. Moreover, it is shown how the BICM capacity can be increased by properly selecting the input distribution, \ie by using so-called \emph{probabilistic shaping} \cite{Barsoum07}. In particular, probabilistic shaping is used to show that PAM input alphabets labeled by the BRGC or the FBC can also be FOO, and to show that the 1~dB gap between the AWGN capacity and the BICM capacity with the BRGC can be almost completely removed.

\section{Preliminaries}\label{Sec.Preliminaries}

\subsection{Notation Convention}\label{Sec.Preliminaries.Notation}

Hereafter we use lowercase letters $x$ to denote a scalar, boldface letters $\mb{x}$ to denote a row vector of scalars, and underlined symbols $\un{\mb{x}}$ to denote a sequence. Blackboard bold letters $\mat{X}$ represent matrices and $x_{i,j}$ represents the entry of $\mat{X}$ at row $i$, column $j$, where all the indices start at zero. The transpose of $\mat{X}$ is denoted by $\mat{X}^\tr{T}$, $\trace{\mat{X}}$ is the trace of $\mat{X}$, and $\|\mat{X}\|^2$ is $\trace{\mat{X}^\tr{T}\mat{X}}$.

We denote random variables by capital letters $Y$, probabilities by $\Pr\set{\cd}$, the probability mass function (pmf) of the random vector $\mb{Y}$ by $P_{\mb{Y}}(\mb{y})$, and the probability density function (pdf) of the random vector $\mb{Y}$ by $p_{\mb{Y}}(\mb{y})$. The joint pdf of the random vectors $\mb{X}$ and $\mb{Y}$ is denoted by $p_{\mb{X},\mb{Y}}(\mb{x},\mb{y})$, and the conditional pdf of $\mb{Y}$ conditioned on $\mb{X}=\mb{x}$ is denoted by $p_{\mb{Y}|\mb{X}=\mb{x}}(\mb{y})$. The same notation applies to joint and conditional pmfs, \ie $P_{\mb{X},\mb{Y}}(\mb{x},\mb{y})$ and $P_{\mb{Y}|\mb{X}=\mb{x}}(\mb{y})$. The expectation of an arbitrary function $f(\mb{X},\mb{Y})$ over the joint pdf of $\mb{X}$ and $\mb{Y}$ is denoted by $\ms{E}_{\mb{X},\mb{Y}}[f(\mb{X},\mb{Y})]$, the expectation over the conditional pdf $p_{\mb{Y}|\mb{X}=\mb{x}}(\mb{y})$ is denoted by $\ms{E}_{\mb{Y}|\mb{X}=\mb{x}}[f(\mb{X},\mb{Y})]$, and $\cov{\mb{X}}$ is the covariance matrix of the random vector $\mb{X}$.

We denote the base-2 representation of the integer $0\le i \le M-1$, where $M=2^m$, by the vector $\mb{b}(i)=[b_{m-1}(i),b_{m-2}(i),\ld,b_0(i)]$, where $b_{m-1}(i)$ is the most significant bit of $i$ and $b_0(i)$ the least significant. To facilitate some of the developments in this paper, we also define the \emph{ordered direct product} as
\begin{align}\label{ODP1}
[\mb{a}_0^\tr{T},\ld,\mb{a}_{p-1}^\tr{T}]^\tr{T} \otimes [\mb{b}_0^\tr{T},\ld,\mb{b}_{q-1}^\tr{T}]^\tr{T} \triangleq
  [\mb{c}_0^\tr{T},\ld,\mb{c}_{p q-1}^\tr{T}]^\tr{T},
\end{align}
where $\mb{c}_{qi + j} = [\mb{a}_i, \mb{b}_j]$ for $i=0,\ldots,p-1$ and $j=0,\ldots,q-1$.
The ordered direct product in \eqref{ODP1} is analogous to the Cartesian product except that it operates on vectors/matrices instead of sets.

\subsection{Binary Labelings}\label{Sec.Preliminaries.Labelings}

A \emph{binary labeling} $\mat{L}$ of order $m\geq 1$ is defined using an $M \times m$ matrix where each row corresponds to one of the $M$ length-$m$ distinct binary codewords, $\mat{L}=[\mb{c}_0^\tr{T}, \ld ,\mb{c}_{M-1}^\tr{T}]^\tr{T}$, where $\mb{c}_i=[c_{i,0},c_{i,1},\ld,c_{i,m-1}]\in\set{0,1}^m$.

In order to recursively define some particular binary labelings, we first define \emph{expansions}, \emph{repetitions}, and \emph{reflections} of binary labelings. To expand a labeling $\mat{L}_m=[\mb{c}_0^\tr{T},\ld,\mb{c}_{M-1}^\tr{T}]^\tr{T}$ into a labeling $\mat{L}_{m+1}$, we repeat each binary codeword once to obtain a new matrix $[\mb{c}_0^\tr{T},\mb{c}_0^\tr{T},\ld,\mb{c}_{M-1}^\tr{T},\mb{c}_{M-1}^\tr{T}]^\tr{T}$, and then we obtain $\mat{L}_{m+1}$ by appending one extra column $[0,1,1,0,0,1,1,0,\ld,0,1,1,0]^\tr{T}$ of length $2M$ \cite{Agrell04}. To generate a labeling $\mat{L}_{m+1}$ from a labeling $\mat{L}_m=[\mb{c}_0^\tr{T},\ld,\mb{c}_{M-1}^\tr{T}]^\tr{T}$ by repetition, we repeat the labeling $\mat{L}_m$ once to obtain a new matrix $[\mb{c}_0^\tr{T},\ld,\mb{c}_{M-1}^\tr{T},\mb{c}_0^\tr{T},\ld,\mb{c}_{M-1}^\tr{T}]^\tr{T}$, and we add an extra column from the left, consisting of $M$ zeros followed by $M$ ones. Finally, to generate a labeling $\mat{L}_{m+1}$ from a labeling $\mat{L}_m=[\mb{c}_0^\tr{T},\ld,\mb{c}_{M-1}^\tr{T}]^\tr{T}$ by reflection, we join $\mat{L}_m$ and a reversed version of $\mat{L}_m$ to obtain a new matrix $[\mb{c}_0^\tr{T},\ld,\mb{c}_{M-1}^\tr{T},\mb{c}_{M-1}^\tr{T},\ld,\mb{c}_{0}^\tr{T}]^\tr{T}$, and we add an extra column from the left, consisting of $M$ zeros followed by $M$ ones \cite{Agrell04}.

In this paper we are particularly interested in the \emph{binary reflected Gray code} (BRGC)\cite{Gray53,Agrell07}, the \emph{natural binary code} (NBC), and the \emph{folded binary code} (FBC) \cite{Lassing03b}. The FBC was analyzed in \cite{Lassing03b} for uncoded transmission and here we will, to our knowledge for the first time, consider it for coded transmission. In Sec.~\ref{Sec.BICM} and Sec.~\ref{Sec.PAM.PSK} it is shown to yield a higher capacity than other labelings under some conditions. We also introduce a new binary labeling denoted \emph{binary semi-Gray code} (BSGC). These binary labelings are generated as follows:
\begin{itemize}
\item The BRGC $\mat{G}_m$ of order $m\geq 1$ is generated by $m-1$ recursive expansions of the trivial labeling $\mat{L}_1=[0,1]^\tr{T}$, or, alternatively, by $m-1$ recursive reflections of $\mat{L}_1$.
\item The NBC $\mat{N}_m$ of order $m\geq 1$ is defined as the codewords $\mb{c}_i$ that are the base-2 representations of the integers $i=0,\ldots,M-1$, \ie $\mat{N}_m =[\mb{b}(0)^\tr{T},\ld,\mb{b}(M-1)^\tr{T}]^\tr{T}$. Alternatively, $\mat{N}_m$ can be generated by $m-1$ recursive repetitions of the trivial labeling $\mat{L}_1$, or as $m-1$ ordered direct products of $\mat{L}_1$ with itself.
\item The BSGC $\mat{S}_m$ of order $m\geq 3$ is generated by replacing the first column of $\mat{G}_m$ by the modulo-2 sum of the first and last columns.
\item The FBC $\mat{F}_m$ of order $m \geq 2$ is generated by one reflection of $\mat{N}_{m-1}$.
\end{itemize}

For any labeling matrix $\mat{L}=[\mb{c}_0^\tr{T}, \ld ,\mb{c}_{M-1}^\tr{T}]^\tr{T}$, where $\mb{c}_i=[c_{i,0},c_{i,1},\ld,c_{i,m-1}]\in\set{0,1}^m$, we define a \emph{modified labeling matrix} $\mat{Q}=\mat{Q}(\mat{L})$ which is obtained by reversing the order of the columns and applying the mapping 
$(0\rightarrow 1, 1\rightarrow-1)$, \ie 
\eqlab{qki_def}{
  q_{i,k} \triangleq 
  \begin{cases}
    -1, & \textrm{if }c_{i,m-1-k}=1 \\
    +1, & \textrm{if }c_{i,m-1-k}=0
  \end{cases}
}
with $i=0,\ld,M-1$ and $k=0,\ld,m-1$.

\begin{example}[Binary labelings of order $m=3$]\label{ExampleG3N3S3}
\begin{align*}
\mat{G}_{3}=\left[
\begin{array}{ccc}
0&0&0	\\	
0&0&1	\\	
0&1&1	\\	
0&1&0	\\	
1&1&0	\\	
1&1&1	\\	
1&0&1	\\	
1&0&0	\\
\end{array}
\right]
,\,
\mat{N}_{3}=\left[
\begin{array}{ccc}
0&0&0	\\	
0&0&1	\\	
0&1&0	\\	
0&1&1	\\	
1&0&0	\\	
1&0&1	\\	
1&1&0	\\	
1&1&1	\\
\end{array}
\right]
,\,
\mat{S}_{3}=\left[
\begin{array}{ccc}
0&0&0	\\	
1&0&1	\\	
1&1&1	\\	
0&1&0	\\	
1&1&0	\\	
0&1&1	\\	
0&0&1	\\	
1&0&0	\\
\end{array}
\right]
,\,
\mat{F}_{3}=\left[
\begin{array}{ccc}
0&0&0	\\	
0&0&1	\\	
0&1&0	\\	
0&1&1	\\	
1&1&1	\\	
1&1&0	\\	
1&0&1	\\	
1&0&0	\\
\end{array}
\right],
\end{align*}
and
\begin{align*}
\mat{Q}(\mat{N}_{3})=\left[
\begin{array}{rrr}
+1&+1&+1\\	
-1&+1&+1	\\	
+1&-1&+1	\\	
-1&-1&+1	\\	
+1&+1&-1	\\	
-1&+1&-1	\\	
+1&-1&-1	\\	
-1&-1&-1	\\
\end{array}
\right].
\end{align*}
\end{example}

\subsection{Constellations and Input Distributions}\label{Sec.SignalSets}

Throughout this paper, we use $\mc{X}$ to represent the set of symbols used for transmission. Each element of $\mc{X}$ is an $N$-dimensional symbol $\mb{x}_i$, $i=0,\ld,M-1$, where $|\mc{X}|=M=2^{m}$ and $\mc{X}\subset \ms{R}^N$. We define the \emph{input alphabet} using an $M\times m$ matrix $\mat{X}=[\mb{x}_0^\tr{T},\ld,\mb{x}_{M-1}^\tr{T}]^\tr{T}$ which contains all the elements of $\mc{X}$.

For practical reasons, we are interested in well structured input alphabets. An $M$-PAM input alphabet is defined by the column vector $\mat{X}_{\tr{PAM}}$ where $x_{i,1}=-(M-2i-1)$ with $i=0,\ld,M-1$. An $M$-PSK input alphabet is the matrix $\mat{X}_{\tr{PSK}}$ where $\mb{x}_i=\left[\cos\frac{(2i+1)\pi}{M},\sin\frac{(2i+1)\pi}{M}\right]$ with $i=0,1,\ld,M-1$. Finally, a rectangular $(M'\times M'')$-QAM input alphabet is the $M'M'' \times 2$ matrix $\mat{X}_\tr{QAM}=\XX'_\tr{PAM}\otimes \XX''_\tr{PAM}$, where $\XX'_\tr{PAM}$ and $\XX''_\tr{PAM}$ are vectors of length $M'$ and $M''$, respectively.

For a given input alphabet $\XX$, the \emph{input distribution} of the symbols is denoted by the pmf $\PP_{\mb{X}}(\mb{x})$, which represents the probabilities of transmitting the symbols $\mb{x}$, \ie $\Pr\set{\mb{X}=\mb{x}}$. We define the matrix $\mat{P}$ as an ordered list containing the probabilities of the symbols, \ie $\mat{P}\triangleq[P_{\mb{X}}(\mb{x}_0),\ld,P_{\mb{X}}(\mb{x}_{M-1})]^\tr{T}$. We use $\mat{U}_M\triangleq[1/M,\ld,1/M]^\tr{T}$ to denote the discrete uniform input distribution.

We define a \emph{constellation} as the list of matrices $\Omega \triangleq [\mat{X},\mat{L},\mat{P}]$, \ie an input alphabet using a given labeling and input distribution. Finally, for a given pair $[\mat{X},\mat{L}]$, we denote with $\mc{I}_{k,u}\subset\set{0,\ld,M-1}$ the set of indexes of the symbols with a binary label $u\in\set{0,1}$ at bit position $k\in\set{0,\ld,m-1}$, \ie $\mc{I}_{k,u}\triangleq\set{i\in\set{0,\ld,M-1}:c_{i,k}=u}$.

\subsection{System Model}\label{Sec.System.Model}

\begin{figure}[!t]
\psfrag{Transmitter}[lB][lB][0.8]{Transmitter}%
\psfrag{BICM}[lB][lB][0.8]{BICM Channel}%
\psfrag{CM}[lB][lB][0.8]{CM Channel}%
\psfrag{w}[rc][rB][0.8]{${\mb{w}}$}%
\psfrag{cpi}[rc][rB][0.8]{$\un{\mb{c}}^\pi$}%
\psfrag{c}[rc][rB][0.8]{$\un{\mb{c}}$}%
\psfrag{enc}[][][0.8]{ENC}%
\psfrag{pi}[][][0.8]{$\pi$}%
\psfrag{mu}[][][0.8]{$\Phi$}%
\psfrag{sn}[lc][lB][0.8]{$\un{\mb{x}}$}%
\psfrag{hn}[lB][lB][0.8]{$\un{\mb{h}}$}%
\psfrag{zn}[lB][lB][0.8]{$\un{\mb{z}}$}%
\psfrag{yn}[lc][lB][0.8]{$\un{\mb{y}}$}%
\psfrag{mu2}[][][0.8]{$\Phi^{-1}$}%
\psfrag{pi2}[][][0.8]{$\pi^{-1}$}%
\psfrag{dec}[][][0.8]{DEC}%
\psfrag{Lkp}[cc][cB][0.8]{$\un{\mb{l}}$}%
\psfrag{Lk}[cc][cB][0.8]{$\un{\mb{l}}^\pi$}%
\psfrag{Receiver}[rB][rB][0.8]{Receiver}%
\psfrag{wh}[lc][lB][0.8]{${\hat{\mb{w}}}$}%
\begin{center}
	\includegraphics[width=1\columnwidth]{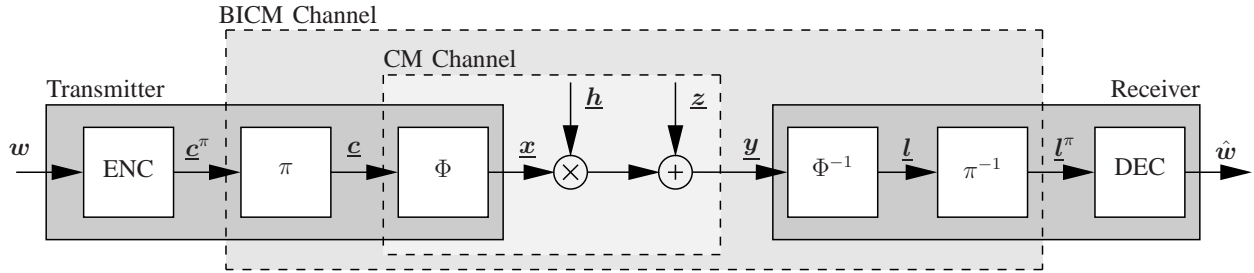}
     \caption{A CM system based on a BICM structure: A binary channel encoder, a bit-level interleaver, a memoryless mapper, the fading channel, and the inverse processes at the receiver side.}
     \label{System_Model_Merged}
\end{center}
\end{figure}

In this paper, we analyze coded modulation schemes (CM) as the one shown in Fig.~\ref{System_Model_Merged}. Each of the $K$ possible messages is represented by the binary vector $\mb{w}\in\set{0,1}^{k_c}$, where $k_c=\log_2{K}$. The transmitter maps each message to a sequence $\un{\mb{x}}=[\mb{x}(0)^\tr{T},\ld,\mb{x}(\Ns-1)^\tr{T}]^\tr{T}\in\mc{X}^{\Ns}$, which corresponds to $\Ns$ $N$-dimensional symbols ($\Ns$ channel uses\footnote{A ``channel use'' corresponds to the transmission of one $N$-dimensional symbol, \ie it can be considered as a ``vectorial channel use''.}). The code $\mbbmss{C}$ is a subset of $\mc{X}^{\Ns}$ such that $|\mbbmss{C}|=K$, which is used for transmission. The transmitter is then defined as a one-to-one function that assigns each information message $\mb{w}$ to one of the $K$ possible sequences $\un{\mb{x}}\in\mbbmss{C}$. The code rate in information bits per coded bits is then given by $R=\dfrac{k_c}{m \Ns}$ or, equivalently, $\Rc=\dfrac{k_c}{\Ns}$ information bits per channel use (information bits per symbol, or information bits per $N$ real dimensions). At the receiver's side, based on the channel observations, a maximum likelihood sequence receiver generates an estimate of the information bits $\mb{\hat{w}}$ selecting the most likely transmitted message.

We consider transmissions over a discrete-time memoryless fast fading channel
\begin{align}\label{Fading_AWGN_channel}
\mb{Y}(t) =\mb{H}(t)\circ\mb{X}(t)+\mb{Z}(t),
\end{align}
where the operator $\circ$ denotes the so-called Schur product (element-wise product) between two vectors, $\mb{X}(t)$, $\mb{H}(t)$, $\mb{Y}(t)$, and $\mb{Z}(t)$ are the underlying random vectors for $\mb{x}(t)$, $\mb{h}(t)$, $\mb{y}(t)$, and $\mb{z}(t)$ respectively, with $t=0,\ld,\Ns-1$ being the discrete time index, and $\mb{Z}(t)$ is a Gaussian noise with zero mean and variance $N_0/2$ in each dimension. The channel is represented by the $N$-dimensional vector $\mb{H}(t)$, and it contains real fading coefficients $H_i$ which are assumed to be random variables, possibly dependent, with same pdf $p_H(h)$. We assume that $\mb{H}(t)$ and $N_0$ are perfectly known at the receiver or can be perfectly estimated. Since the channel is memoryless, from now on we drop the index $t$. 

The conditional transition pdf of the channel in \eqref{Fading_AWGN_channel} is given by
\begin{align}\label{ctpdf_AWGN}
p_{\mb{Y}|\mb{X}=\mb{x},\mb{H}=\mb{h}}(\mb{y}) 	& = \frac{1}{(N_0\pi)^{N/2}}\exp{\biggl(-\frac{\|\mb{y}-\mb{h}\circ\mb{x}\|^2}{N_0}\biggr)}.
\end{align}
We assume that both $\mb{H}$ and $\mb{X}$ have finite and nonzero second moments, that $\mb{X}$, $\mb{H}$, and $\mb{Z}$ are mutually independent, and that there exists a constant $\omega >0$ such that for all sufficiently large $\Delta>0$ the vector $\mb{H}$ satisfies 
\begin{align}\label{H_condition_new}
\Pr\set{\|\mb{H}\|^2>\Delta}	& \leq \exp(-\Delta^\omega).
\end{align}
This condition will be used in the proof of Theorem~\ref{PrelovTheo3} in Sec.~\ref{FOOasymptotics}.

Each transmitted symbol conveys $\Rc$ information bits and thus, the relation between the average symbol energy $\Es\triangleq\ms{E}_{\mb{X}}[\|\mb{X}\|^2]$ and the average information bit energy $\Eb$ is given by $\Es=\Rc\Eb$. We define the average signal-to-noise ratio (SNR) as
\begin{align}\label{EsN0_and_EbN0}
\SNR \triangleq\frac{\ms{E}_{\mb{H},\mb{X}}[\|\mb{H}\circ\mb{X}\|^2]}{N_0} = \ms{E}_{H}[H^2]\frac{\Es}{N_0}= \ms{E}_{H}[H^2]\Rc\frac{\Eb}{N_0}.
\end{align}

The AWGN channel is obtained as a special case of \eqref{Fading_AWGN_channel} by taking $\mb{H}$ as the all-one vector. Another particular case is obtained when $H_0=H_1=\ldots =H_{N-1}=A$, which particularizes to the Rayleigh fading channel when $A=\sqrt{A_1^2+A_2^2}$ and $A_1,A_2$ are independent zero-mean Gaussian random variables. In this case, the instantaneous SNR defined by $\ms{E}_{\mb{X}}[\|\mb{H}\circ\mb{X}\|^2]/ N_0=A^2\Es/N_0$ follows a chi-square distribution with one degree of freedom (an exponential distribution). Similarly, the Nakagami-$m$ fading channel is obtained when $A$ follows a Nakagami-$m$ distribution. It can be shown that the condition \eqref{H_condition_new} is fulfilled in all the cases above.

In a BICM system \cite{Zehavi92,Caire98}, the transmitter in Fig.~\ref{System_Model_Merged} is realized using a serial concatenation of a binary encoder of rate $R=\Rc/m$, a bit level interleaver, and a memoryless mapper $\Phi$. The mapper $\Phi$ is defined as a one-to-one mapping rule that maps the length-$m$ binary random vector $\mb{C}=[C_0,\ld,C_{m-1}]$ to one symbol $\mb{X}$, \ie $\Phi:\set{0,1}^{m}\rightarrow \mc{X}$. At the receiver's side, the demapper computes soft information on the coded bits, which are then deinterleaved and passed to the channel decoder. The a posteriori L-values for the $k$th bit in the symbol and for a given fading realization are given by
\begin{align}
l_{k}(\mb{y})
		& \triangleq  \log_\tr{e}\frac{\Pr\set{\mb{Y}=\mb{y}|C_k=1,\mb{H}=\mb{h}}}{\Pr\set{\mb{Y}=\mb{y}|C_k=0,\mb{H}=\mb{h}}}		\label{LLR2}\\
		& =  \sum_{u\in\set{0,1}}(-1)^{u+1} \log_\tr{e}{\sum_{i \in\mc{I}_{k,u}} \exp\left(-\frac{\|\mb{y}-\mb{h}\circ\mb{x}_i\|^2}{N_0}\right) } \label{LLR3},\\
		& \approx  \frac{1}{N_0}\sum_{u\in\set{0,1}}(-1)^{u} 
\min_{i \in\mc{I}_{k,u}}\|\mb{y}-\mb{h}\circ\mb{x}_i\|^2 \label{LLR4},
\end{align}
where to pass from \eqref{LLR3} to \eqref{LLR4}, the so-called max-log \cite{Viterbi98} approximation was used.

The max-log metric in \eqref{LLR4} (already proposed in\cite{Zehavi92,Caire98}) is suboptimal; however, it is very popular in practical implementations because of its low complexity, e.g., in the 3rd generation partnership project (3GPP) working groups \cite{Tsgr1_15_1093}. It is also known that when Gray-labeled constellations are used, the use of this simplification results in a negligible impact on the receiver's performance \cite[Fig.~9]{Classon02} \cite[Fig.~6]{Alvarado06c}. The max-log approximation also allows BICM implementations which do not require the knowledge of $N_0$, for example, when a Viterbi decoder is used, or when the demapper passes hard decisions to the decoder. Moreover, the use of the max-log approximation transforms the nonlinear relation $l_{k}(\mb{y})$ in \eqref{LLR3} into a piecewise linear relation. This has been used to develop expressions for the pdf of the L-values in \eqref{LLR4} using arbitrary input alphabets \cite{Szczecinski08b} (based on an algorithmic approach), closed-form expressions for QAM input alphabets labeled by the BRGC for the AWGN channel \cite{Benjillali06b,Alvarado07d}, and for fading channels \cite{Szczecinski07f}. Recently, closed-form approximations for the pdf of the L-values in \eqref{LLR4} for arbitrary input alphabets and binary labeling in fading channels have been presented \cite{Kenarsari10}.

\subsection{The Hadamard Transform}

The Hadamard transform (HT) is a discrete, linear, orthogonal transform, like for example the Fourier transform, but its coefficients take values in $\pm 1$ only. Among the different applications that the HT has, one that is often overlooked is as an analysis tool for binary labelings \cite{Knagenhjelm96,Khan10}. The HT is defined by means of an $M \times M$ matrix, the Hadamard matrix, which is defined recursively as follows when $M$ is a power of two \cite[pp.~53--54]{MacWilliams77}.
\eq{
  \mat{H}_1 \triangleq 1 
  \qquad
  \mat{H}_{2M} \triangleq \begin{bmatrix}
    \mat{H}_M & \mat{H}_M \\
    \mat{H}_M & -\mat{H}_M
  \end{bmatrix}, \qquad M \ge 1
}

\begin{example}[Hadamard matrix $\mat{H}_8$]\label{ExampleH8}
\eqlab{h8}{
  \mat{H}_8 = \left[
\begin{array}{rrrrrrrr}
    +1& +1& +1& +1& +1& +1& +1& +1 \\
    +1& -1& +1&-1& +1&-1& +1&-1 \\
    +1& +1& -1&-1& +1& +1&-1&-1 \\
    +1& -1& -1& +1& +1&-1&-1& +1 \\
    +1& +1& +1& +1&-1&-1&-1&-1 \\
    +1& -1& +1&-1&-1& +1&-1& +1 \\
    +1& +1& -1&-1&-1&-1& +1& +1 \\
    +1& -1& -1& +1&-1& +1& +1&-1
  \end{array}
  \right]
.}
\end{example}

In the following, we will drop the index, letting $\mat{H}$ represent a Hadamard matrix of any size $M=2^m$. Hadamard matrices have the following appealing properties.
\eqlab{hinv}{
  \mat{H}^\tr{T} = \mat{H}, \quad   \mat{H}^{-1} = \frac{1}{M} \mat{H}
.}
It can be shown \cite[Sec.~1.1]{Sutter91}\cite[Sec.~III]{Pratt69} that the elements of a Hadamard matrix are $h_{i,j} = \prod_{k=0}^{m-1} (-1)^{b_k(i) b_k(j)}$, from which we observe for future use that for all $i=0,\ldots,M-1$ and $l=0,\ldots,m-1$,
\eqlab{p2}{
  h_{i,0} = 1, \qquad
  h_{i,2^l} = \prod_{k=0}^{m-1} (-1)^{b_k(i) b_k(2^l)}= (-1)^{b_l(i)}
,}
where $b_l(i)$ is the $l$th bit of the base-2 representation of the integer $i$.

At this point it is interesting to note the close relation between the columns of the matrix $\mat{Q}(\mat{N}_3)$ in Example~\ref{ExampleG3N3S3} and the columns $2^l$ of $\mat{H}_8$ in \eqref{h8} for $l=0,1,2$. Its generalization is given by the following lemma, whose proof follows immediately from \eqref{qki_def}, the definition of the NBC in Sec.~\ref{Sec.Preliminaries.Labelings}, and \eqref{p2}.
\begin{lemma}\label{q_NBC}
Let $\mat{Q}=\mat{Q}(\NN_m)$ be the modified labeling matrix for the NBC of order $m$, and let $\mat{H}$ be the Hadamard matrix. For any $m$, and for $k=0,\ldots,m-1$ and $i=0,\ld, M-1$,
\begin{align}\label{xqv}
q_{i,k}=h_{i,2^k}.
\end{align}
\end{lemma}

The HT operates on a vector of length $M=2^m$, for any integer $m$, or in a more general case, on a matrix with $M=2^m$ rows. The transform of a matrix $\mat{X}$ is denoted $\tilde{\mat{X}}$ and has the same dimensions as $\mat{X}$. It is defined as
\eqlab{HT}{
  \tilde{\mat{X}} \triangleq \frac{1}{M}\mat{H} \mat{X}
}
and the inverse transform is $\mat{X} = \mat{H} \tilde{\mat{X}}$. Equivalently, 
\eqlab{p3}{
  \tilde{\mb{x}}_j = \frac{1}{M} \sum_{i=0}^{M-1} h_{j,i} \mb{x}_i, \qquad \mb{x}_i = \sum_{j=0}^{M-1} h_{i,j} \tilde{\mb{x}}_j,
}
where from \eqref{hinv} we have that $h_{j,i}=h_{i,j}$, and where we have introduced the row vectors $\mb{x}_i$ and $\tilde{\mb{x}}_j$ such that
\eq{
\mat{X} = \bigl[
\mb{x}_0^\tr{T}, \ld , \mb{x}_{M-1}^\tr{T}
\bigr]^\tr{T},
\qquad
\tilde{\mat{X}} = \bigl[
\tilde{\mb{x}}_0^\tr{T}, \ld , \tilde{\mb{x}}_{M-1}^\tr{T}
\bigr]^\tr{T}
.}
Because of \eqref{p2}, the first element of the transform is simply $\tilde{\mb{x}}_0 = \frac{1}{M} \sum_{i=0}^{M-1}\mb{x}_i$.

Finally, using $\sum_{j=0}^{M-1} \|\tilde{\mb{x}}_j\| ^2=\tr{trace}\bigl(\tilde{\mat{X}}^\tr{T}\tilde{\mat{X}}\bigr)$, \eqref{HT}, and \eqref{hinv}, we note that a variant of Parseval's theorem holds:
\eqlab{p5}{
  \sum_{j=0}^{M-1} \|\tilde{\mb{x}}_j\| ^2 = 
  \frac{1}{M} \sum_{i=0}^{M-1} \|\mb{x}_i\| ^2
.}

\section{Capacity of Coded Modulation Systems}\label{Sec.Capacity_CM_Systems}

In this section we analyze the capacity of CM schemes, \ie the so-called CM and BICM capacities. We review their relation and we analyze how the selection of the constellation influences them. We pay special attention to the selection of the binary labeling and the use of probabilistic shaping for BICM.

\subsection{AMI and Channel Capacity}

In this subsection, we assume the use of a continuous input alphabet, \ie $\mc{X}=\ms{R}^N$, which upperbounds the performance of finite input alphabets.

The \emph{average mutual information} (AMI) in bits\footnote{Throughout this paper all the AMIs are given in bits.} per channel use between the random vectors $\mb{X}$ and $\mb{Y}$ when the channel is perfectly known at the receiver is defined as
\begin{align}
I_{\bX}(\bX;\bY) 	& \triangleq \ms{E}_{\bX,\bY}\left[\log_2{\frac{p_{\bX,\bY}(\bX,\bY)}{p_{\bY}(\bY)p_{\bX}(\bX)}}\right]\label{AMIdefinition1}\\
				& = \ms{E}_{\bX,\bY}\left[\log_2{\frac{p_{\bY|\bX}(\bY)}{p_{\bY}(\bY)}}\right] ,
\end{align}
where we use $\bX$ as the index of $I_{\bX}(\bX;\bY)$ to emphasize the fact that the AMI depends on the input PDF $p_{\bX}(\bx)$. 
For an arbitrary channel parameter $\bH$, the AMI in \eqref{AMIdefinition1} can be expressed as\footnote{We note that the AMI with perfect channel state information is usually denoted by $I_{\mb{X}}(\mb{X};\mb{Y}|\mb{H})$, however, and for notation simplicity, we use $I_{\mb{X}}(\mb{X};\mb{Y})$.}
\begin{align}
I_{\bX}(\bX;\bY) 	& = \ms{E}_{\bX,\bY,\bH}\left[\log_2{\frac{p_{\bY|\bX,\bH}(\bY)}{p_{\bY|\bH}(\bY)}}\right], 	\label{AMIdefinition2}
\end{align}
where $p_{\bY|\bX=\bx,\bH=\bh}(\by)$ is given by \eqref{ctpdf_AWGN}.

The \emph{channel capacity} of a continuous-input continuous-output memoryless channel is defined as the maximum AMI between its input and output \cite[Ch.~4]{Cover06_Book}\cite[eq.~(3)]{Cover02}
\begin{align}\label{Channel_Capacity}
\mf{C}\left(\SNR\right) 	& \triangleq \max_{p_{\mb{X}}(\mb{x})} I_{\mb{X}}(\mb{X};\mb{Y}),
\end{align}
where the maximization is over all possible input distributions. The capacity in \eqref{Channel_Capacity} has units of [bit/channel use] (or equivalently [bit/symbol]), and it is an upper bound on the number of bits per symbol that can be reliably transmitted through the channel, where a symbol consists of $N$ real dimensions. Shannon's channel coding theorem states that it is not possible to transmit information reliably above this fundamental limit, \ie
\begin{align}\label{Capacity_Inequality}
\Rc \leq  \mf{C}\left(\SNR\right) =  \mf{C}\left(\Rc \ms{E}_{H}[H^2] \frac{\Eb}{N_0}\right).
\end{align}

The \emph{AWGN capacity}, denoted by $\mf{C}^\tr{AW}\left(\SNR\right)$, is defined as the channel capacity of the AWGN channel (obtained from \eqref{Fading_AWGN_channel} using $\mb{H}(t)=\mb{1}$), and it is given by \cite[Sec.~9.4]{Cover06_Book}
\begin{align}\label{AWGN_Capacity}
\mf{C}^\tr{AW}\left(\SNR\right) 	& = \frac{N}{2}\log_2 \left(1+\frac{2}{N} \SNR \right).
\end{align}
This capacity is attained when $\mb{X}$ are i.i.d. zero-mean Gaussian random variables with variance $\Es/N$ in each dimension and it follows from the fact that the noise is independent in each dimension, and thus, the transmission of $\mb{X}$ can be considered as a transmission through $N$ parallel independent Gaussian channels.  

We define the \emph{conditional AMI} for discrete input alphabets as the AMI between $\mb{X}$ and $\mb{Y}$ conditioned on the outcome of a third random variable $U$, \ie
\begin{align}
I_{\mb{X}|U=u}(\mb{X};\mb{Y}) 	& \triangleq \ms{E}_{\bX,\bY|U=u}\left[\log_2{\frac{p_{\bY|\bX,U=u}(\bY)}{p_{\bY|U=u}(\bY)}}\right]\\
						& = \ms{E}_{\mb{X},\mb{Y},\mb{H}|U=u}\left[\log_2{\frac{p_{\mb{Y}|\mb{X},\mb{H},U=u}(\mb{Y})}{p_{\mb{Y}|\mb{H},U=u}(\mb{Y})}}\right]
\label{AMIdefinition5},
\end{align}
which is valid for any random $\bH$.

\subsection{CM Capacity}

The \emph{CM capacity} is defined as the AMI between $\mb{X}$ and $\mb{Y}$ for a given constellation $\Omega$, \ie
\begin{align}
\mf{I}^{\tr{CM}}_\Omega\left(\SNR\right) 	& \triangleq I_{\mb{X}}(\mb{X};\mb{Y}) 	\label{CMCapacity1} \\
									& = I_{\mb{X}}(\mb{C};\mb{Y})			\label{CMCapacity2}	\\
									& = \sum_{k=0}^{m-1}I_{\mb{X}}(C_k;\mb{Y}|C_0,\ld,C_{k-1}), \label{CMCapacity3} 
\end{align}
where to pass from \eqref{CMCapacity1} to \eqref{CMCapacity2}, we used the fact that the mapping rule between $\mb{C}$ and $\mb{X}$ is one-to-one. To pass from \eqref{CMCapacity2} to \eqref{CMCapacity3} we have used the chain rule of mutual information \cite[Sec.~2.5]{Cover06_Book}, where $I_{\mb{X}}(C_k;\mb{Y}|C_0,\ld,C_{k-1})$ represents a \emph{bit level} AMI which represents the maximum rate that can be used at the $(k+1)$th bit position, given a perfect knowledge of the previous $k$ bits.

The CM capacity in \eqref{CMCapacity1} corresponds to the capacity of the memoryless ``CM channel'' in Fig.~\ref{System_Model_Merged} for a given constellation $\Omega$. We note that different binary labelings will produce different values of $I_{\mb{X}}(C_k;\mb{Y}|C_0,\ld,C_{k-1})$ in \eqref{CMCapacity3}; however, the overall sum will remain constant, \ie the CM capacity does not depend on the binary labeling. We use the name ``CM capacity'' for $\mf{I}^{\tr{CM}}_\Omega\left(\SNR\right)$ in \eqref{CMCapacity1} following the standard terminology\footnote{Sometimes, this is also called joint capacity \cite{Barsoum07}, or (constellation) constrained capacity \cite{Schreckenbach07_Thesis,Brannstrom09}.} used in the literature (\cf \cite{Caire98,Martinez08b,Martinez09,Stierstorfer09a,Stierstorfer09_Thesis}), although we recognize the misusage of the word capacity since no optimization over the input distribution is performed (cf.~\eqref{Channel_Capacity}). Moreover, it is also possible to optimize the input alphabet in order to obtain an increase in the AMI (so-called signal shaping \cite{Fischer02_Book}). Nevertheless, throughout this paper we will refer to the AMI for a given $\Omega$ in \eqref{CMCapacity1} as the CM capacity.

In this paper we are interested in optimal constellations, and therefore, we define the \emph{maximum CM capacity} as
\begin{align}
\Imax^\tr{CM}\left(\SNR\right)	& \triangleq \max_{\Omega} \mf{I}^{\tr{CM}}_\Omega\left(\SNR\right) \label{CMCapacity4} \\
							& =  \max_{[\mat{X},\mat{P}]} \sum_{k=0}^{m-1}I_{\mb{X}}(C_k;\mb{Y}|C_0,\ld,C_{k-1}) \label{CMCapacity5}.
\end{align}
As mentioned before, the CM capacity \emph{does not} depend on the binary labeling, \ie it does not depend on how the mapping rule $\Phi$ is implemented, and therefore, in \eqref{CMCapacity5} we only show two optimization parameters: the input alphabet and the input distribution. 

The CM capacity in \eqref{CMCapacity1} (for a given constellation $\Omega$) is an upper bound on the number of bits per symbol that can be reliably transmitted using for example TCM \cite{Ungerboeck82} or MLC with multistage decoding (MLC-MSD)\cite{Imai77,Wachsmann99}. MLC-MSD is in fact a direct application of the summation in \eqref{CMCapacity3}, \ie $m$ parallel encoders are used, each of them having a rate $R_{k}=I_{\mb{X}}(C_k;\mb{Y}|C_0,\ld,C_{k-1})$. At the receiver's side, the first bit level is decoded and the decisions are passed to the second decoder, which then passes the decisions to the third decoder, and so on. Other design rules can also be applied in MLC, \cf \cite{Wachsmann99}. The maximum CM capacity $\Imax^\tr{CM}\left(\SNR\right)$ in \eqref{CMCapacity5} represents an upper bound on the number of bits per symbol that can be reliably transmitted using a fully optimized system, \ie a system where for each SNR value $\SNR$, the input alphabet and the input distribution are selected in order to maximize the CM capacity $\mf{I}^{\tr{CM}}_\Omega\left(\SNR\right)$.

\subsection{BICM with Arbitrary Input Distributions}\label{Sec.BICM-NU} %

It is commonly assumed that the sequence generated by the binary encoder in Fig.~\ref{System_Model_Merged} is infinitely long and symmetric, and also that the interleaver ($\pi$) operates over this infinite sequence, simply permuting it in a random way. Under these standard assumptions, it follows that the input symbol distribution will be always $\mat{P}=\mat{U}_M$. Since in this paper we are interested in analyzing a more general setup where the input symbol distribution can by modified, we develop a more general model in which we relax the equiprobable input distribution assumption.

Let $C_k\in\set{0,1}$ the binary random variable representing the bits at the $k$th modulator's input, where the pmf $P_{C_k}(u)$ represents the probability of transmitting a bit $u$ at bit position $k$. We assume that in general $\sum_{k=0}^{m-1}P_{C_k}(0)\neq \sum_{k=0}^{m-1}P_{C_k}(1)$, \ie the coded and interleaved sequence could have more zeros than ones (or vice-versa). Note that since $P_{C_k}(u)$ is a pmf, $P_{C_k}(0)+P_{C_k}(1)=1$.

Let $\mb{c}_i=[c_{i,0},\ld,c_{i,m-1}]$ be the binary label of the symbol $\mb{x}_i$. We assume that the bits at the input of the modulator are independent, and therefore, the input symbol probabilities are
\begin{align}\label{Pxuncond}
P_{\mb{X}}(\mb{x}_i)=\prod_{k=0}^{m-1}P_{C_k}(c_{i,k}).
\end{align}

The independence condition on the coded bits that results in \eqref{Pxuncond} can be obtained if the interleaver block in Fig.~\ref{System_Model_Merged} completely breaks the temporal correlation of the coded bits. The condition that the coded and interleaved sequence could be asymmetric can be obtained for example by using an encoder with nonuniform outputs, or by a particular puncturing scheme applied to the coded bits. This can be combined with the use of multiple interleavers and multiplexing \cite{Alvarado09c}, which would allow $P_{C_k}(u)\neq 1/2$. Examples of how to construct a BICM scheme where nonuniform input symbol distributions are obtained include the ``shaping encoder'' of \cite{Legoff05,LeGoff07} and the nonuniform signaling scheme based on a Huffman code of \cite{Raphaeli04}.

For future use, we also define the conditional input symbol probabilities, conditioned on the $k$th bit being $u$, as
\begin{align}
\nonumber
P_{\mb{X}|C_k=u}(\mb{x}_i)
&=
\begin{cases}
\displaystyle{\prod_{\substack{k'=0\\k'\neq k}}^{m-1}P_{C_{k'}}(c_{i,k'}),} & \text{if $c_{i,k}=u$}\\
0, & \text{if $c_{i,k}\neq u$}\\
\end{cases}\\
&=
\begin{cases}
\ds{\frac{P_{\mb{X}}(\mb{x}_i)}{P_{C_k}(u)}}, & \text{if $i\in \mc{I}_{k,u}$}\\
0, & \text{if $i\notin \mc{I}_{k,u}$}\\
\end{cases} \label{Pxcond},
\end{align}
where $\mc{I}_{i,k}$ is defined in Sec.~\ref{Sec.SignalSets}.

\subsection{BICM Capacity}\label{Sec.BICM}

The ``BICM channel'' in Fig.~\ref{System_Model_Merged} was introduced in \cite{Martinez06} and it is what separates the encoder and decoder in a BICM system. The \emph{BICM capacity} is then defined as the capacity of the BICM channel. Using the definitions in Sec.~\ref{Sec.BICM-NU} and the equivalent channel model in \cite[Fig.~6]{Caire98}, which replaces the BICM channel by $m$ parallel binary-input continuous-output channels, the BICM capacity for a given constellation $\Omega$ is defined as
\begin{align}
\mf{I}^\tr{BI}_{\Omega}\left(\SNR\right) 	& \triangleq  \sum_{k=0}^{m-1}I_{C_k}(C_k;\mb{Y}) \label{BICMCapacity1}\\
	& = \sum_{k=0}^{m-1}\ms{E}_{C_k,\mb{H},\mb{Y}}\left[\log_2{\frac{p_{\mb{Y}|C_k,\mb{H}}(\mb{Y})}{p_{\mb{Y}|\mb{H}}(\mb{Y})}}\right] \label{BICMCapacity2}\\
	& = \sum_{k=0}^{m-1}\sum_{u\in\set{0,1}}P_{C_k}(u)\ms{E}_{\mb{H},\mb{Y}|C_k=u}\left[\log_2{\frac{p_{\mb{Y}|\mb{H},C_k=u}(\mb{Y})}{p_{\mb{Y}|\mb{H}}(\mb{Y})}}\right] \label{BICMCapacity3}\\
	& = \int_{\ms{R}^N} p_{\mb{H}}(\mb{h}) \sum_{k=0}^{m-1}\sum_{u\in\set{0,1}}\sum_{i\in\mc{I}_{k,u}} \PP_{\mb{X}}(\mb{x}_i) \int_{\ms{R}^N}p_{\mb{Y}|\mb{X}=\mb{x}_i,\mb{H}=\mb{h}}(\mb{y}) \cd \nonumber \\
& \qquad \qquad \qquad \log_2{ \frac{ \frac{1}{P_{C_k}(u)}\sum_{j\in\mc{I}_{k,u}}P_{\mb{X}}(\mb{x}_j) p_{\mb{Y}|\mb{X}=\mb{x}_j,\mb{H}=\mb{h}}(\mb{y}) }{ \sum_{\mb{x}\in\mc{X}} \PP_{\mb{X}}(\mb{x}) p_{\mb{Y}|\mb{X}=\mb{x},\mb{H}=\mb{h}}(\mb{y}) }}\, d\mb{y}\, d\mb{h},	\label{BICMCapacity4}		\end{align}
where \eqref{BICMCapacity4} follows from \eqref{BICMCapacity3}, \eqref{Pxcond}, and the fact that the value of $C_k$ does not affect the conditional channel transition probability, \ie $p_{\mb{Y}|\mb{X}=\mb{x},\mb{H}=\mb{h},C_k=u}(\mb{y})=p_{\mb{Y}|\mb{X}=\mb{x},\mb{H}=\mb{h}}(\mb{y})$.  The BICM capacity in \eqref{BICMCapacity4} is a general expression that depends on all the constellation parameters $\Omega$. This can be numerically implemented using Gauss--Hermite quadratures, or alternatively, by using a one-dimensional integration based on the pdf of the L-values developed in \cite{Szczecinski08b, Alvarado07d,Szczecinski07f,Kenarsari10}. 

The AMIs $I_{C_k}(C_k;\mb{Y})$ in \eqref{BICMCapacity1} are, in contrast to the ones in \eqref{CMCapacity5}, not conditioned on the previous bit values. Because of this, and unlike the CM capacity, the binary labeling strongly affects the BICM capacity $\mf{I}^\tr{BI}_{\Omega}\left(\SNR\right)$ in \eqref{BICMCapacity1}. Note that the BICM capacity is equivalent to the capacity achieved by MLC with (suboptimal) parallel decoding of the individual bit levels, because in BICM, the bits are treated as independent \cite{Wachsmann99}. The differences are that BICM uses only one encoder, and that in BICM the equivalent channels are not used in parallel, but time multiplexed. Again, following the standard terminology\footnote{It is also called parallel decoding capacity in \cite{Barsoum07}, or receiver constrained capacity in \cite{Schreckenbach07_Thesis}.} used in the literature (\cf \cite{Caire98,Martinez08b,Martinez09,Stierstorfer09a,Stierstorfer09_Thesis}), we use the name ``BICM capacity'' even though no optimization over the input distribution is performed.

If all the bits at the input of the modulator are equally likely, \ie $\PP_{C_k}(u)=1/2$ for $k=0,\ld,m-1$ and $u\in\set{0,1}$, we obtain from \eqref{Pxuncond} $\PP_{\mb{X}}(\mb{x})=1/M$. Under these constraints, and assuming an AWGN channel ($\mb{H}=\mb{1}$), the BICM capacity in \eqref{BICMCapacity4} is given by
\begin{align}
\mf{I}_{\Omega}^\tr{BI}\left(\SNR\right)	& = \frac{1}{M}\sum_{k=0}^{m-1}\sum_{u\in\set{0,1}} \sum_{i\in\mc{I}_{k,u}} \int_{\ms{R}^N} p_{\mb{Y}|\mb{X}=\mb{x}_i}(\mb{y})\log_2{\frac{ 2\sum_{j\in\mc{I}_{k,u}}p_{\mb{Y}|\mb{X}=\mb{x}_j}(\mb{y}) }{\sum_{\mb{x}\in\mc{X}}p_{\mb{Y}|\mb{X}=\mb{x}}(\mb{y})}} \,d\mb{y}\label{BICMCapacity6},
\end{align}
where the constellation is $\Omega=[\mat{X},\mat{L},\mat{U}_M]$. This expression coincides  with the ``standard'' BICM capacity formula (\cf \cite[Sec.~3.2.1]{Fabregas08_Book}, \cite[eq.~(15)]{Caire98}, \cite[eq.~(11)]{Martinez09}).

One relevant question here is what is the optimum labeling from a capacity maximization point of view. Once this question is answered, approaching the fundamental limit will depend only on a good design of the channel encoder/decoder. Caire \emph{et al.} conjectured the optimality of the BRGC, which, as the next example shows, is not correct at all SNR. This was first disproved in \cite{Stierstorfer07a} for PAM input alphabets based on an exhaustive search of binary labelings up to $M=8$.

\begin{figure}[!t]
\begin{center}
\begin{tabular}{@{}c@{}}%
\psfrag{xlabel}[cc][cB][0.70]{$\SNR$~[dB]}%
\psfrag{ylabel}[cc][ct][0.70]{$\Rc$~[bit/symbol]}%
\psfrag{Gaussian-In}[l][l][0.55]{$\mf{C}^\tr{AW}(\SNR)$}%
\psfrag{CM-BRGC}[l][l][0.55]{$\mf{I}_{\Omega}^\tr{CM}(\SNR)$}%
\psfrag{BICM-BRGC-C}[l][l][0.55]{$\mf{I}^\tr{BI}_{\Omega}(\SNR)$ (BRGC)}%
\psfrag{BICM-NBC-C}[l][l][0.55]{$\mf{I}^\tr{BI}_{\Omega}(\SNR)$ (NBC)}%
\psfrag{BICM-FBC-C}[l][l][0.55]{$\mf{I}^\tr{BI}_{\Omega}(\SNR)$ (FBC)}%
\psfrag{BICM-BSGC-C}[l][l][0.55]{$\mf{I}^\tr{BI}_{\Omega}(\SNR)$ (BSGC)}%
\psfrag{kcleq}[b][r][0.55]{$\Rc\leq  \mf{I}^\tr{BI}_{\Omega}\left(\SNR\right)$}%
\includegraphics[width=0.50\columnwidth]{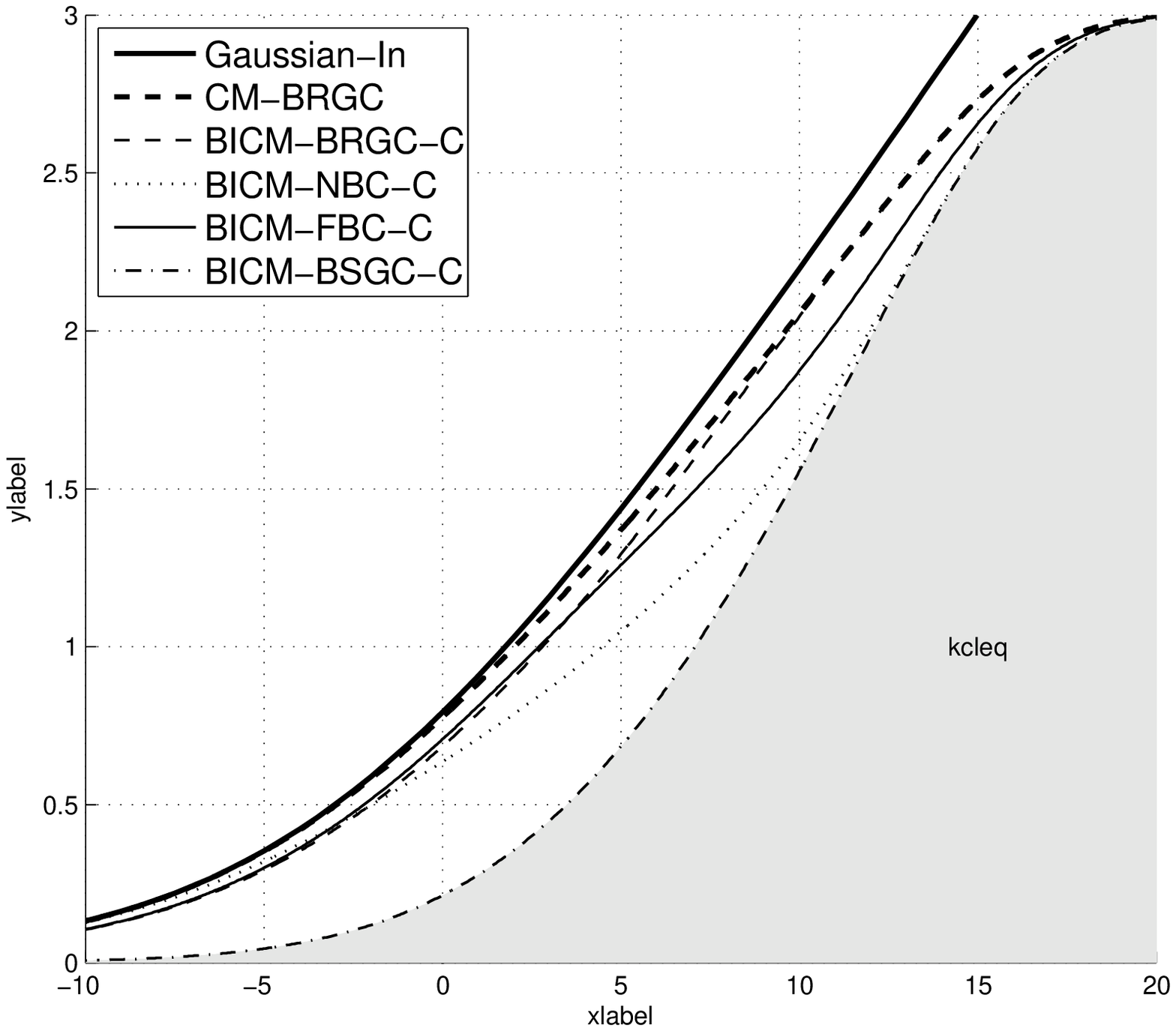}%
\\{\footnotesize (a)}%
\end{tabular}\hfill\begin{tabular}{@{}c@{}}%
\psfrag{xlabel}[cc][cB][0.70]{${\Eb}/{N_0}$~[dB]}%
\psfrag{ylabel}[cc][ct][0.70]{$\Rc$~[bit/symbol]}%
\psfrag{Gaussian-In}[l][l][0.55]{$f^\tr{AW}(\Rc)$}%
\psfrag{CM-BRGC}[l][l][0.55]{$f_\Omega^\tr{CM}(\Rc)$}%
\psfrag{BICM-BRGC}[l][l][0.55]{$f_\Omega^\tr{BI}(\Rc)$ (BRGC)}%
\psfrag{BICM-NBC}[l][l][0.55]{$f_\Omega^\tr{BI}(\Rc)$ (NBC)}%
\psfrag{BICM-FBC}[l][l][0.55]{$f_\Omega^\tr{BI}(\Rc)$ (FBC)}%
\psfrag{BICM-BSGC}[l][l][0.55]{$f_\Omega^\tr{BI}(\Rc)$ (BSGC)}%
\psfrag{kcleq}[l][l][0.55]{$\dfrac{\Eb}{N_0} \geq \dfrac{ \mf{C}^{-1}(\Rc)}{\Rc}$}%
\psfrag{Constant-Eb}[l][l][0.55]{Constant $\Eb/N_0$}%
\psfrag{EbNo03}[l][l][0.8][90]{$\Eb/N_0=7~\tr{dB}$}
\psfrag{SL}[l][l][0.55]{$-1.59$~dB}%
\includegraphics[width=0.50\columnwidth]{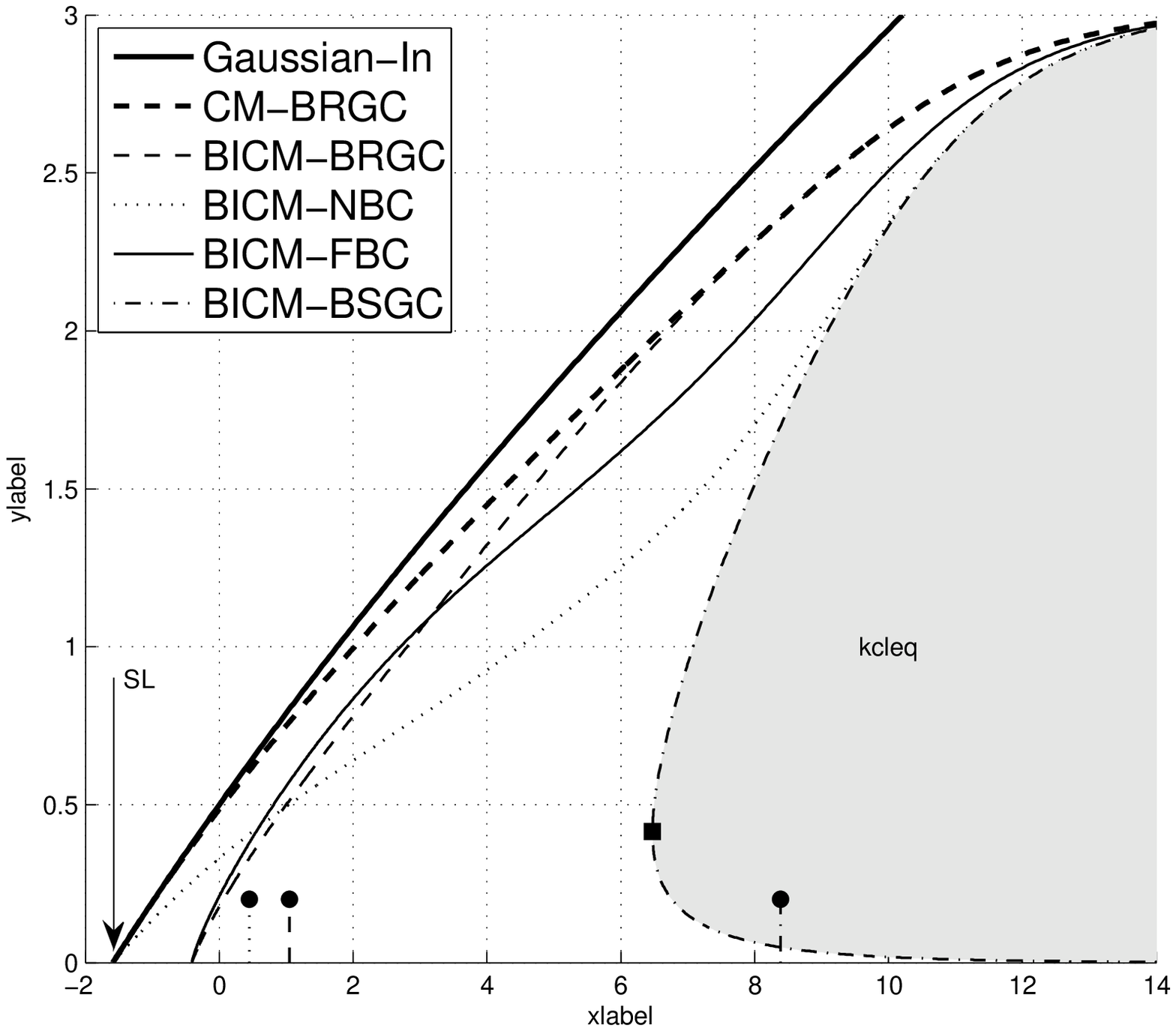}%
\\{\footnotesize (b)}%
\end{tabular}%
\psfrag{ylabel}[c][t][0.7]{$g(\Rc)$}%
\psfrag{xlabel}[c][b][0.7]{$\Rc$~[bit/symbol]}%
\psfrag{Gaussian-In}[l][l][0.55]{$g^\tr{AW}(\Rc)$}%
\psfrag{CM-BRGC}[l][l][0.55]{$g_\Omega^\tr{CM}(\Rc)$}%
\psfrag{BICM-BRGC}[l][l][0.55]{$g_\Omega^\tr{BI}(\Rc)$ (BRGC)}%
\psfrag{BICM-NBC}[l][l][0.55]{$g_\Omega^\tr{BI}(\Rc)$ (NBC)}%
\psfrag{BICM-FBC}[l][l][0.55]{$g_\Omega^\tr{BI}(\Rc)$ (FBC)}%
\psfrag{BICM-BSGC}[l][l][0.55]{$g_\Omega^\tr{BI}(\Rc)$ (BSGC)}%
\begin{center}
\includegraphics[width=0.50\columnwidth]{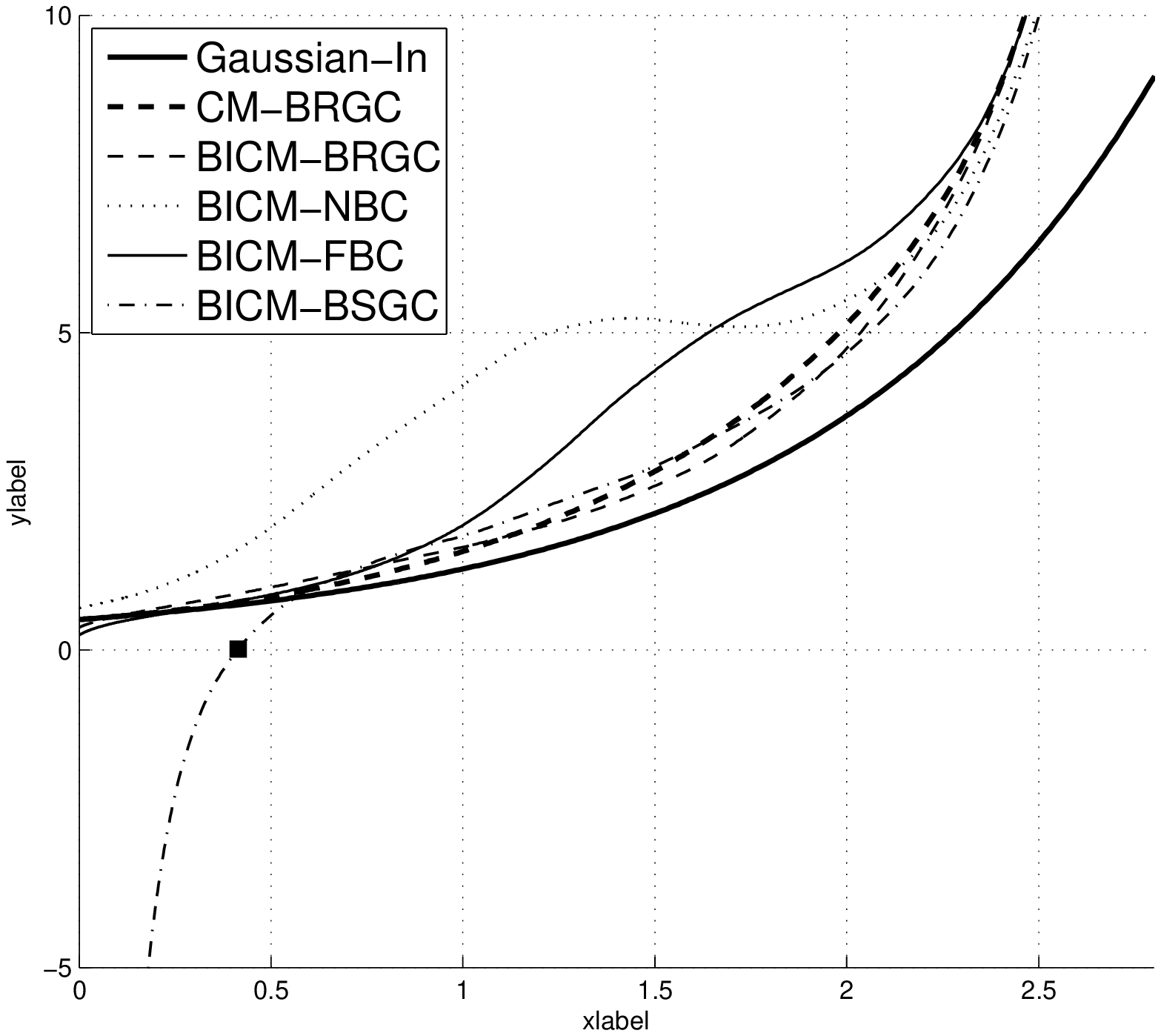}%
\\{\footnotesize (c)}%
\end{center}
\caption{CM capacity and BICM capacity for 8-PAM with $\mat{U}_8$ using the four labelings defined in Sec.~\ref{Sec.Preliminaries.Labelings}: plotted vs.~$\SNR$ (a) and $\Eb/N_0$ (b), and their corresponding functions $g(\Rc)$ (c). The shadowed regions represent the achievable rates using the BSGC. The black squares represent the minimum ${\Eb}/{N_0}$ for the BSGC. The black circles represents the ${\Eb}/{N_0}$ needed for a rate $R = 1/15$ turbo code to reach $\tr{BER}=10^{-6}$ (\cf Sec.~\ref{Sec.TurboExamples}).}
    \label{AW_CM_BI_Capacities_8PAM_BRGC_NBC_BSGC}
\end{center}
\end{figure}
\begin{example}[CM and BICM Capacities for the AWGN channel, $8$-PAM, and $\mat{U}_8$]\label{Example-CM-BICM-8PAM}
In Fig.~\ref{AW_CM_BI_Capacities_8PAM_BRGC_NBC_BSGC}, we show the BICM capacity in \eqref{BICMCapacity6} and the CM capacity in \eqref{CMCapacity1} for 8-PAM, $\mat{P}=\mat{U}_8$, and the four binary labelings in Example~\ref{ExampleG3N3S3}. Fig.~\ref{AW_CM_BI_Capacities_8PAM_BRGC_NBC_BSGC} (a) illustrates that the difference between the CM capacity and the BICM capacity is small if the binary labeling is properly selected. The best of the four binary labelings is the NBC for low SNR ($\Rc\leq 0.43$~bit/symbol), the FBC for medium SNR ($0.43\leq \Rc\leq1.09$~bit/symbol), and the BRGC for high SNR  ($\Rc\geq1.09$~bit/symbol). Hence, the BRGC is suboptimal in at least 36\% of the $\Rc$ range. The gap between the CM capacity and the BICM capacity for the BSGC is quite large at low to moderate SNR. The low-SNR behavior is better elucidated in Fig.~\ref{AW_CM_BI_Capacities_8PAM_BRGC_NBC_BSGC} (b), where the same capacity curves are plotted versus $\Eb/N_0$ instead of $\SNR$. Interestingly, the CM capacity and the BICM capacity using the NBC achieve the SL at asymptotically low rates; Gaussian inputs are not necessary, \cf \cite[Sec.~I]{Verdu02}.

\end{example}

Formally, $\Eb/N_0$ is bounded from below by $f(\Rc)$, where
\begin{align}\label{AWGN_Capacity_Inequality_inverse}
f(\Rc) \triangleq \frac{ \mf{C}^{-1}(\Rc)}{\ms{E}_{H}[H^2]\Rc}.
\end{align}
This function always exists, because the capacity%
\footnote{From now on we will refer to ``capacity'' using the notation $\mf{C}(\SNR)$ in a broad sense. $\mf{C}(\SNR)$ can be the AWGN capacity $\mf{C}^\tr{AW}(\SNR)$ in \eqref{AWGN_Capacity}, the CM capacity $\mf{I}_\Omega^\tr{CM}(\SNR)$ in \eqref{CMCapacity1}, or the BICM capacity $\mf{I}_\Omega^\tr{BI}(\SNR)$ in \eqref{BICMCapacity1}.}
$\mf{C}(\SNR)$ is a strictly increasing%
\footnote{This can be proved using the relation between the AMI and the minimum mean square error (MMSE) presented in \cite{Guo05}, \ie that the derivative of the AMI with respect to $\SNR$ is proportional to the MMSE for any $\SNR$. Since the MMSE is a strictly decreasing function of $\SNR$, the AMI is a strictly increasing function of $\SNR$.}
function of $\SNR$ and thus invertible, while in contrast $f(\Rc)$ is in general not monotone. This is the reason why a given $\Eb/N_0$ for some labelings maps to more than one capacity value, as shown in \cite{Martinez08b}. The phenomenon can be understood by considering the function $\mf{I}^\tr{BI}_{\Omega}(\SNR)$ in a linear $\SNR$ scale, instead of logarithmic as in Fig.~\ref{AW_CM_BI_Capacities_8PAM_BRGC_NBC_BSGC} (a). If plotted, the function would pass through the origin for all labelings. Furthermore, any straight line through the origin represents a constant $\ms{E}_{H}[H^2]\Eb/N_0$ by \eqref{EsN0_and_EbN0}, where the slope is determined by the value of $\ms{E}_{H}[H^2]\Eb/N_0$. Such a line cannot intersect $\mf{I}^\tr{BI}_{\Omega}(\SNR)$ more than once for $\SNR>0$, if $\mf{I}^\tr{BI}_{\Omega}(\SNR)$ is concave. This is the case for the BRGC, NBC, and FBC, and therefore the function $\Rc=f^{-1}(\ms{E}_{H}[H^2]\Eb/N_0)$ exists, as illustrated for in Fig.~\ref{AW_CM_BI_Capacities_8PAM_BRGC_NBC_BSGC} (b). However, for some labelings such as the BSGC (and many others shown in \cite[Fig.~3.5]{Stierstorfer09_Thesis}), $\mf{I}^\tr{BI}_{\Omega}(\SNR)$ is not concave and $f(\Rc)$ is not invertible. This phenomenon has also been observed for linear precoding for BICM with iterative demapping and decoding \cite[Fig.~3]{Simoens08}, punctured turbo codes \cite[Fig.~3]{Brannstrom09b}, and incoherent $M$-ary PSK \cite[Figs.~2 and 5]{Peleg98} and frequency-shift keying channels \cite[Figs.~1 and 6]{Stark85}.

Since analytical expressions for the inverse function of the capacity are usually not available,  expressions for $f(\Rc)$ are rare in the literature. One well-known exception is the capacity of the Gaussian channel given by \eqref{AWGN_Capacity}, for which
\begin{align}\label{rho_of_C_AWGN}
f^\tr{AW}(\Rc) & = \frac{N}{2\Rc}(2^{2\Rc/N}-1),
\end{align}
which results in the SL
\begin{align}\label{rho_of_C_AWGN2}
\lim_{\Rc\rightarrow 0^+}{f^\tr{AW}(\Rc)} = \log_\tr{e}(2)=-1.59~\tr{dB}.
\end{align}
Analogously, we will use the notation $f_\Omega^\tr{CM}(\Rc)$ and $f_\Omega^\tr{BI}(\Rc)$ when the capacity considered is the CM and the BICM capacity, respectively.\footnote{The same notation convention will be used for other functions that will be introduced later in the paper.}

The results in Fig.~\ref{AW_CM_BI_Capacities_8PAM_BRGC_NBC_BSGC} (a)--(b) suggest a more general question: What are the optimal constellations for BICM at a given $\SNR$? To formalize this question, and in analogy to the maximum CM capacity in \eqref{CMCapacity4}, we define the \emph{maximum BICM capacity} as
\begin{align}
\Imax^\tr{BI}\left(\SNR\right)	& \triangleq \max_{\Omega} \mf{I}^{\tr{BI}}_\Omega\left(\SNR\right) \label{BICMCapacity5},
\end{align}
where the optimization is in this case over the three parameters defining $\Omega$.
In analogy to the maximum CM capacity, the maximum BICM capacity represents an upper bound on the number of bits per symbol that can be reliably transmitted using a fully optimized BICM system, \ie a system where for each $\SNR$, the constellation is selected to maximize the BICM capacity.

We conclude this subsection by expressing the BICM capacity as a difference of AMIs and conditional AMIs , which will facilitate the analysis in Sec.~\ref{Sec.asym.low}. The following result is a somehow straightforward generalization of \cite[Proposition~1]{Martinez08b}, \cite[eq.~(65)]{Brannstrom09} to $N$-dimensional input alphabets, nonuniform input distributions, and fading channels.
\begin{theorem}\label{BICMCapacityDiff.Theorem}
The BICM capacity can be expressed as
\begin{align}
\mf{I}_{\Omega}^\tr{BI}\left(\SNR\right) 	&=  \sum_{k=0}^{m-1}\sum_{u\in\set{0,1}}\PP_{C_k}(u)\bigl[I_{\mb{X}}(\mb{X};\mb{Y})-I_{\mb{X}|C_k=u}(\mb{X};\mb{Y})\bigr]. \label{BICMCapacityDiff}
\end{align}
\end{theorem}
\begin{IEEEproof}
For any function $e(\mb{X},\mb{Y},\mb{H})$,
\eq{
\ms{E}_{\mb{H},\mb{Y}|C_k=u}\left[\log_2{\frac{p_{\mb{Y}|\mb{H}C_k=u}(\mb{Y})}{p_{\mb{Y}|\mb{H}}(\mb{Y})}}\right] =
\ms{E}_{\mb{X},\mb{Y},\mb{H}|C_k=u}\left[
\log_2{\frac{e(\mb{X},\mb{Y},\mb{H})}{p_{\mb{Y}|\mb{H}}(\mb{Y})}} -
\log_2{\frac{e(\mb{X},\mb{Y},\mb{H})}{p_{\mb{Y}|\mb{H},C_k=u}(\mb{Y})}}
\right] 
.}
Using this relation in \eqref{BICMCapacity3}, letting $e(\mb{X},\mb{Y},\mb{H}) \triangleq p_{\mb{Y}|\mb{X},\mb{H}}(\mb{Y}) = p_{\mb{Y}|\mb{X},\mb{H},C_k=u}(\mb{Y})$, observing that the first term is independent of $u$, and utilizing \eqref{AMIdefinition2} and \eqref{AMIdefinition5} yields the theorem.
\end{IEEEproof}

\subsection{Minimum $\Eb/N_0$ for Reliable Transmission}
\label{minEbN0}

In this section, we determine the minimum $\Eb/N_0$ that permits reliable transmission, for a given input alphabet and labeling. As observed in Fig.~\ref{AW_CM_BI_Capacities_8PAM_BRGC_NBC_BSGC} (b), this minimum does not necessarily occur at rate $\Rc=0$. 
\begin{theorem}[Minimum $\Eb/N_0$]
The minimum $\Eb/N_0$ is given by $f(\tilde{R}_c)$, where $\tilde{R}_c=0$ or $\tilde{R}_c$ is one of the solutions of $g(\Rc) = 0$, where
\begin{align}\label{g_of_Rc_definition}
g(\Rc) \triangleq \frac{df(\Rc)}{d\Rc} = \frac{1}{\Rc}\frac{d \mf{C}^{-1}(\Rc)}{d\Rc}-\frac{ \mf{C}^{-1}(\Rc)}{\Rc^2}.
\end{align}
\end{theorem}
\begin{IEEEproof}
Any smooth function has a minimum given by the solution of its first derivative equal to zero or at the extremes of the considered interval. Since in general $0\leq \Rc< \infty$, two extreme cases should be considered. However, $\lim_{\Rc\rightarrow\infty}f^\tr{AW}(\Rc)=\lim_{\Rc\rightarrow m^-}f_\Omega(\Rc)=\infty$, and therefore, the only extreme point of interest is $\tilde{R}_c=0$.
\end{IEEEproof}

Since $f(\Rc)$ is in general not known analytically, the function $g(\Rc)$ must be numerically evaluated using $ \mf{C}(\SNR)$. An exception to this is the capacity of the AWGN channel, where $g^\tr{AW}(\Rc)$ can be calculated analytically. Moreover, it can be proved that in this case, a minimum $\Eb/N_0$ for nonzero rates does not exist.

\begin{corollary}[Minimum $\Eb/N_0$ for the AWGN channel]\label{Minum_EbNo_Theo}
The minimum $\Eb/N_0$ for the AWGN channel is unique, and it is obtained for zero-rate transmissions.
\end{corollary}
\begin{IEEEproof}
The derivative of $f^\tr{AW}(\Rc)$ in \eqref{rho_of_C_AWGN} is given by
\begin{align}\label{g_of_kc_AW}
g^\tr{AW}(\Rc) 	& = \frac{N+(2\Rc\log_\tr{e} 2-N)2^{2\Rc/N}}{2\Rc^2} \triangleq \frac{g_\tr{num}^\tr{AW}(\Rc)}{g_\tr{den}^\tr{AW}(\Rc)},
\end{align}
To prove that a zero for a nonzero rate does not exit, we need to prove that $g_\tr{num}^\tr{AW}(\Rc)>0$ for $\Rc>0$, since $g_\tr{den}^\tr{AW}(\Rc)>0$ for $\Rc>0$. This follows because $\lim_{\Rc\rightarrow 0^+}g_\tr{num}^\tr{AW}(\Rc)=0$ and the first derivative of $g_\tr{num}^\tr{AW}(\Rc)$ is strictly positive: 
\begin{align*}
\frac{dg_\tr{num}^\tr{AW}(\Rc)}{d\Rc}=\frac{4}{N} \Rc(\log_\tr{e}{2})^22^{2\Rc/N}>0.
\end{align*}
\end{IEEEproof}

In Fig.~\ref{AW_CM_BI_Capacities_8PAM_BRGC_NBC_BSGC} (c), we present the function $g(\Rc)$ in \eqref{g_of_Rc_definition} for the same constellations presented in Fig.~\ref{AW_CM_BI_Capacities_8PAM_BRGC_NBC_BSGC} (a)--(b). If $g(\Rc) = 0$  has at least one solution for $\Rc>0$, the capacity curve will have a local minimum (shown with a filled square in Fig.~\ref{AW_CM_BI_Capacities_8PAM_BRGC_NBC_BSGC} (b)--(c) for the BSGC). Note also that the BSGC has an interesting property, namely, $\lim_{\Rc\rightarrow 0^+} g^\tr{BI}_{\Omega}(\Rc)= -\infty$, and consequently, $\lim_{\Rc\rightarrow 0^+} f^\tr{BI}_{\Omega}(\Rc) = +\infty$. In this sense, the BSGC is an extremely bad labeling for $M$-PAM input alphabets and asymptotically low rates.

\subsection{Probabilistic Shaping}
\label{probshaping}

The maximum BICM capacity in \eqref{BICMCapacity5} is an optimization problem for which analytical solutions are unknown. In this subsection, we study the solution of \eqref{BICMCapacity5} when the input alphabet and the binary labeling are kept constant, \ie we study the so-called probabilistic shaping. Formally, we want to solve $\mat{P}^*(\SNR) \triangleq \argmax_{\mat{P}} \mf{I}^{\tr{BI}}_\Omega\left(\SNR\right)$, where $\Omega=[\mat{X},\mat{L},\mat{P}]$, for a given input alphabet $\mat{X}$ and labeling $\mat{L}$. Since this optimization problem turns out to have multiple local minima and no analytical methods are known for solving it, we perform a grid search with steps of 0.01 based on Gauss-Hermite quadratures. The optimization is performed over the three variables defining the input distribution: $P_{C_1}(0)$, $P_{C_2}(0)$, and $P_{C_3}(0)$. For each SNR value, the input distribution that maximizes the BICM capacity is selected.

\begin{figure}[!t]
\begin{tabular}{@{}c@{}}%
\psfrag{xlabel}[cc][cB][0.70]{$\SNR$~[dB]}%
\psfrag{ylabel}[cc][ct][0.70]{$\Rc$~[bit/symbol]}%
\psfrag{Gaussian-In}[l][l][0.55]{$\mf{C}^\tr{AW}(\SNR)$}%
\psfrag{CM-BRGC}[l][l][0.55]{$\mf{I}_{\Omega}^\tr{CM}(\SNR)$}%
\psfrag{BICM-BRGC}[l][l][0.55]{$\mf{I}^\tr{BI}_{\Omega}(\SNR)$ (BRGC)}%
\psfrag{BICM-NBC}[l][l][0.55]{$\mf{I}^\tr{BI}_{\Omega}(\SNR)$ (NBC)}%
\psfrag{BRGC-PShapin}[l][l][0.55]{$\mf{I}^\tr{BI}_{\Omega^*}(\Rc)$ (BRGC)}%
\psfrag{NBC-PShapin}[l][l][0.55]{$\mf{I}^\tr{BI}_{\Omega^*}(\Rc)$ (NBC)}%
\psfrag{SL}[l][l][0.85]{$-1.59$~dB}%
	\includegraphics[width=0.50\columnwidth]{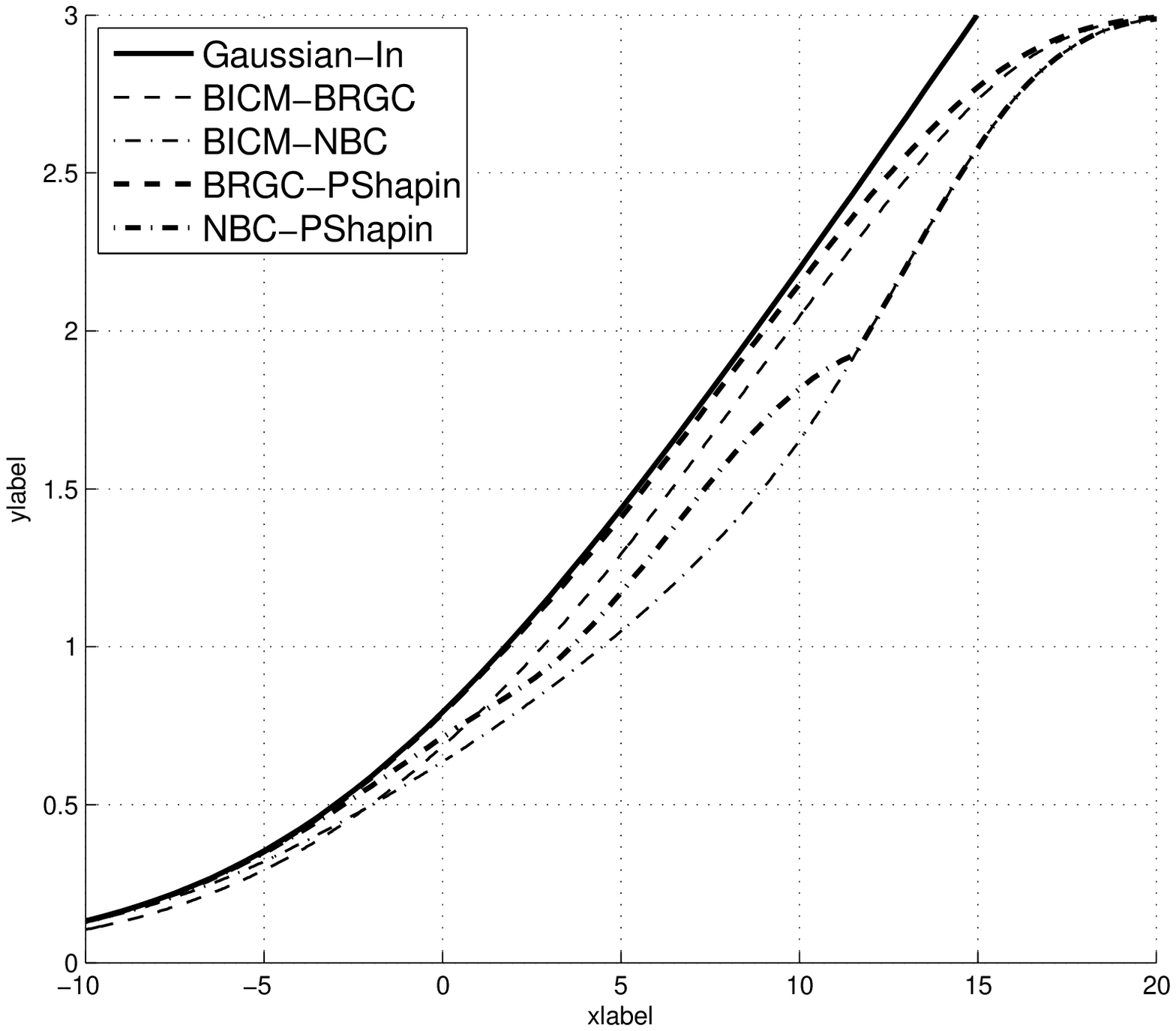}%
	\\{\footnotesize (a)}%
\end{tabular}
\hfill
\begin{tabular}{@{}c@{}}%
\psfrag{xlabel}[c][b][0.7]{${\Eb}/{N_0}$~[dB]}%
\psfrag{ylabel}[c][t][0.7]{$\Rc$~[bit/symbol]}%
\psfrag{Gaussian-In}[l][l][0.55]{$f^\tr{AW}(\Rc)$}%
\psfrag{BICM-BRGC}[l][l][0.55]{$f_\Omega^\tr{BI}(\Rc)$ (BRGC)}%
\psfrag{BICM-NBC}[l][l][0.55]{$f_\Omega^\tr{BI}(\Rc)$ (NBC)}%
\psfrag{BRGC-PShapin}[l][l][0.55]{$\mf{I}^\tr{BI}_{\Omega^*}(\Rc)$ (BRGC)}%
\psfrag{NBC-PShapin}[l][l][0.55]{$\mf{I}^\tr{BI}_{\Omega^*}(\Rc)$ (NBC)}%
\psfrag{SL}[l][l][0.55]{$-1.59$~dB}%
	\includegraphics[width=0.50\columnwidth]{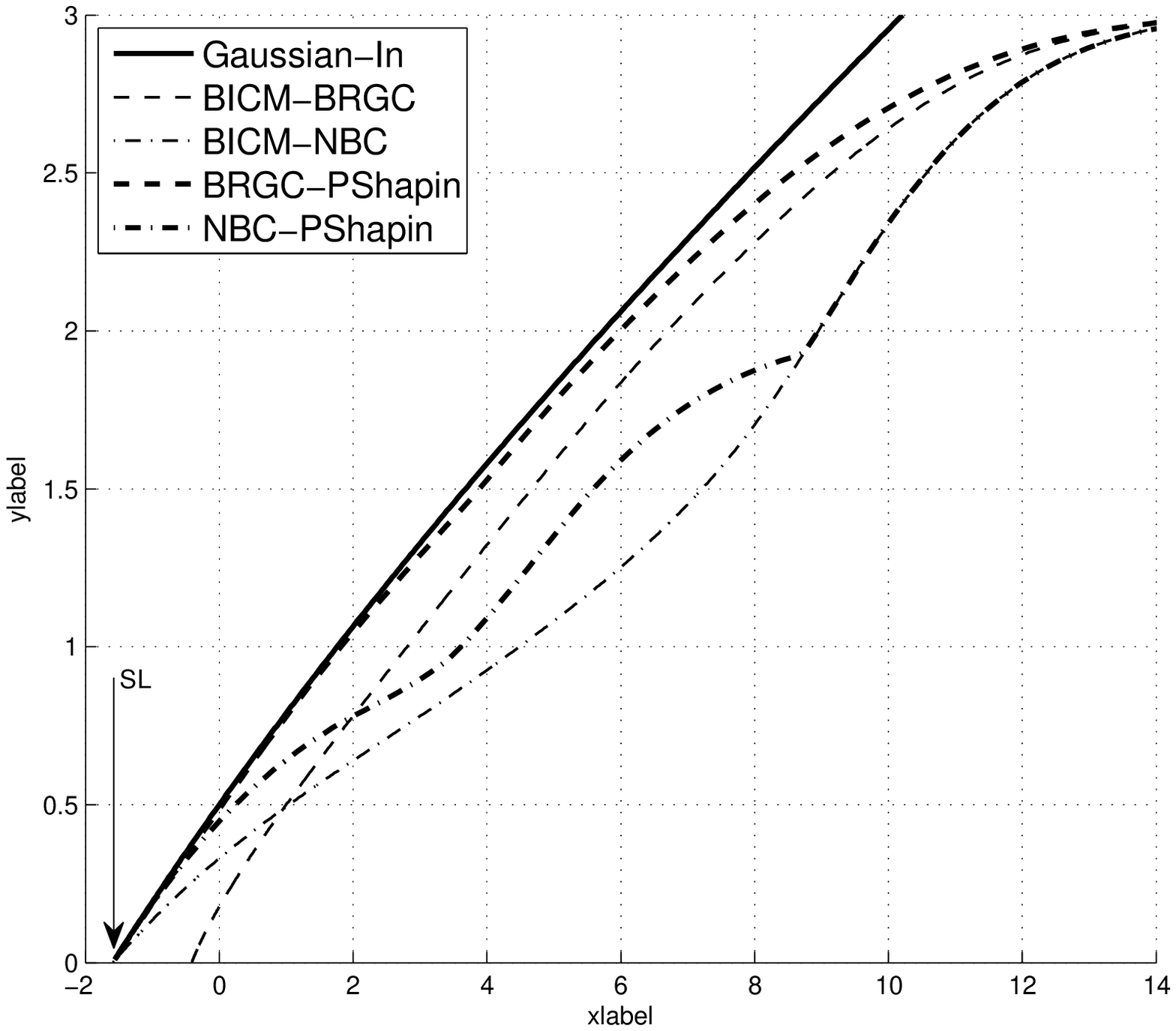}%
	\\{\footnotesize (b)}%
\end{tabular}
	\caption{BICM capacity for 8-PAM using $\mat{U}_8$ and $\mat{P}^*$ for the BRGC and the NBC versus (a) $\SNR$ and (b) ${\Eb}/{N_0}$.}
    	\label{f_BI_8PAM_BRGC_NBC_PShap_100}
\end{figure}

In Fig.~\ref{f_BI_8PAM_BRGC_NBC_PShap_100}, we show the BICM capacity for an 8-PAM input alphabet labeled by the BRGC and the NBC, when the optimized input distributions are used. We use the notation $\Omega^*=[\mat{X},\mat{L},\mat{P}^*]$. The results in this figure show how, by properly selecting the input distribution, the BICM capacity can be increased. The gap between the BICM capacity and the AWGN capacity is almost completely eliminated for $\Rc\leq 2~\tr{bit/symbol}$ (in contrast to a gap of approximately 1~dB in Fig.~\ref{AW_CM_BI_Capacities_8PAM_BRGC_NBC_BSGC} (b)). Similar results have been presented recently in \cite{Fabregas10a} for 4-PAM. Interestingly, Fig.~\ref{f_BI_8PAM_BRGC_NBC_PShap_100} shows that if the input distribution is optimized, the NBC is not the optimal binary labeling for low SNR anymore, but the BRGC with an optimized input distribution achieves the SL. This is also the case for the FBC, but we do not show those results not to overcrowd the figure.

\section{BICM for Asymptotically Low Rates}\label{Sec.asym.low}

In this subsection, we are interested in finding an asymptotic expansion for the CM and the BICM capacities when $\SNR\rightarrow 0$. 

\subsection{Relation between AWGN and BICM capacity}

We start by proving that the BICM capacity can be optimal in the sense of being equal to the AWGN capacity only for zero rate. This very simple result motivates the developments in order to characterize the behavior of BICM for asymptotically low SNR.

\begin{theorem}\label{Lemma.CM.BICM.AWGN}
The AWGN capacity, the CM capacity, and the BICM capacity are related through the following two inequalities.
\begin{enumerate}[i)]
\item $\mf{I}_{\Omega}^{\tr{CM}}\left(\SNR\right) \leq \mf{C}^{\tr{AW}}\left(\SNR\right)$ with equality if and only if $\SNR=0$, and
\item $\mf{I}_{\Omega}^{\tr{BI}}\left(\SNR\right) \leq \mf{I}_{\Omega}^{\tr{CM}}\left(\SNR\right)$.
\end{enumerate}
\end{theorem}
\begin{IEEEproof}
We start by proving that $\mf{I}_{\Omega}^{\tr{AW}}\left(\SNR\right) < \mf{C}^{\tr{AW}}\left(\SNR\right)$ for $\SNR>0$, where $\mf{I}_{\Omega}^{\tr{AW}}\left(\SNR\right)$ is the CM capacity of the AWGN channel. From \eqref{CMCapacity1} and $I_{\mb{X}}(\mb{X};\mb{Y})=h(\mb{Y})-h(\mb{Z})$, we express the CM capacity for a given $\Omega$ in terms of differential entropies as $\mf{I}^\tr{AW}_{\Omega}(\SNR) = h(\mb{Y})-{N}/{2}\log_2(2\pi N_0\tr{e})$. Since the differential entropy $\ds{h(\mb{Y})=-\int_{\ms{R}^N}p_{\mb{Y}}(\mb{y})\log_2{p_{\mb{Y}}(\mb{y})}\,d\mb{y}}$ is maximized if and only if $\mb{Y}$ is Gaussian distributed \cite[Theorem~8.6.5]{Cover06_Book}, the use of any constellation $\Omega$ (discrete input alphabet) will give a smaller differential entropy $h(\mb{Y})$ than for a Gaussian $\mb{Y}$, which proves that $\mf{I}_{\Omega}^{\tr{AW}}\left(\SNR\right) < \mf{C}^{\tr{AW}}\left(\SNR\right)$ for $\SNR>0$.

We now prove that $\mf{I}_{\Omega}^{\tr{CM}}\left(\SNR\right) < \mf{I}_{\Omega}^{\tr{AW}}\left(\SNR\right)$ for any fading channel and $\SNR>0$. To do this, we note that the CM capacity for fading channels is equal to the CM capacity for the AWGN channel averaged over the distribution of the instantaneous SNR. Furthermore, $\mf{I}_{\Omega}^{\tr{AW}}\left(\SNR\right)$ is a strictly concave function of $\SNR$ for $\SNR>0$, because the second derivative of the AMI as a function of the SNR (the first derivative of the MMSE, see footnote~8) is strictly negative for $\SNR>0$ \cite[Proposition~5]{Guo08} \cite[Proposition~7]{Guo11}. Therefore, Jensen's inequality holds, which yields $\mf{I}_{\Omega}^{\tr{CM}}\left(\SNR\right)=\ms{E}_H[\mf{I}_{\Omega}^{\tr{AW}}\left(H^2\Es/N_0\right)] < \mf{I}_{\Omega}^{\tr{AW}}\left(\ms{E}_H[H^2]\Es/N_0\right)= \mf{I}_{\Omega}^{\tr{AW}}\left(\SNR\right) $ for $\SNR>0$. This and the fact that $\mf{I}_{\Omega}^{\tr{CM}}\left(0\right)=\mf{C}^{\tr{AW}}\left(0\right)=0$ proves item i). The proof of item ii) was presented in \cite[Sec.~III]{Caire98}.
\end{IEEEproof}

\begin{corollary}\label{Lemma.BICM.vs.AWGN}
The BICM capacity \emph{and} the maximum BICM capacity can be equal to the AWGN capacity \emph{only} for zero
rates, \ie
$\mf{I}_{\Omega}^{\tr{BI}}\left(\SNR\right)=\Imax^{\tr{BI}}\left(\SNR\right)=\mf{C}^{\tr{AW}}\left(\SNR\right)$
only if $\SNR=0$.
\end{corollary}
\begin{IEEEproof}
From Theorem~\ref{Lemma.CM.BICM.AWGN}, we know that for any $\SNR>0$, the inequality $\mf{I}_{\Omega}^{\tr{BI}}\left(\SNR\right) \leq\mf{I}_{\Omega}^{\tr{CM}}\left(\SNR\right) < \mf{C}^{\tr{AW}}\left(\SNR\right)$ holds. Therefore, for any $\SNR>0$, $\mf{I}_{\Omega}^{\tr{BI}}\left(\SNR\right)< \mf{C}^{\tr{AW}}\left(\SNR\right)$. The proof for the BICM capacity is completed noting that $\mf{I}_{\Omega}^{\tr{BI}}\left(0\right)=\mf{C}^{\tr{AW}}\left(0\right)=0$. The proof for the maximum BICM capacity follows from the fact that Theorem~\ref{Lemma.CM.BICM.AWGN} holds also when an optimization over $\Omega$ is applied.
\end{IEEEproof}

Corollary~\ref{Lemma.BICM.vs.AWGN} simply states that the only rates for which the AWGN will be equal to the BICM capacity and the maximum BICM capacity is $\Rc=0$ (or equivalently $\SNR=0$). In the following subsections, we analyze the asymptotic behavior of the BICM capacity when $\SNR=0$.

\subsection{A Linear Approximation of the Capacity and the SL}\label{Sec.LinearApp}

Any capacity function $\mf{C}(\SNR)$ can be approximated using a Taylor expansion around $\SNR=0$ as $ \mf{C}(\SNR)=\alpha\SNR+O(\SNR^2)$. By inversion of power series \cite[Sec.~1.3.4.5]{Zwillinger03_Book}, we find
\begin{align*}
\mf{C}^{-1}(\Rc) 	& = \frac{1}{\alpha} \Rc+O(\Rc^2),
\end{align*}
and using \eqref{AWGN_Capacity_Inequality_inverse}, it is possible to obtain a linear approximation of the function $f(\Rc)$
\begin{align}\label{f_Rc.Linear}
f(\Rc) 			& = \frac{1}{\alpha}+O(\Rc).
\end{align}
For asymptotically low rates, \eqref{f_Rc.Linear} results in
\begin{align}\label{fRc_Asymptotic}
\lim_{\Rc\rightarrow 0^+} f(R_C) = \frac{1}{\alpha},
\end{align}
and since from \eqref{rho_of_C_AWGN2} ${1}/{\alpha}\geq\log_\tr{e}(2)$, we obtain
\begin{align}\label{alpha_ineq}
\alpha\leq\log_2\tr{e}.
\end{align}
It is clear from \eqref{alpha_ineq} that a capacity function $\mf{C}(\SNR)$ that has a coefficient $\alpha=\log_2\tr{e}$ achieves the SL $-1.59~\tr{dB}$\footnote{Or equivalently, if we measure the AMI in nats, $\alpha=\log_\tr{e}\tr{e}=1$.}. Moreover, based on the results for the BSGC in Fig.~\ref{AW_CM_BI_Capacities_8PAM_BRGC_NBC_BSGC} (b), the coefficient $\alpha$ can be as low as zero.

\subsection{First-Order Asymptotics of the BICM Capacity}\label{FOOasymptotics}

\begin{theorem}[Linear approximation of the AMI]\label{PrelovTheo3}
When the channel is perfectly known at the receiver, and for any input distribution $P_{\mb{X}}(\mb{x})$, the AMI between $\mb{X}$ and $\mb{Y}$ in \eqref{Fading_AWGN_channel} can be expressed as
\begin{align*}
I_{\mb{X}}(\mb{X};\mb{Y}) = \alpha\SNR+O(\SNR^2)
\end{align*}
when $\SNR\rightarrow 0$, where
\begin{align}\label{PrelovTheo3.2}
\alpha = \log_2\tr{e}\biggl(1-\frac{\|\ms{E}_{\mb{X}}[\mb{X}]\|^2}{\Es}\biggr).
\end{align}
\end{theorem}
\begin{IEEEproof}
The proof is given in Appendix~\ref{Appendix.PrelovTheo3}.
\end{IEEEproof}

Theorem~\ref{PrelovTheo3} shows how to calculate the first-order asymptotics of an AMI with arbitrary input distribution. The following corollary follows directly from the definition of the CM capacity in \eqref{CMCapacity1}, where the input distribution is given by \eqref{Pxuncond}.

\begin{corollary}[Coefficient $\alpha_\Omega^\tr{CM}$]\label{alpha_CM_Theorem}
The CM capacity can be expressed as
\begin{align*}
\mf{I}_{\Omega}^\tr{CM}\left(\SNR\right) = \alpha_\Omega^\tr{CM} \SNR + O(\SNR^2)
\end{align*}
when $\SNR\rightarrow 0$, where $\alpha_\Omega^\tr{CM}$ is given by \eqref{PrelovTheo3.2}.
\end{corollary}

The next theorem gives the first-order asymptotics for the BICM capacity.
\begin{theorem}[Coefficient $\alpha_\Omega^\tr{BI}$]\label{a1.a2.BICM}
The coefficient $\alpha$ for the BICM capacity $\mf{I}_\Omega^\tr{BI}(\SNR)$ is given by
\begin{align}\label{a1.BICM.general}
\alpha_ \Omega^\tr{BI}=\frac{\log_2{\tr{e}}}{\Es}\left[\sum_{k=0}^{m-1}\sum_{u\in\set{0,1}}\PP_{C_k}(u)\|\ms{E}_{\mb{X}|C_k=u}[\mb{X}]\|^2-m\|\ms{E}_{\mb{X}}[\mb{X}]\|^2\right],
\end{align}
\end{theorem}

\begin{IEEEproof}
Reordering the result of Theorem~\ref{BICMCapacityDiff.Theorem}, we have that
\begin{align*}
\mf{I}_{\Omega}^\tr{BI}\left(\SNR\right) 	&=  \sum_{k=0}^{m-1}\biggl\{I_{\mb{X}}(\mb{X};\mb{Y})-\sum_{u\in\set{0,1}}\PP_{C_k}(u)I_{\mb{X}|C_k=u}(\mb{X};\mb{Y})\biggr\}.
\end{align*}
Since $I_{\mb{X}}(\mb{X};\mb{Y})$ and $I_{\mb{X}|C_k=u}(\mb{X};\mb{Y})$ are AMIs, we can apply \eqref{PrelovTheo3.2} to each of them, which gives
\begin{align*}
\alpha_{\Omega}^\tr{BI}	&= \frac{\log_2\tr{e}}{\Es}\sum_{k=0}^{m-1}\biggl\{\Es-\|\ms{E}_{\mb{X}}[\mb{X}]\|^2-\sum_{u\in\set{0,1}}\PP_{C_k}(u)\bigl(\ms{E}_{\mb{X}|C_k=u}[\|\mb{X}\|^2]-\|\ms{E}_{\mb{X}|C_k=u}[\mb{X}]\|^2\bigr)\biggr\}.
\end{align*}
We recognize $\sum_{u\in\set{0,1}}\PP_{C_k}(u)\ms{E}_{\mb{X}|C_k=u}[\|\mb{X}\|^2]$ as the average symbol energy $\Es$, which completes the proof.
\end{IEEEproof}

The first-order coefficients of the expansion of the CM and BICM capacities in Corollary~\ref{alpha_CM_Theorem} and Theorem~\ref{a1.a2.BICM} do not depend on the fading. This simply states that, under the constraints imposed on $\mb{H}$, the fading has no effect on the first-order behavior of the BICM capacity. Consequently, the analysis of the optimal constellations for fading channels at low SNR can be reduced, without loss of generality, to the AWGN case.

Corollary~\ref{alpha_CM_Theorem} and Theorem~\ref{a1.a2.BICM} generalize the results in \cite{Martinez08b, Alvarado10c} by considering constellations with nonuniform input distributions and arbitrary dimensions, mean, and variance. This generalization will allow us to analyze optimal constellations $\Omega$ in the next section.

In general, we know from \eqref{alpha_ineq} that $\alpha^\tr{BI}_\Omega\leq\log_2\tr{e}$, which can be interpreted as the penalty of a certain BICM system over an optimal CM system (without interleaving). In the following section we analyze $\alpha^\tr{BI}_\Omega$ for PAM and PSK input alphabets with different binary labelings and $\mat{P}=\mat{U}_M$ and we also show how to obtain $\alpha^\tr{BI}_\Omega=\log_2\tr{e}$ for general constellations.

\section{First-Order Optimal Constellations for BICM}\label{Sec.OptimalConstellations}

Shannon stated in 1959, ``There is a curious and provocative duality between the properties of a source with a distortion measure and those of a channel'' \cite{Shannon59b}. Many instances of this duality have been observed during the last 50 years of communications research. A good summary of this is presented in \cite[Sec.~V]{Cover02}. The coefficient $\alpha$ is mathematically similar to the so-called \emph{linearity index} \cite{Knagenhjelm96}, which was used to indicate the approximative performance of labelings in a source coding application at high SNR. The usage of the HT in this section was inspired by the analysis in \cite{Knagenhjelm96}.

\subsection{FOO Constellations}
In view of the SL \eqref{alpha_ineq}, we define a \emph{first-order optimal (FOO) constellation} for BICM\footnote{A similar first-order optimality criterion for the CM capacity can be defined. In this case, based on \eqref{PrelovTheo3.2}, any constellation based on a zero-mean input alphabet is an FOO constellation for the CM capacity, regardless of the input distribution $P_{\mb{X}}(\mb{x})$. Conversely, no FOO constellation can have nonzero mean.} as a constellation $\Omega$ that results in a coefficient $\alpha_\Omega^\tr{BI}=\log_2\tr{e}$.

\begin{theorem}[Coefficient $\alpha_\Omega^\tr{BI}$ for arbitrary constellations]\label{theorem.alpha.bicm.general}
For any constellation $\Omega$
\begin{align}\label{alpha_BICM_general}
  \alpha_\Omega^\tr{BI} = \frac{\log_2{\tr{e}}}{2\Es} 
    \sum_{k=0}^{m-1} 
    \left\{
    \biggl\|\sum_{i=0}^{M-1} \frac{q_{i,k} \mb{x}_i P_{\mb{X}}(\mb{x}_i)}{\sqrt{\PP_{C_k}(c_{i,k})}} \biggr\|^2+
    \biggl\|\sum_{i=0}^{M-1} \frac{\mb{x}_i P_{\mb{X}}(\mb{x}_i)}{\sqrt{\PP_{C_k}(c_{i,k})}} \biggr\|^2-
    2\bigl\|\ms{E}_{\mb{X}}[\mb{X}]\bigr\|^2
    \right\},
\end{align}
where $q_{i,k}$ are the elements of the modified labeling matrix in \eqref{qki_def}.
\end{theorem}

\begin{IEEEproof}
The proof is given in Appendix~\ref{Appendix.theorem.alpha.bicm.general}.
\end{IEEEproof}

Theorem~\ref{theorem.alpha.bicm.general} is a very general theorem valid for any constellation $\Omega$. From this theorem, it is clear that the problem of designing FOO constellations for BICM has three degrees of freedom: the input alphabet $\mat{X}$, the binary labeling $\mat{L}$, and the input distribution $\mat{P}$.

From now on, we restrict our attention to uniform input distributions $\mat{P}$. This restriction can be justified from the fact that due to the digital implementation of the transceivers, changing the input alphabet or the binary labeling can be implemented without complexity increase. On the other hand, implementation of probabilistic shaping requires a modification of the channel encoder and/or the interleaver. If $\mat{P}=\mat{U}_M$, then $P_{C_k}(u)=1/2$ for $k=0,1,\ld,m-1$ and $u\in\set{0,1}$, and \eqref{alpha_BICM_general} simplifies into
\begin{align} \label{alpha.bicm.uniform.zm}
  \alpha_\Omega^\tr{BI} = \frac{\log_2{\tr{e}}}{\Es} 
    \sum_{k=0}^{m-1}
    \biggl\|\frac{1}{M}\sum_{i=0}^{M-1} q_{i,k} \mb{x}_i \biggr\|^2.
\end{align}

Keeping $\XX$ fixed and changing the labeling $\LL$ is equivalent to fixing $\LL$ and reordering the rows of $\XX$. Therefore, a joint optimization of $\Omega = [\XX, \LL,\mat{U}_M]$ over $\XX$ and $\LL$ can without loss of generality be reduced to an optimization over $\XX$ only, for an arbitrary $\LL$. In the following analysis, we will hence sometimes fix the labeling to be the NBC, without loss of generality.

The expression for $\alpha_\Omega^\tr{BI}$ in \eqref{alpha.bicm.uniform.zm} can be simplified further using the HT, as elaborated in the next theorem.

\begin{theorem}[The HT and $\alpha_\Omega^\tr{BI}$]\label{thB}
The coefficient $\alpha_\Omega^\tr{BI}$ for a constellation $\Omega=[\mat{X},\mat{N}_m,\mat{U}_M]$ is given by
\eq{
  \alpha_\Omega^\tr{BI} = \frac{\log_2{\tr{e}}}{\Es} \sum_{k=0}^{m-1} \| \tx_{2^k} \|^2,
}
where $\tx_{2^k}$ are elements of the HT of $\mat{X}$ defined by \eqref{HT}.
\end{theorem}

\begin{IEEEproof}
Using Lemma~\ref{q_NBC} and \eqref{p3} in \eqref{alpha.bicm.uniform.zm}, we obtain
\eq{
  \alpha_\Omega^\tr{BI} &= \frac{\log_2{\tr{e}}}{\Es} \sum_{k=0}^{m-1} 
    \left\| \frac{1}{M}\sum_{i=0}^{M-1} h_{i,2^k} \mb{x}_i \right\|^2 = \frac{\log_2{\tr{e}}}{\Es} \sum_{k=0}^{m-1} \| \tx_{2^k} \|^2
.}
\end{IEEEproof}

It follows from Theorem \ref{thB} and \eqref{p5} that 
\eqlab{e8}{
  \alpha_\Omega^\tr{BI} &\le \frac{\log_2{\tr{e}}}{\Es} \sum_{j=0}^{M-1} \| \tx_j \|^2 = \frac{\log_2{\tr{e}}}{M \Es} \sum_{i=0}^{M-1} \| \mb{x}_i \|^2 = \log_2{\tr{e}}
}
for any constellation, which is in perfect agreement with \eqref{alpha_ineq}. We now proceed to determine the class of input alphabets and labelings for which the bound \eqref{e8} is tight.

\begin{theorem}[Linear projection of a hypercube]\label{thD}
A constellation $\Omega=[\mat{X},\mat{L},\mat{U}_M]$ is FOO if and only if there exists an $m\times N$ matrix $\mat{V}=[\mb{v}_0^\tr{T},\ldots,\mb{v}_{m-1}^\tr{T}]^\tr{T}$ such that
\eqlab{eq:thD}{
\mat{X}=\mat{Q}(\mat{L})\mat{V}.
}
\end{theorem}

\begin{IEEEproof}
Consider first the NBC. Equality holds in \eqref{e8} if and only if $\tx_j=0$ for all $j=0,\ldots,M-1$ except $j=1,2,4,\ldots,2^{m-1}$. For such input alphabets, \eqref{p3} yields
\eq{
  \mb{x}_i = \sum_{k=0}^{m-1} h_{i,2^k} \tx_{2^k}
.}
Letting $\mb{v}_k \triangleq \tx_{2^k}$ for $k=0,\ldots,m-1$ and using \eqref{xqv}, we obtain
\eqlab{e7}{
  \mb{x}_i = \sum_{k=0}^{m-1} q_{i,k} \mb{v}_k, \qquad i=0,\ldots,M-1
.}
Letting $\mat{V} = [\mb{v}_0^\tr{T},\ldots,\mb{v}_{m-1}^\tr{T}]^\tr{T}$ completes the proof for $\mat{L}=\mat{N}_m$. That the theorem also holds for an arbitrary labeling follows by synchronously reordering the rows of $\mat{X}$ and $\mat{L}$, as explained before Theorem~\ref{thB}.
\end{IEEEproof}

\begin{figure*}
\psfrag{2v0}[r][r][0.8]{$2\mb{v}_0$}
\psfrag{2v1}[][][0.8]{$2\mb{v}_1$}
\psfrag{2v2}[r][r][0.8]{$2\mb{v}_2$}
\psfrag{000}[][][0.8]{$000$}
\psfrag{001}[][][0.8]{$001$}
\psfrag{010}[][][0.8]{$010$}
\psfrag{011}[][][0.8]{$011$}
\psfrag{100}[][][0.8]{$100$}
\psfrag{101}[][][0.8]{$101$}
\psfrag{110}[][][0.8]{$110$}
\psfrag{111}[][][0.8]{$111$}
\centerline{%
\subfigure[OTTO constellation]
{\includegraphics[width=0.45\columnwidth]{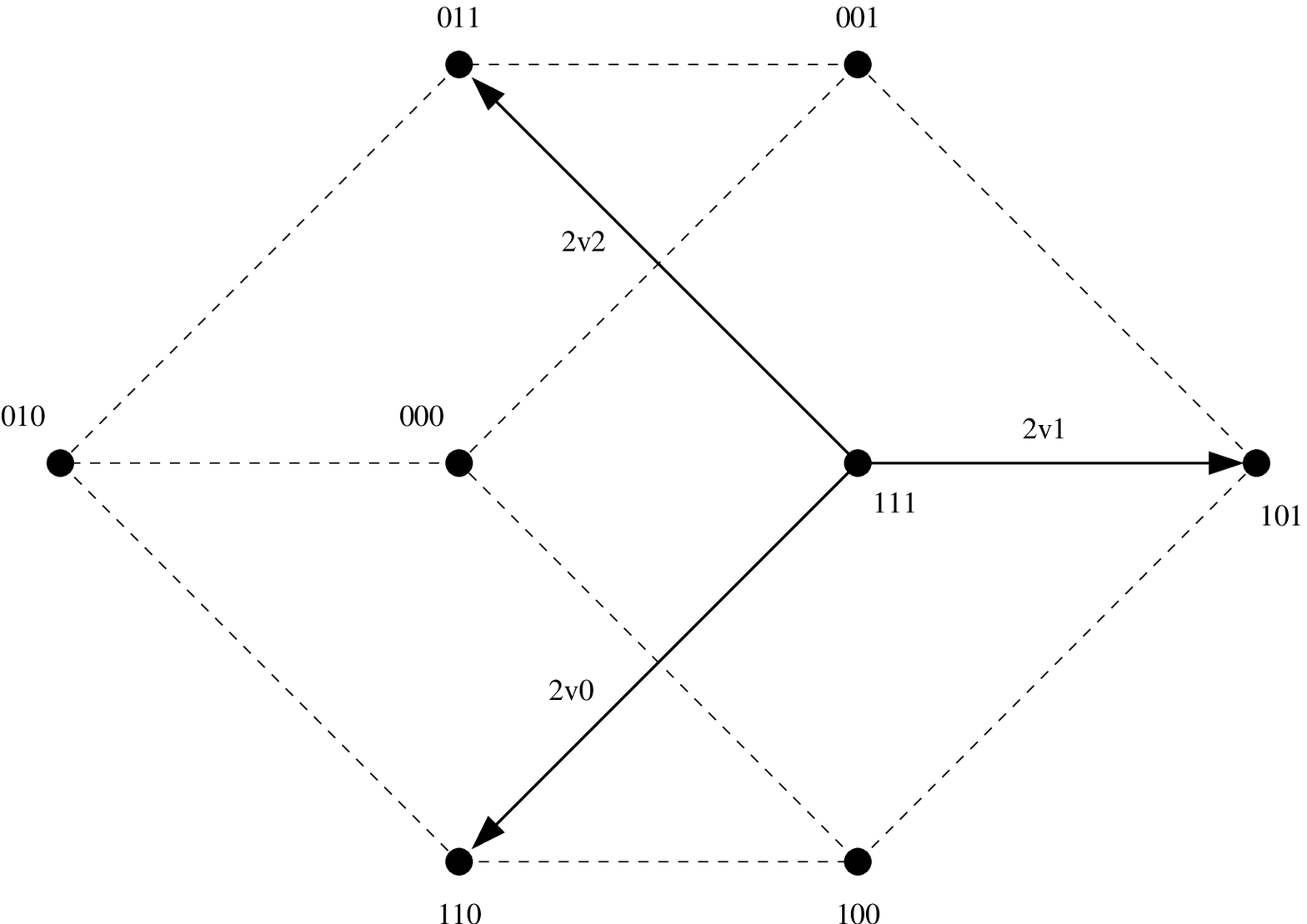}
\label{OTTO}} 
\hfil
\subfigure[OTOTO constellation]
{\includegraphics[width=0.35\columnwidth]{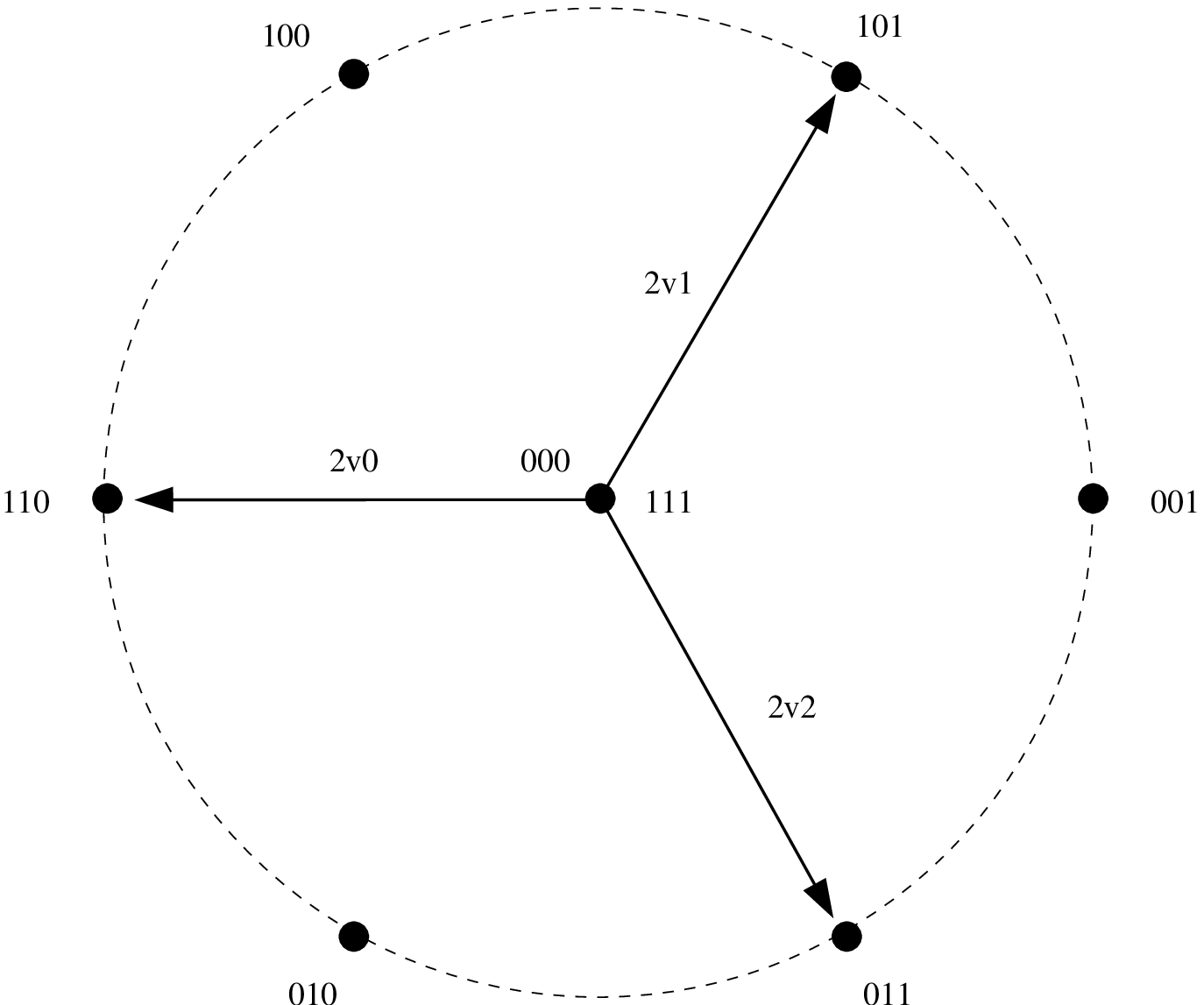}
\label{OTOTO}}
}
\caption{The two FOO constellations defined in Example~\ref{OTTO_OTOTO_Example} ($m=3$ and $N=2$). Graphically, the OTTO constellation in (a) gives the impression of a projected cube. The OTOTO constellation in (b) gives the impression of a 6-PSK input alphabet with two extra points located at the origin.}
\label{OTTO_and_OTOTO}
\end{figure*}

Theorem \ref{thD} has an appealing geometrical interpretation. Writing the set of constellation points as in \eqref{eq:thD}, each row of $\mat{Q}$ can be interpreted as a vertex of an $m$-dimensional hypercube, and $\mat{V}$ as an $m\times N$ projection matrix. Hence, a constellation for BICM is FOO if and only if its constellation is a \emph{linear projection of a zero-mean hypercube.}
This interpretation, as well as all theorems presented so far, holds for an arbitrary dimension $N$. In the rest of this section, we will exemplify the results for $N=1$ and $2$, because such input alphabets are easily visualized (Figs.~\ref{OTTO_and_OTOTO}--\ref{Hierarchical_8PAM}) and often used in practice (PAM, QAM, and PSK).

\begin{example}[OTTO and OTOTO constellations]\label{OTTO_OTOTO_Example}
To exemplify the concept of Theorem~\ref{thD}, we present two constellations that are FOO. The projection matrices for the ``one-three-three-one'' (OTTO) and the ``one-two-one-two-one'' (OTOTO) constellations are defined as
\begin{align*}
\mat{V}_\tr{OTTO}=
\left[
\begin{array}{rr}
	-1&-1\\
	+1&0\\
	-1&+1
  \end{array}
  \right]
  ,\qquad
\mat{V}_\tr{OTOTO}=
\left[
\begin{array}{rr}
	-1&0\\
	\cos{(\pi/3)}&\sin{(\pi/3)}\\
	\cos{(\pi/3)}&-\sin{(\pi/3)}
  \end{array}
  \right].
\end{align*}
Both constellations are shown in Fig.~\ref{OTTO_and_OTOTO}. The figure illustrates that the minimum Euclidean distance, which is an important figure-of-merit at high SNR, plays no role at all when constellations are optimized for low SNR.
\end{example}

A particular case of Theorem~\ref{thD} are the nonequally spaced (NES) $M$-PAM input alphabets, as specified in the following corollary.

\begin{corollary}\label{Theo.NESpacedPAM}
If a NES $M$-PAM input alphabet $\XX$ consists of the points $\pm v_0 \pm v_1 \pm \cdots \pm v_{m-1}$, there exists a binary labeling $\LL$ such that the constellation $[\XX,\LL,\mat{U}_M]$ is FOO.
\end{corollary}

\begin{example}[Hierarchical constellations]\label{Hierarchical_Example}
The so-called ``hierarchical constellations'' \cite{Vitthaladevuni03,Morimoto95,Hossain06b} are defined by the one-dimensional input alphabet \cite[eq.~(3)]{Vitthaladevuni03}
\eqlab{x.hier}{
x_i=\sum_{k=0}^{m-1}(2b_{k}(i)-1)d_k,
}
where $b_{k}(i)$ is the base-2 representation of the integer $i$ with $i=0,\ld,M-1$, and where $d_k>0$ for $k=0,\ld,m-1$ are the distances defining the input alphabet. The additional condition $x_i<x_{i+1}$ for $i=0,\ld,M-2$ is usually imposed so that overlapping points in the input alphabet are avoided. This condition also keeps the labeling of the input alphabet unchanged.

In Fig.~\ref{Hierarchical_8PAM}, we show a hierarchical 8-PAM input alphabet. In this figure, the $M$ constellation points are shown with black circles, while the white squares/triangles 
represent 2- and 4-PAM input alphabets from which the 8-PAM input alphabet can be recursively (hierarchically) constructed.

The binary labeling used in hierarchical constellations is usually assumed to be the BRGC. In this case, we find that when $\XX$ is given by \eqref{x.hier}, the system in \eqref{eq:thD} has no solutions for $\mat{V}$, and therefore, the constellation is not FOO. However, if the NBC is used instead (as in Fig.~\ref{Hierarchical_8PAM}), all hierarchical constellations are FOO, because $\XX=\mat{Q}(\mat{N}_m)\mat{V}$ gives a projection matrix $\mat{V}=[-d_0,-d_1,\ld,-d_{m-1}]^\tr{T}$.

\begin{figure}
\begin{center}
\psfrag{000}[cc][][0.80]{000}
\psfrag{001}[cc][][0.80]{001}
\psfrag{010}[cc][][0.80]{011}
\psfrag{011}[cc][][0.80]{010}
\psfrag{100}[cc][][0.80]{111}
\psfrag{101}[cc][][0.80]{110}
\psfrag{110}[cc][][0.80]{100}
\psfrag{111}[cc][][0.80]{101}
\psfrag{a1}[cc][][0.80]{$x_{0}$}
\psfrag{a2}[cc][][0.80]{$x_{1}$}
\psfrag{a3}[cc][][0.80]{$x_{2}$}
\psfrag{a4}[cc][][0.80]{$x_{3}$}
\psfrag{a5}[cc][][0.80]{$x_{4}$}
\psfrag{a6}[cc][][0.80]{$x_{5}$}
\psfrag{a7}[cc][][0.80]{$x_{6}$}
\psfrag{a8}[cc][][0.80]{$x_{7}$}
\psfrag{2d0}[cc][][0.80]{$2d_{0}$}
\psfrag{2d1}[cc][][0.80]{$2d_{1}$}
\psfrag{2d2}[cc][][0.80]{$2d_{2}$}
 \includegraphics[width=0.93\columnwidth]{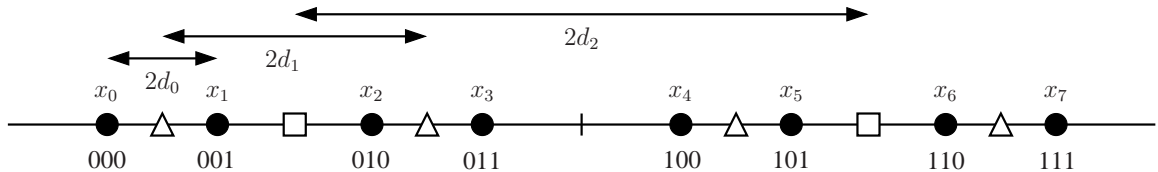}
  \caption{Hierarchical 8-PAM constellation. The constellation is FOO and $\mat{V}=[-d_0,-d_1,\ld,-d_{m-1}]^\tr{T}$.}
  \label{Hierarchical_8PAM}
  \end{center}
 \end{figure}
 \end{example}

\subsection{Labelings for PAM, QAM, and PSK}\label{Sec.PAM.QAM.PSK}

While we have so far kept the labeling fixed and searched for good input alphabets, we now take the opposite approach and search for good labelings for a given input alphabet. In this section we analyze the practically relevant input alphabets PAM, QAM, and PSK defined in Sec.~\ref{Sec.Preliminaries.Labelings}. Throughout this section, we assume $\mat{P}=\mat{U}_M$.


\begin{example}[NBC for $M$-PAM]\label{ex:mpam}
Let $\mat{V}=[v_0,v_1,\ldots,v_{m-1}]^\tr{T} = [-1,-2,-4,\ldots,-2^{m-1}]^\tr{T}$ and let $\mat{L}=\mat{N}_m$. With $q_{i,k}$ given by \eqref{xqv}, we obtain from \eqref{eq:thD} the constellation $\mat{X}_{\tr{PAM}}\triangleq[-M+1,-M+3,\ldots,M-1]^\tr{T}$, which shows that the constellation $[\mat{X}_\tr{PAM},\mat{N}_m,\mat{U}_M]$ is FOO. In view of Theorem \ref{thB}, the optimality of $M$-PAM input alphabets comes from the fact that the HT of $\mat{X}_{\tr{PAM}}$ has its only nonzero elements in the $m$ positions $1,2,4,\ldots,2^{m-1}$.
\end{example}

It follows from Example \ref{ex:mpam} that the constellation $[\mat{X}_\tr{PAM},\mat{N}_m,\mat{U}_M]$ is FOO, which has also been shown in \cite{Stierstorfer09a}. The following theorem states that the NBC is the unique labeling with this property, apart from trivial bit operations that do not alter the characteristics of the labeling.

\begin{theorem}\label{Sec.PAM.QAM.PSK.PAMTheorem}
The constellation $[\mat{X}_\tr{PAM},\mat{L},\mat{U}_M]$ is FOO if and only if $\mat{L}=\mat{N}_m$, or any other binary labeling that can be derived from the NBC by inverting the bits in certain positions or by permuting the sequence of bits in all codewords.
\end{theorem}

\begin{IEEEproof}
The proof is given in Appendix~\ref{Appendix.PAM.QAM.PSK.PAMTheorem}.
\end{IEEEproof}


In order to extend this result to rectangular QAM constellations, we first state a theorem about product constellations in general.

\begin{theorem}\label{Sec.PAM.QAM.PSK.QAMTheorem}
A two-dimensional constellation $[\mat{X},\mat{L},\mat{U}_M]$, where $\mat{X}=\XX'\otimes \XX''$ is the ordered direct product of two one-dimensional input alphabets $\XX'$ and $\XX''$ and all symbols $\mb{x}_i$ are distinct, is FOO if and only if both the following items hold.
\begin{itemize}
\item There exist labelings $\mat{L}'$ and $\mat{L}''$ such that $[\XX',\mat{L}',\mat{U}_{M'}]$ and $[\XX'',\mat{L}'',\mat{U}_{M''}]$ are both FOO (where $M'$ and $M''$ are the sizes of $\XX'$ and $\XX''$, resp.).
\item $\LL=\Pi_\tr{C}(\LL'\otimes \LL'')$, where $\Pi_\tr{C}$ is an arbitrary column permutation.
\end{itemize}

\end{theorem}

\begin{IEEEproof}
The proof is given in Appendix~\ref{Appendix.PAM.QAM.PSK.QAMTheorem}.
\end{IEEEproof}

As a special case, the theorem applies to rectangular QAM constellations since they are defined as the ordered direct product of two PAM input alphabets. In view of Theorem \ref{Sec.PAM.QAM.PSK.PAMTheorem}, and since $\NN_{m'}\otimes \NN_{m''}=\NN_{m'+m''}$, the following corollary gives necessary and sufficient conditions for a rectangular $(M'\times M'')$-QAM constellation to be FOO.

\begin{corollary}
A constellation $[\mat{X}_\tr{QAM},\mat{L},\mat{U}_M]$, where $\mat{X}_\tr{QAM}$ is an $(M'\times M'')$-QAM input alphabet and $M=M'M''=2^m$, is FOO if and only if $\mat{L}=\mat{N}_{m}$, or any other binary labeling that can be derived from $\mat{N}_{m}$ by inverting the bits in certain positions or by permuting the sequence of bits in all codewords.
\end{corollary}

Can a constellation based on an $M$-PSK input alphabet be FOO with a suitably chosen labeling? What about constant-energy constellations in higher dimensions? A complete answer to these questions is given by the following theorem. An intuitive interpretation is that a constellation based on a constant-energy input alphabet is FOO if and only if it forms the vertices of an orthogonal parallelotope, or ``hyperrectangle.''

\begin{theorem} \label{Sec.PAM.QAM.PSK.constant-energy}
A constellation $[\mat{X},\mat{L},\mat{U}_M]$, where $\|\mb{x}_i\|^2$ is constant for all $i=0,\ldots,M-1$, is FOO if and only if $\mat{X}$ can be written in the form \eqref{eq:thD} with orthogonal vectors $\mb{v}_0,\ldots,\mb{v}_{m-1}$.
\end{theorem}

\begin{IEEEproof}
The proof is given in Appendix~\ref{Appendix.PAM.QAM.PSK.constant-enery}.
\end{IEEEproof}

The case of PSK input alphabets follows straightforwardly as a special case of Theorem \ref{Sec.PAM.QAM.PSK.constant-energy}. Indeed, the fact that a set of $m$ orthogonal vectors cannot exist in fewer than $m$ dimensions leads to the following conceptually simple corollaries.

\begin{corollary}
FOO constellations based on constant-energy input alphabets in $N$ dimensions cannot have more than $2^N$ points.
\end{corollary}

\begin{corollary}
No FOO constellations based on $M$-PSK input alphabets exist for $M>4$.
\end{corollary}

Observe that the criterion in Theorem \ref{Sec.PAM.QAM.PSK.constant-energy} is that $\mb{v}_0,\ldots,\mb{v}_{m-1}$ should be orthogonal, not necessarily orthonormal. Thus, FOO constellations based on constant-energy input alphabets are not necessarily hypercubes. In particular, a 4-PSK input alphabet does not have to be equally spaced to give an FOO constellation. Indeed, any rotationally symmetric but nonequally spaced 4-PSK input alphabet (\ie a rectangular one) gives an FOO constellation.

\subsection{$M$-PAM and $M$-PSK Input Alphabets}\label{Sec.PAM.PSK}
In this subsection, we particularize the results in Sec.~\ref{Sec.LinearApp} to practically relevant BICM schemes, \ie $M$-PAM and $M$-PSK input alphabets with uniform input distributions using the four binary labelings defined in Sec.~\ref{Sec.Preliminaries.Labelings}.

\begin{theorem}[Coefficient $\alpha_\Omega^\tr{BI}$ for $\Omega=\text{$[$}\mat{X}_\tr{PAM},\mat{L}_m,\mat{U}_M\text{$]$}$]\label{Theorem.alpha.PAM}
For $M$-PAM input alphabets using $\mat{U}_M$, the coefficient $\alpha_\Omega^\tr{BI}$ for the binary labelings defined in Sec~\ref{Sec.Preliminaries.Labelings} is given by
\begin{align}\label{alpha_BI-PAM}
\alpha^\tr{BI}_\Omega &=
\begin{cases}
\dfrac{3M^2}{4(M^2-1)}\log_2\tr{e},	& \text{if } \mat{L}_m=\mat{G}_m \text{ or } \mat{L}_m=\mat{F}_m \\
\log_2\tr{e},					& \text{if } \mat{L}_m=\mat{N}_m\\
0,							& \text{if } \mat{L}_m=\mat{S}_m\\
\end{cases}.
\end{align}
\end{theorem}

\begin{IEEEproof}
The proof is given in Appendix \ref{Appendix.Theorem.alpha.PAM}.
\end{IEEEproof}

\begin{theorem}[Coefficient $\alpha_\Omega^\tr{BI}$ for $\Omega=\text{$[$}\mat{X}_\tr{PSK},\mat{L}_m,\mat{U}_M\text{$]$}$]\label{Theorem.alpha.PSK}
For $M$-PSK input alphabets using $\mat{U}_M$, the coefficient $\alpha_\Omega^\tr{BI}$ for the binary labelings defined in Sec~\ref{Sec.Preliminaries.Labelings} is given by
\begin{align}\label{alpha_BI-PSK}
\alpha^\tr{BI}_\Omega &=
\begin{cases}
\dfrac{8\log_2\tr{e}}{M^2\sin^2{(\pi/M)}},										& \text{if } \mat{L}_m=\mat{G}_m\\
\dfrac{4\log_2\tr{e}}{M^2\sin^2{(\pi/M)}},										& \text{if } \mat{L}_m=\mat{N}_m\\
\dfrac{4\log_2\tr{e}}{M^2\sin^2{(\pi/M)}} \left[1+\bigl(1-\sec\frac{2\pi}{M}\bigr)^2\right],		& \text{if } \mat{L}_m=\mat{S}_m\\
\dfrac{4\log_2\tr{e}}{M^2\sin^2{(\pi/M)}}\left[1+\ds{\sum_{k=2}^{m}\tan^2{(\pi/2^k)}}\right],	& \text{if } \mat{L}_m=\mat{F}_m\\
\end{cases},
\end{align}
\end{theorem}
where $\sec x = 1/\cos x$ is the secant function.

\begin{IEEEproof}
The proof is given in Appendix \ref{Appendix.Theorem.alpha.PSK}.
\end{IEEEproof}

\begin{figure}
\begin{center}
\begin{tabular}{@{}c@{}}%
\psfrag{ylabel}[c][t][0.7]{$\Pr\set{\alpha_\Omega^\tr{BI}=\alpha}$}%
\psfrag{xlabel}[c][b][0.7]{$\alpha/\log_2\tr{e}$}%
\psfrag{BRGC/FBC}[l][l][0.55]{BRGC/FBC}%
\psfrag{NBC}[l][l][0.55]{NBC}%
\psfrag{BSGC}[l][l][0.55]{BSGC}%
\includegraphics[width=0.50\columnwidth]{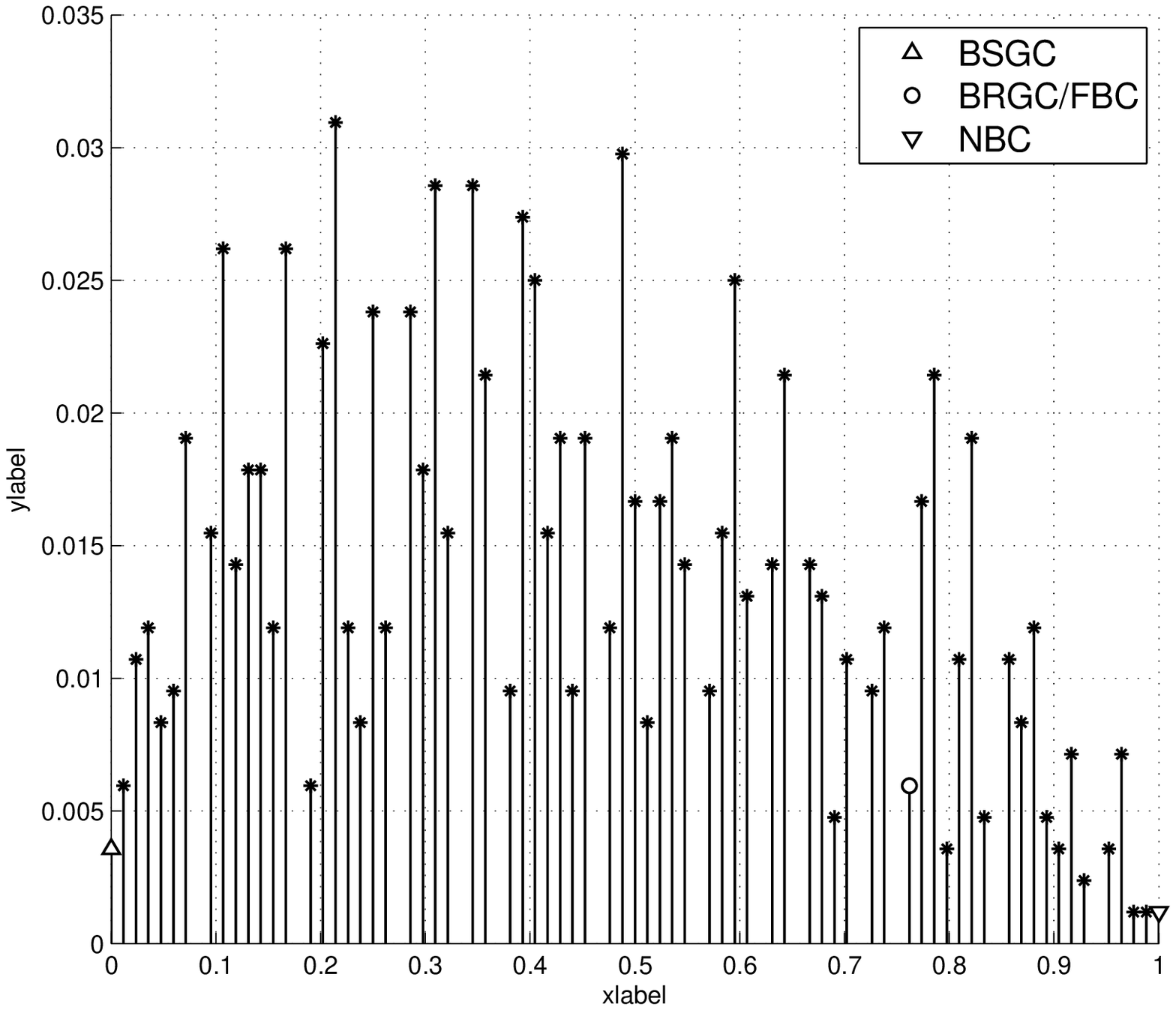}%
\\{\footnotesize (a)}%
\end{tabular}\hfill\begin{tabular}{@{}c@{}}%
\psfrag{ylabel}[c][t][0.7]{$\Pr\set{\alpha_\Omega^\tr{BI}=\alpha}$}%
\psfrag{xlabel}[c][b][0.7]{$\alpha/\log_2\tr{e}$}%
\psfrag{BRGC}[l][l][0.55]{BRGC}%
\psfrag{BSGC}[l][l][0.55]{BSGC}%
\psfrag{NBC}[l][l][0.55]{NBC}%
\psfrag{FBC}[l][l][0.55]{FBC}%
\includegraphics[width=0.50\columnwidth]{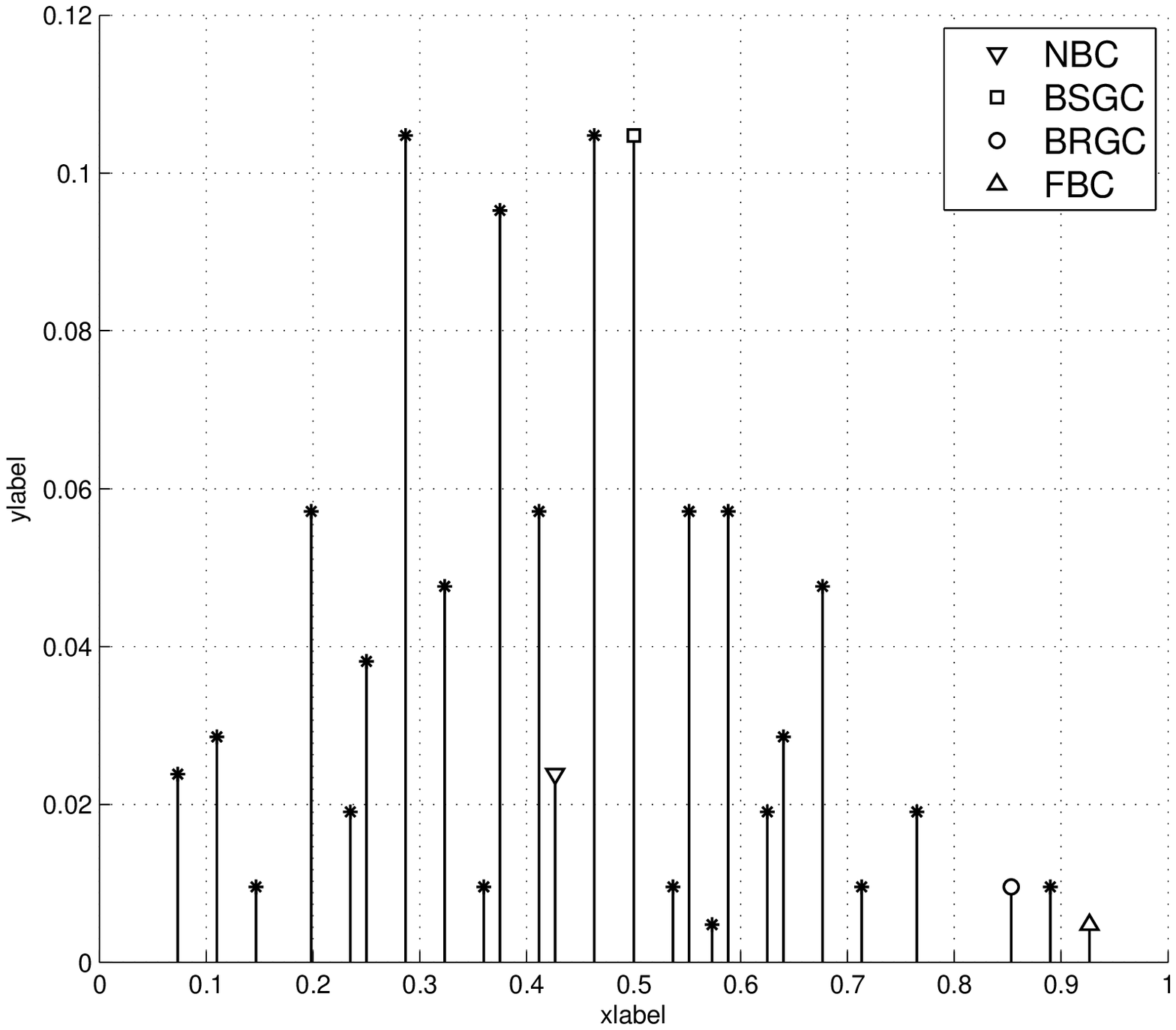}%
\\{\footnotesize (b)}%
\end{tabular}%
	\caption{The pmf of $\alpha_\Omega^\tr{BI}$ for 8-PAM (a) and 8-PSK (b) with $\mat{U}_8$. The four labelings defined in Sec.~\ref{Sec.Preliminaries.Labelings} are shown with white markers.}
	\label{PMF_8PAM}
\end{center}
\end{figure}

In Fig.~\ref{PMF_8PAM}, we present the pmf of the coefficient $\alpha_\Omega^\tr{BI}$ obtained via an exhaustive enumeration of the $8!=40320$ different binary labelings (without discarding trivial operations) for 8-PAM and 8-PSK with $\mat{U}_8$. For 8-PAM, Fig.~\ref{PMF_8PAM} (a) shows that many binary labelings are better than the BRGC at low SNR, the best one being the NBC as found in \cite{Stierstorfer09a}. On the other extreme we find the BSGC, which gives a coefficient equal to zero, reflecting the inferior performance in Fig.~\ref{AW_CM_BI_Capacities_8PAM_BRGC_NBC_BSGC} (b). Based on \eqref{fRc_Asymptotic}, we obtain that the $\Eb/N_0$ for reliable transmission at asymptotically low rates in this case is $\infty$, and it is independent of $M$. We find that among the $8!$ possible binary labelings, there exist 72 \emph{classes} of binary labelings that have a different $\alpha_\Omega^\tr{BI}$, and therefore, a different first-order asymptotic behavior. We also note that the BICM capacity for the BRGC and the FBC in Fig.~\ref{AW_CM_BI_Capacities_8PAM_BRGC_NBC_BSGC} (b) are different for $\SNR>0$. However, their coefficient $\alpha^\tr{BI}_\Omega$ in \eqref{alpha_BI-PAM} is the same, and thus, the curves for these labelings in Fig.~\ref{AW_CM_BI_Capacities_8PAM_BRGC_NBC_BSGC} (b) merge at low rates.

Fig.~\ref{PMF_8PAM} (b) shows that for 8-PSK, there exist only 26 classes of binary labelings with different coefficients $\alpha_\Omega^\tr{BI}$. In particular, the NBC and the BSGC result in a moderate coefficient, and the BRGC in a quite high coefficient. We found that the FBC is the asymptotically optimal binary labeling for 8-PSK, unique up to trivial operations, and we conjecture it to be optimal for any $M$-PSK input alphabet and $m\geq 2$. Interestingly, there are no binary labelings for 8-PSK that give a coefficient zero or one, and the number of distinct pmf values is only ten (25 for 8-PAM).

From \eqref{fRc_Asymptotic} we know that $\alpha_\Omega^\tr{BI}$ determines the behavior of the function $f_\Omega^\tr{BI}(\Rc)$ for asymptotically low rates.  Following the idea introduced in \cite{Martinez08b}, we analyze how the values of  $\alpha^\tr{BI}_\Omega$ for PAM and PSK input alphabets behave when $M\rightarrow\infty$. A summary of the values of $\ds{\lim_{M\rightarrow \infty}\alpha_\Omega^\tr{BI}}$ and $\ds{\lim_{\Rc\rightarrow 0^+} f_\Omega^\tr{BI}(R_C)}$ for $M$-PAM and $M$-PSK input alphabets using $\mat{U}_M$ are presented in Table~\ref{alpha_and_f_asymptotic}, for the four labelings previously analyzed\footnote{The limits $\ds{\lim_{M\rightarrow \infty}\alpha_\Omega^\tr{BI}}$ for $M$-PSK are obtained based on $\lim_{M\rightarrow\infty}\frac{1/M^2}{\sin^2{\pi/M}}=\frac{1}{\pi^2}$ (obtained by L'H\^opital's rule). For the NBC, we obtain numerically that $\sum_{k=2}^{\infty}\tan^2{(\pi/2^k)}\approx 1.2240$, which gives the coefficient 8.89 in Table~\ref{alpha_and_f_asymptotic}.}. For most of the constellations, there is a bounded loss with respect to the SL when $M\rightarrow\infty$. For the BRGC, this difference is 1.25~dB for $M$-PAM and 0.91~dB for $M$-PSK. On the other hand, for the NBC and $M$-PAM, the difference is zero for any $M$. Note that all the coefficients $\alpha_\Omega^\tr{BI}$ in \eqref{alpha_BI-PAM} and in \eqref{alpha_BI-PSK} are nonincreasing functions of $M$.
 
\begin{table}[t]
\renewcommand{\arraystretch}{1.2}
\caption{First-order asymptotics of $M$-PAM and $M$-PSK input alphabets using $\mat{U}_M$ for different binary labelings.}%
\label{alpha_and_f_asymptotic}
\begin{center}
\begin{tabular}{ccccc}
\hline %

\hline%
						& 	\multicolumn{2}{c}{PAM} 			& \multicolumn{2}{c}{PSK}\\

 $\mat{L}_m$ 				& 	$\ds{\lim_{M\rightarrow \infty}\alpha_\Omega^\tr{BI}}$	& $\ds{\lim_{\Rc\rightarrow 0^+} f_\Omega^\tr{BI}(R_C)}$ & $\ds{\lim_{M\rightarrow \infty}\alpha_\Omega^\tr{BI}}$	& $\ds{\lim_{\Rc\rightarrow 0^+} f_\Omega^\tr{BI}(R_C)}$\\
\hline%

\hline%
$\mat{G}_m$ 					&   $\frac{3}{4}\log_2\tr{e}$ 	& $-0.34~\tr{dB}$ 	& $\frac{8}{\pi^2}\log_2\tr{e}$	& $-0.68~\tr{dB}$ \\
$\mat{N}_m$ 					&   $\log_2\tr{e}$ 			& $-1.59~\tr{dB}$ 	& $\frac{4}{\pi^2}\log_2\tr{e}$	& $2.33~\tr{dB}$ \\
$\mat{S}_m$ 					&   $0$ 					& $\infty$ 			& $\frac{4}{\pi^2}\log_2\tr{e}$	& $2.33~\tr{dB}$ \\
$\mat{F}_m$ 					&   $\frac{3}{4}\log_2\tr{e}$ 	& $-0.34~\tr{dB}$ 	& $\frac{8.89}{\pi^2}\log_2\tr{e}$& $-1.14~\tr{dB}$ \\
\hline%
 
\hline
\end{tabular}
\end{center}
\end{table}

\section{Numerical Examples}\label{Sec.NumExamples}

\subsection{Turbo-coded System Simulation}\label{Sec.TurboExamples}

\begin{figure}[t]
\psfrag{ylabel}[c][t][0.8]{BER}%
\psfrag{xlabel}[c][b][0.8]{${\Eb}/{N_0}$~[dB]}%
\psfrag{NBC-Exact-4}[l][l][0.75]{$\mat{N}_{2}$ (4-PAM)}%
\psfrag{BRGC-Exact-4}[l][l][0.75]{$\mat{G}_{2}$ (4-PAM)}%
\psfrag{NBC-Exact-8}[l][l][0.75]{$\mat{N}_{3}$ (8-PAM)}%
\psfrag{BRGC-Exact-8}[l][l][0.75]{$\mat{G}_{3}$ (8-PAM)}%
\begin{center}
	\includegraphics[width=0.55\columnwidth]{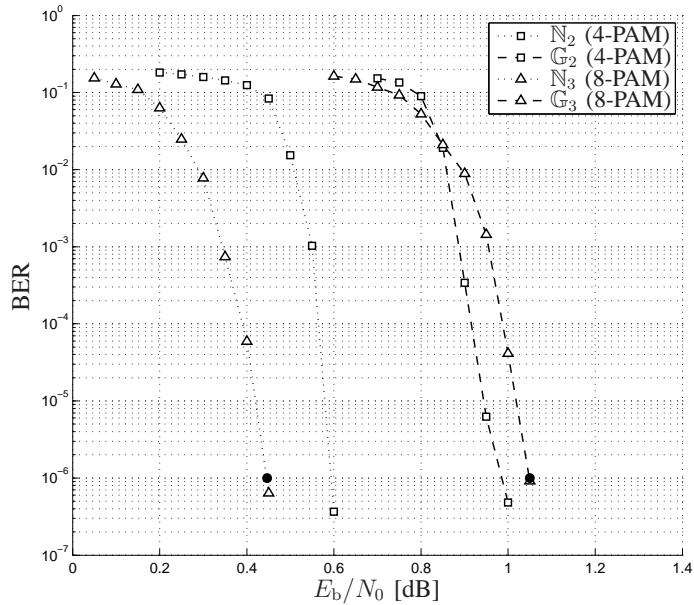}
    \caption{BER for the rate-1/15 turbo code with 4-PAM and 8-PAM for the BRGC and the NBC ($\Rc\approx0.13~\tr{bit/symbol}$ and $\Rc\approx0.2~\tr{bit/symbol}$ respectively). The metrics' computation is based on \eqref{LLR3}, the interleaver size is $N=16384$, the decoder is based on the Log-MAP algorithm, and it performs 12 turbo iterations. The filled circles represent the ${\Eb}/{N_0}$ needed for the configuration to reach a $\tr{BER}=10^{-6}$, which are also shown for 8-PAM in Fig.~\ref{AW_CM_BI_Capacities_8PAM_BRGC_NBC_BSGC} (b).}
    \label{TC_4PAM_and_8PAM_BRGC_NBC}
\end{center}
\end{figure}

In order to validate the analysis presented in the previous sections, we are interested in corroborating if the use of the NBC instead of the BRGC for PAM input alphabets actually translates into a real gain when capacity-approaching codes are used. To this end, we simulate a BICM scheme which combines a very low rate capacity-approaching code with $M$-PAM input alphabets. We use Divsalar's rate-1/15 turbo code, formed by a parallel concatenation of two identical 16-state rate-1/8 recursive systematic convolutional (RSC) convolutional codes defined by their polynomial generators $(1,21/23,25/23,27/23,31/23,33/23,35/23,37/23)_8$ \cite{Divsalar95b}. The two RSCs are separated by a randomly generated interleaver of length $N=16384$, and 64 tail bits are added to terminate the trellis, giving an effective code rate of $R = 16384/(15\cd16384+64)$. We combine this turbo code (via a randomly generated interleaver) with 4-PAM and 8-PAM using NBC or BRGC, yielding $\Rc\approx0.13~\tr{bit/symbol}$ and $\Rc\approx0.2~\tr{bit/symbol}$ respectively. The constellation symbols are equally likely, the decoder is based on the Log-MAP algorithm, and it performs 12 turbo iterations. In Fig.~\ref{TC_4PAM_and_8PAM_BRGC_NBC}, the bit error rate (BER) performance of such a system is presented. 

We study the ${\Eb}/{N_0}$ needed for the four different constellations to reach a BER target $\tr{BER}=10^{-6}$. For 4-PAM, the values for the BRGC and the NBC are, respectively, ${\Eb}/{N_0}=0.99~\tr{dB}$ and ${\Eb}/{N_0}=0.29~\tr{dB}$, \ie the NBC offers a gain of $0.4~\tr{dB}$ compared to the BRGC. For 8-PAM, the obtained values are ${\Eb}/{N_0}=1.05~\tr{dB}$ and ${\Eb}/{N_0}=0.45~\tr{dB}$, which again demonstrate the suboptimality of the BRGC in the low SNR regime. Moreover, we also simulated an 8-PAM input alphabet labeled by the BSGC. We obtained in this case ${\Eb}/{N_0}=8.40~\tr{dB}$, \ie a degradation of 7.95~dB is caused by a bad selection of the binary labeling. The values of ${\Eb}/{N_0}$ obtained for these last three cases are shown in Fig.~\ref{AW_CM_BI_Capacities_8PAM_BRGC_NBC_BSGC} (b). These results show that the turbo-coded system performs within 1~dB of capacity, and that the losses of $0.6~\tr{dB}$ and $7.95~\tr{dB}$ can be observed from the capacity curves as well. This indicates that the results obtained from Fig.~\ref{AW_CM_BI_Capacities_8PAM_BRGC_NBC_BSGC} for different labelings can be used as an \emph{a priori} estimate of the system performance when capacity-approaching codes are used.

\subsection{Capacity vs.~$\Eb/N_0$}

\begin{figure}[!t]
\begin{center}
\begin{tabular}{@{}c@{}}%
\psfrag{xlabel}[cc][cB][0.70]{${\Eb}/{N_0}$~[dB]}%
\psfrag{ylabel}[cc][ct][0.70]{$\Rc$~[bit/symbol]}%
\psfrag{Gaussian-In}[l][l][0.55]{$f^\tr{AW}(\Rc)$}%
\psfrag{CM-BRGC-8}[l][l][0.55]{$f_\Omega^\tr{CM}(\Rc)$}%
\psfrag{BICM-BRGC-8}[l][l][0.55]{$f_\Omega^\tr{BI}(\Rc)$ (BRGC)}%
\psfrag{BICM-NBC-8}[l][l][0.55]{$f_\Omega^\tr{BI}(\Rc)$ (NBC)}%
\psfrag{BICM-FBC-8}[l][l][0.55]{$f_\Omega^\tr{BI}(\Rc)$ (FBC)}%
\psfrag{CM-BRGC-4}[l][l][0.55]{$f_\Omega^\tr{CM}(\Rc)$}%
\psfrag{BICM-NBC-4-As}[l][l][0.55]{$f_\Omega^\tr{BI}(\Rc)$ (Hierarchical)}%
\psfrag{SL}[l][l][0.55]{--1.59~dB}%
\psfrag{4-PAM}[l][l][0.55]{4-PAM}%
\psfrag{8-PAM}[l][l][0.55]{8-PAM}%
\includegraphics[width=0.50\columnwidth]{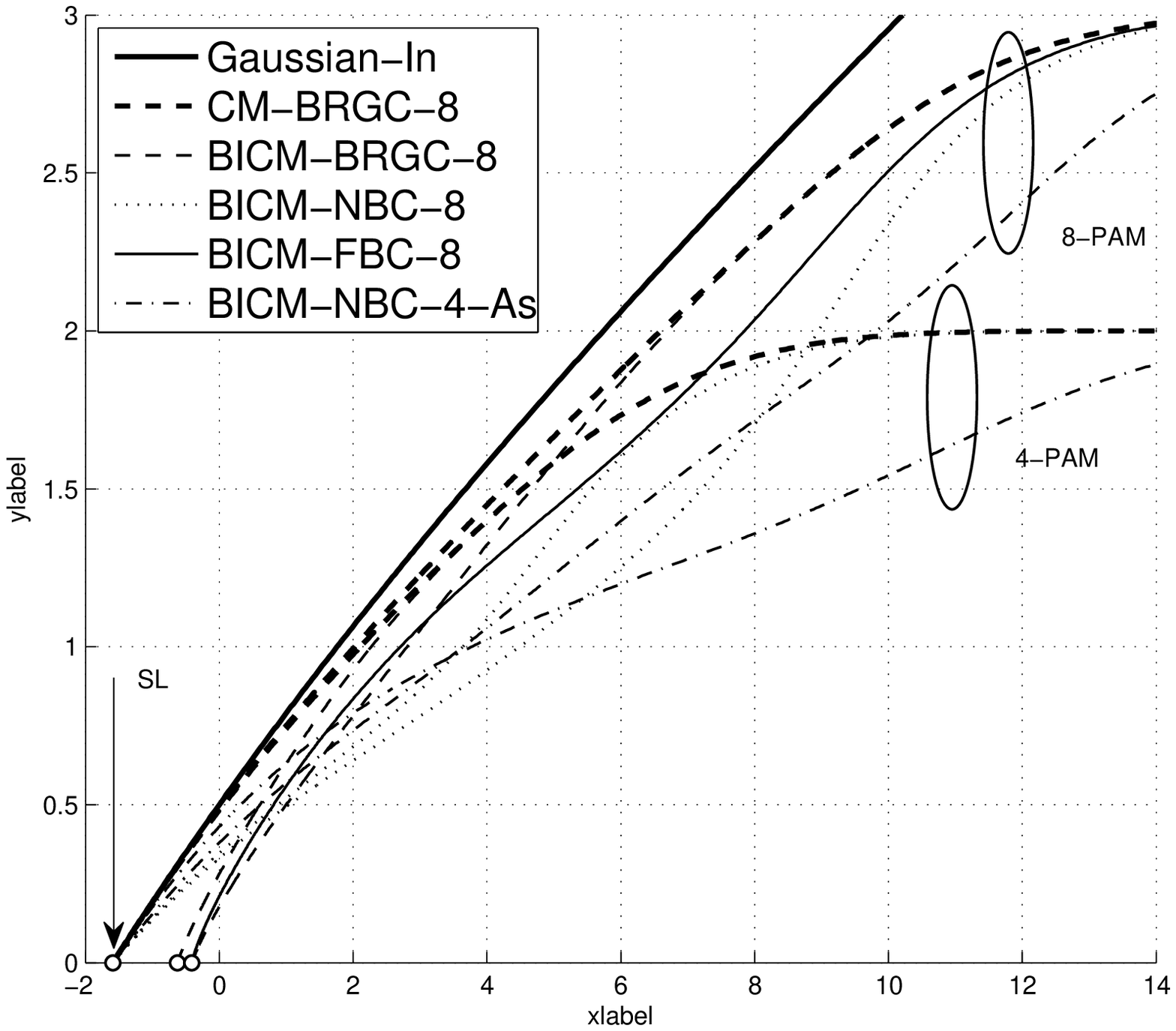}%
\\{\footnotesize (a)}%
\end{tabular}\hfill\begin{tabular}{@{}c@{}}%
\psfrag{xlabel}[cc][cB][0.70]{$\Rc$~[bit/symbol]}%
\psfrag{ylabel}[cc][ct][0.70]{$\Delta_\Omega(\Rc)$~[dB]}%
\psfrag{CM-BRGC-8}[l][l][0.55]{$\Delta_\Omega^\tr{CM}(\Rc)$}%
\psfrag{BICM-BRGC-8}[l][l][0.55]{$\Delta_\Omega^\tr{BI}(\Rc)$ (BRGC)}%
\psfrag{BICM-NBC-8}[l][l][0.55]{$\Delta_\Omega^\tr{BI}(\Rc)$ (NBC)}%
\psfrag{BICM-FBC-8}[l][l][0.55]{$\Delta_\Omega^\tr{BI}(\Rc)$ (FBC)}%
\psfrag{BICM-NBC-4-Asy}[l][l][0.55]{$\Delta_\Omega^\tr{BI}(\Rc)$ (Hierarchical)}%
\psfrag{4PAM}[l][l][0.55][90]{4-PAM}%
\psfrag{8PAM}[l][l][0.55][90]{8-PAM}%
\psfrag{16PAM}[l][l][0.55][90]{16-PAM}%
\includegraphics[width=0.50\columnwidth]{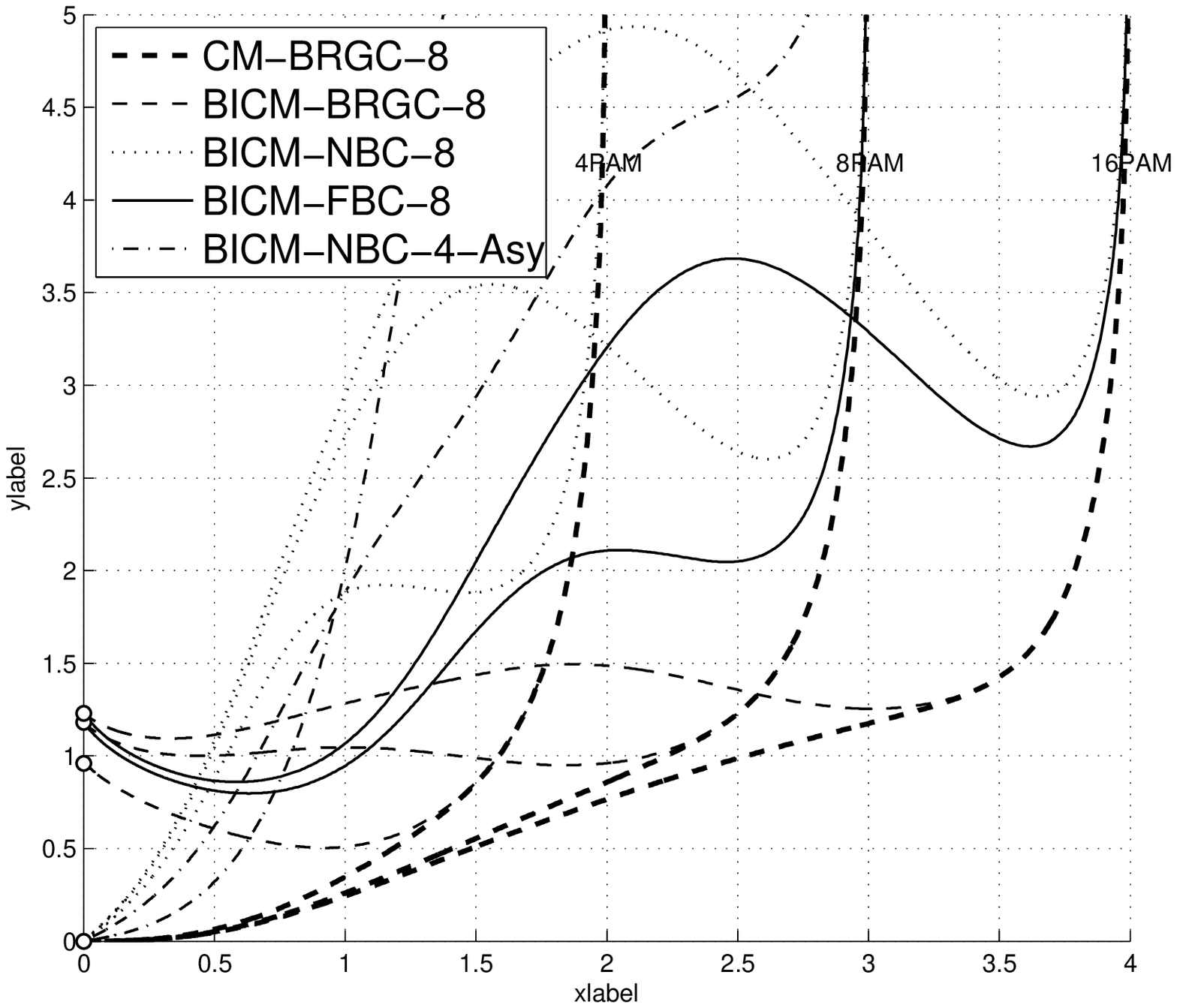}%
\\{\footnotesize (b)}%
\end{tabular}%
\caption{(a) AWGN capacity, CM capacity, and BICM capacities for $M$-PAM with the BRGC, NBC and FBC. The BICM capacity for hierarchical 4-PAM with $\mat{V}=[-1,-5]^\tr{T}$ and 8-PAM with $\mat{V}=[-1,-2,-6]^\tr{T}$ is also shown. The white circles give the performance at $\Rc=0$, where $\alpha_\Omega^\tr{BI}$ determines the BICM capacity. The BRGC and FBC are equivalent for $M=4$. (b) SNR gap $\Delta_\Omega(\Rc)$ in \eqref{delta_rho1} for the same capacities and for 16-PAM.}
    \label{AW_CM_BI_Capacities_PAM_EbN0}
\end{center}
\end{figure}

In Fig.~\ref{AW_CM_BI_Capacities_PAM_EbN0} (a), we show the function $f^\tr{AW}(\Rc)$ and $f_\Omega^\tr{CM}(\Rc)$, defined in Sec.~\ref{Sec.BICM}, using 4-PAM and 8-PAM input alphabets. We also show $f_\Omega^\tr{BI}(\Rc)$ for 4-PAM and 8-PAM input alphabets for different binary labelings and for hierarchical 4-PAM and 8-PAM constellations (Example~\ref{Hierarchical_Example}). The curves in Fig.~\ref{AW_CM_BI_Capacities_PAM_EbN0} (a) intersect the horizontal axis at ${\Eb}/{N_0}=1/\alpha_\Omega$, where $1/\alpha^\tr{AW}=1/\log_2{\tr{e}}=-1.59~\tr{dB}$ represents the SL. From this figure, we observe that for CM both constellations are FOO, while for BICM only four of them are FOO, the ones labeled by the NBC.

\begin{figure}[!t]
\begin{center}
\begin{tabular}{@{}c@{}}%
\psfrag{xlabel}[cc][cB][0.70]{${\Eb}/{N_0}$~[dB]}%
\psfrag{ylabel}[cc][ct][0.70]{$\Rc$~[bit/symbol]}%
\psfrag{Gaussian-In}[l][l][0.55]{$f^\tr{AW}(\Rc)$}%
\psfrag{CM-BRGC-8P}[l][l][0.55]{$f_\Omega^\tr{CM}(\Rc)$}%
\psfrag{BI-BRGC-8P}[l][l][0.55]{$f_\Omega^\tr{BI}(\Rc)$ (BRGC)}%
\psfrag{BI-NBC-8P}[l][l][0.55]{$f_\Omega^\tr{BI}(\Rc)$ (NBC)}%
\psfrag{BI-FBC-8P}[l][l][0.55]{$f_\Omega^\tr{BI}(\Rc)$ (FBC)}%
\psfrag{BI-BSGC-8P}[l][l][0.55]{$f_\Omega^\tr{BI}(\Rc)$ (BSGC)}%
\psfrag{BI-OTTO}[l][l][0.55]{$f_\Omega^\tr{BI}(\Rc)$ (OTTO)}%
\psfrag{BI-OTOTO}[l][l][0.55]{$f_\Omega^\tr{BI}(\Rc)$ (OTOTO)}%
\psfrag{SL}[l][l][0.55]{--1.59~dB}%
\psfrag{OTTO}[cc][cc][0.55]{OTTO}%
\psfrag{OTOTO}[cc][cc][0.55]{OTOTO}%
\includegraphics[width=0.50\columnwidth]{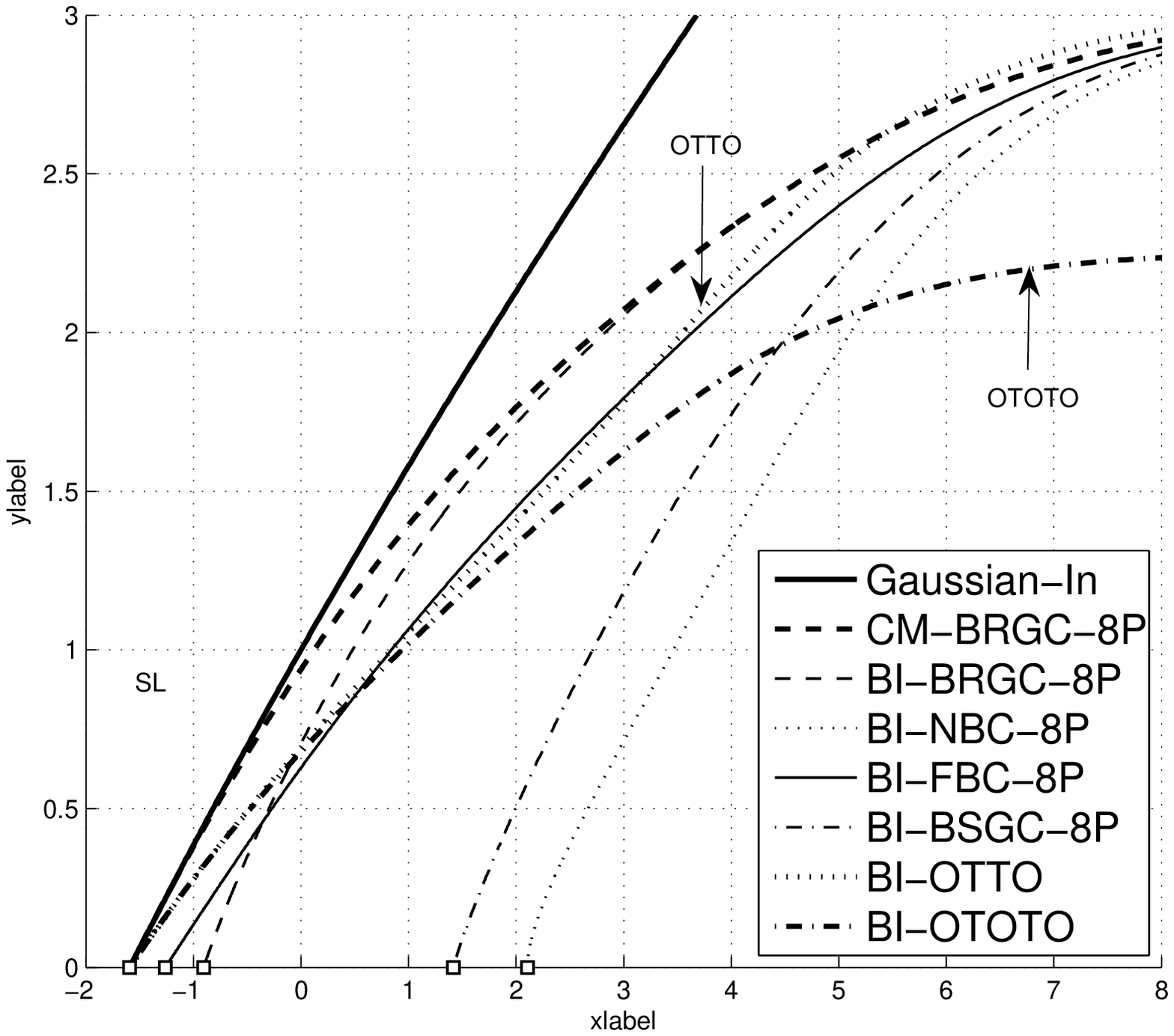}%
\\{\footnotesize (a)}%
\end{tabular}\hfill\begin{tabular}{@{}c@{}}%
\psfrag{xlabel}[cc][cB][0.70]{$\Rc$~[bit/symbol]}%
\psfrag{ylabel}[cc][ct][0.70]{$\Delta_\Omega(\Rc)$~[dB]}%
\psfrag{CM-BRGC-8P}[l][l][0.55]{$\Delta_\Omega^\tr{CM}(\Rc)$}%
\psfrag{BI-BRGC-8P}[l][l][0.55]{$\Delta_\Omega^\tr{BI}(\Rc)$ (BRGC)}%
\psfrag{BI-NBC-8P}[l][l][0.55]{$\Delta_\Omega^\tr{BI}(\Rc)$ (NBC)}%
\psfrag{BI-FBC-8P}[l][l][0.55]{$\Delta_\Omega^\tr{BI}(\Rc)$ (FBC)}%
\psfrag{BI-BSGC-8P}[l][l][0.55]{$\Delta_\Omega^\tr{BI}(\Rc)$ (BSGC)}%
\psfrag{BI-OTTO}[l][l][0.55]{$\Delta_\Omega^\tr{BI}(\Rc)$ (OTTO)}%
\psfrag{OTTO}[cc][cc][0.55]{OTTO}%
\psfrag{BI-OTOTO}[l][l][0.55]{$\Delta_\Omega^\tr{BI}(\Rc)$ (OTOTO)}%
\psfrag{OTOTO}[cc][cc][0.55]{OTOTO}%
\includegraphics[width=0.50\columnwidth]{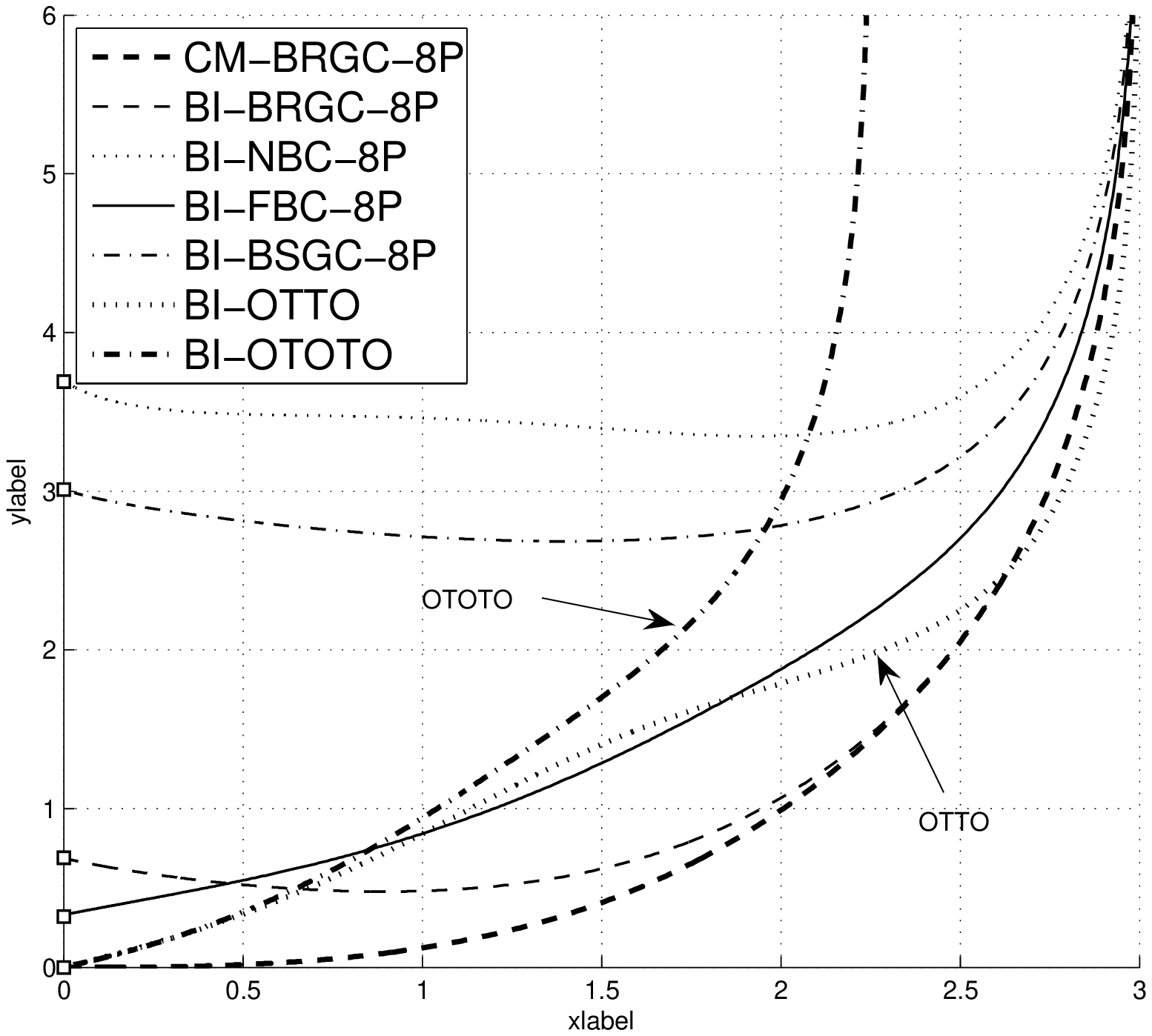}%
\\{\footnotesize (b)}%
\end{tabular}%
\caption{AWGN capacity, CM capacity, and BICM capacities for 8-PSK with the BRGC, NBC, FBC and BSGC. The BICM capacity for the OTTO and OTOTO constellations are also shown. The white circles give the performance at $\Rc=0$, where $\alpha_\Omega^\tr{BI}$ determines the BICM capacity. (b) SNR gap $\Delta_\Omega(\Rc)$ for the same capacities.}
    \label{AW_CM_BI_Capacities_PSK_EbN0}
\end{center}
\end{figure}

In Fig.~\ref{AW_CM_BI_Capacities_PSK_EbN0} (a), similar results for 8-PSK are shown. We also include the results for the OTTO and OTOTO constellations in Fig.~\ref{OTTO_and_OTOTO} (Example~\ref{OTTO_OTOTO_Example}). From this figure, we observe that for the CM capacity the 8-PSK input alphabet gives an FOO constellation, and for BICM, the OTTO and the OTOTO constellations are FOO. Moreover, for high SNR, the OTTO constellation results in a higher capacity than the constellations based on 8-PSK input alphabets.

\subsection{The SNR Gap}

Borrowing the idea from \cite{Barsoum07}, we define the \emph{SNR gap} as the horizontal difference%
\footnote{The gap is the same regardless of whether the horizontal axis represents $\Eb/N_0$ or $\SNR$.}
between the CM and BICM capacity and the capacity of the AWGN channel for a given $\Rc$, \ie 
\begin{align}
\Delta_\Omega^\tr{CM}(\Rc) 	=  \frac{f_\Omega^\tr{CM}(\Rc)}{f^\tr{AW}(\Rc)}, \qquad
\Delta_\Omega^\tr{BI}(\Rc) 	=  \frac{f_\Omega^\tr{BI}(\Rc)}{f^\tr{AW}(\Rc)}
.\label{delta_rho1}
\end{align}
These expressions, which represent the additional energy needed for a given constellation to achieve the same $\Rc$ as the optimal scheme (the AWGN capacity), are evaluated numerically in Figs.~\ref{AW_CM_BI_Capacities_PAM_EbN0} (b) and \ref{AW_CM_BI_Capacities_PSK_EbN0} (b). In Table~\ref{Delta_asymptotic}, we present a summary of the SNR gap at asymptotically low rates for different constellations. This asymptotic SNR gap is given by $\log_2{\tr{e}}/\alpha_\Omega^\tr{BI}$, and is a scaled special case of the results presented in Sec.~\ref{Sec.PAM.PSK}.

\begin{table}[t]
\renewcommand{\arraystretch}{1.2}
\caption{The SNR gap at asymptotically low rates for BICM and different constellations.}%
\label{Delta_asymptotic}
\begin{center}
\begin{tabular}{c|c|c}
\hline %

\hline%
\multicolumn{2}{c}{Constellation}		&	$\log_2{\tr{e}}/\alpha_\Omega^\tr{BI}$~[dB]\\
\hline %

\hline%
\multirow{2}{*}{4-PAM}				&	BRGC/FBC 	& 0.96\\
								&	NBC 		& 0\\
\hline
\multicolumn{2}{c|}{4-PAM Hierarchical}	&	0 \\
\hline
\multirow{3}{*}{8-PSK}				&	BRGC		& 0.69\\
								&	NBC 		& 3.69\\
								&	FBC 			& 0.32\\
								&	BSGC 		& 3.01\\
\hline
\multicolumn{2}{c|}{OTTO}			&	0 \\
\hline
\multicolumn{2}{c|}{OTOTO}			&	0 \\
\hline
\multicolumn{2}{c|}{8-PAM Hierarchical}	&	0 \\
\hline
\multirow{2}{*}{8-PAM}				&	BRGC/FBC 	& 1.18\\
								&	NBC 		& 0\\
								&	BSGC 		& $\infty$\\
\hline
\multirow{2}{*}{16-PAM}				&	BRGC/FBC 	& 1.23\\
								&	NBC 		& 0\\
								&	BSGC 		& $\infty$\\
\hline%
 
\hline
\end{tabular}
\end{center}
\end{table}

\section{Conclusions}\label{Sec.Conclusions}

In this paper, we introduced a general model for BICM which considers arbitrary input alphabets, input distributions, and binary labelings, and we analyzed different aspects of the BICM capacity. Probabilistic shaping for BICM was analyzed and the relation between the BICM capacity and $\Eb/N_0$ was studied. Four binary labelings (BRGC, NBC, BSGC, and FBC) were analyzed in detail, and for 8-PAM with uniform input distribution, the results showed that the BICM capacity is maximized by the NBC and the FBC for 36\% of the rates.

First-order asymptotic of the BICM capacity for arbitrary constellations were presented, which allowed us to analyze the behavior of the BICM capacity for low rates. The $\Eb/N_0$ required for reliable transmission at asymptotically low rates was found to take values between the SL $-1.59$~dB and infinity. The asymptotic analysis was used to compare binary labelings for PAM and PSK input alphabets, as well as to predict the actual system performance at low rates when capacity-approaching codes are used. The asymptotically best labelings for $M$-PAM and $M$-PSK with uniform input distributions appear to be the NBC and FBC, respectively.

Using the first-order asymptotic of the BICM capacity, we analyzed the problem of FOO constellations for BICM. We showed that, under some mild conditions, the fading does not change the analysis of FOO constellations made for the AWGN channel. Interpreting the codewords of a binary labeling as the vertices of a hypercube, a constellation for BICM with uniform input distributions is FOO if and only if the input alphabet forms a linear projection of this hypercube. Important special cases of this result are that constellations based on equally spaced $M$-PAM and $M$-QAM input alphabets are FOO if and only if the NBC is used. Another particular case are the hierarchical (nonequally spaced) $M$-PAM input alphabets labeled by the NBC. We also showed that constellations based on constant-energy $M$-PSK input alphabets can never be FOO if $M>4$, regardless of the binary labeling.

In this paper, we focused on asymptotically low rates, and we answered the question about FOO constellations for this case. The analysis of second-order optimal constellations for BICM, and the dual problem for asymptotically high rates, or more generally, for any rate, is still an open research problem.

\appendices

\section{Proof of Theorem~\ref{PrelovTheo3}}\label{Appendix.PrelovTheo3}

In \cite[Theorem~3]{Prelov04}, the model $\mb{Y}=\tilde{\mb{H}}\mb{X}+\mb{Z}$ is considered, where $\tilde{\mb{H}}$ is a matrix. This theorem states that the AMI between $\mb{X}$ and $\mb{Y}$ when $\tilde{\mb{H}}$ is known at the receiver can be expressed as
\begin{align}\label{Prelov_expansion_KnownH}
I_{\mb{X}}(\mb{X};\mb{Y}) = \frac{\log_2{\tr{e}}}{N_0}\trace{\ms{E}_{\tilde{\mb{H}}}[\tilde{\mb{H}}\cov{\mb{X}}\tilde{\mb{H}}^\tr{T}]} + O(N_0^{-2})
\end{align}
when $N_0\rightarrow \infty$, if the two following conditions are fulfilled:
\begin{itemize}
\item There exist finite constants $c>0$ and $d>0$ such that $\ms{E}_{\mb{X}}[\|\mb{X}\|^{4+d}]<c$.
\item There exists a constant $\nu>0$ such that the matrix $\tilde{\mb{H}}$ satisfies $\Pr\set{\|\tilde{\mb{H}}\|>\delta}\leq\exp(-\delta^\nu)$ for all sufficiently large $\delta>0$.
\end{itemize}

Since we consider real-valued vectors only, we have replaced the Hermitian conjugates in \cite{Prelov04} by transpositions in \eqref{Prelov_expansion_KnownH}. Moreover, \cite[Theorem~3]{Prelov04} requires $\mb{Z}$, $\mb{X}$, and $\tilde{\mb{H}}\mb{X}$ to be ``proper complex''. Nevertheless, the results are still valid if the two conditions in the items above are fulfilled, as explained in \cite[Remark~6]{Prelov04}.

The first condition is fulfilled since $\mb{x}_0,\ld,\mb{x}_{M-1}$ are all finite, and therefore, $\ms{E}_{\mb{X}}[\|\mb{X}\|^d]< \infty$ for all $d>0$. The second condition is fulfilled because $\tilde{\mb{H}}=\diag{\mb{H}}$ and because of the condition \eqref{H_condition_new} imposed on $\mb{H}$. Moreover, since $\tilde{\mb{H}}=\diag{\mb{H}}$ and $\mb{H}$ contains i.i.d. elements, $\ms{E}_{\tilde{\mb{H}}}[\tilde{\mb{H}}\cov{\mb{X}} \tilde{\mb{H}}^\tr{T}]=\ms{E}_H[H^2]\cov{\mb{X}}$. The use of the identity $\trace{\cov{\mb{X}}}=\ms{E}_{\mb{X}}[\|\mb{X}\|^2]-\|\ms{E}_{\mb{X}}[\mb{X}]\|^2$, the definition of $\Es$, and the relation $\SNR=\ms{E}_H[H^2]\frac{\Es}{N_0}$ in \eqref{Prelov_expansion_KnownH}, gives \eqref{PrelovTheo3.2}.

\section{Proof of Theorem~\ref{theorem.alpha.bicm.general}}\label{Appendix.theorem.alpha.bicm.general}

Expanding the inner sum in \eqref{a1.BICM.general}, we obtain
\begin{align*}
\sum_{u\in\set{0,1}}\PP_{C_k}(u)\|\ms{E}_{\mb{X}|C_k=u}[\mb{X}]\|^2=
\PP_{C_k}(0)\biggl\|\sum_{i\in\mc{I}_{k,0}}\mb{x}_i P_{\mb{X}|C_k=0}(\mb{x}_i) \biggr\|^2
+
\PP_{C_k}(1)\biggl\|\sum_{i\in\mc{I}_{k,1}}\mb{x}_i P_{\mb{X}|C_k=1}(\mb{x}_i) \biggr\|^2.
\end{align*}
Using the identity $\|\mb{a}\|^2+\|\mb{b}\|^2 = \frac{1}{2}\|\mb{a}-\mb{b}\|^2+\frac{1}{2}\|\mb{a}+\mb{b}\|^2$ and \eqref{Pxcond}, we obtain
\begin{align*}
\sum_{u\in\set{0,1}}\PP_{C_k}(u)&\|\ms{E}_{\mb{X}|C_k=u}[\mb{X}]\|^2 \\
&=
\frac{1}{2}
\biggl\|\sqrt{\PP_{C_k}(0)}\sum_{i\in\mc{I}_{k,0}}\mb{x}_i P_{\mb{X}|C_k=0}(\mb{x}_i) - \sqrt{\PP_{C_k}(1)}\sum_{i\in\mc{I}_{k,1}}\mb{x}_i P_{\mb{X}|C_k=1}(\mb{x}_i) \biggr\|^2\\
&\qquad+
\frac{1}{2}
\biggl\|\sqrt{\PP_{C_k}(0)}\sum_{i\in\mc{I}_{k,0}}\mb{x}_i P_{\mb{X}|C_k=0}(\mb{x}_i)+\sqrt{\PP_{C_k}(1)} \sum_{i\in\mc{I}_{k,1}}\mb{x}_i P_{\mb{X}|C_k=1}(\mb{x}_i) \biggr\|^2\\
&=
\frac{1}{2}
\biggl\|\frac{1}{\sqrt{\PP_{C_k}(0)}}\sum_{i\in\mc{I}_{k,0}}\mb{x}_i P_{\mb{X}}(\mb{x}_i) -\frac{1}{\sqrt{\PP_{C_k}(1)}} \sum_{i\in\mc{I}_{k,1}}\mb{x}_i P_{\mb{X}}(\mb{x}_i) \biggr\|^2\\
&\qquad+
\frac{1}{2}
\biggl\|\frac{1}{\sqrt{\PP_{C_k}(0)}}\sum_{i\in\mc{I}_{k,0}}\mb{x}_i P_{\mb{X}}(\mb{x}_i) +\frac{1}{\sqrt{\PP_{C_k}(1)}} \sum_{i\in\mc{I}_{k,1}}\mb{x}_i P_{\mb{X}}(\mb{x}_i) \biggr\|^2.
\end{align*}
In this expression, and based on the definition of $q_{i,k}$ in \eqref{qki_def}, we recognize the first term as the first term inside the outer sum in \eqref{alpha_BICM_general}, and the second term as the second term inside the outer sum in \eqref{alpha_BICM_general}. This used in \eqref{a1.BICM.general} completes the proof.

\section{Proof of Theorem~\ref{Sec.PAM.QAM.PSK.PAMTheorem}}\label{Appendix.PAM.QAM.PSK.PAMTheorem}

Consider any FOO constellation $[\mat{X}_\tr{PAM},\mat{L},\mat{U}_M]$, where the binary labeling is defined by $q_{i,k}$ for $k=0,\ldots,m-1$ and $i=0,\ldots,M-1$. From Theorem \ref{thD}, there exist real values $v_k$ for $k=0,\ldots,m-1$ such that \eqlab{xi-pam}{
   x_i = \sum_{k=0}^{m-1}q_{i,k}v_k
}
for $i=0,\ldots,M-1$. We wish to find all combinations of $q_{i,k}$ and $v_k$ that satisfy \eqref{xi-pam}.

We start by giving two properties of the column vector $\mat{V}=[v_0,v_1, \ldots,v_{m-1}]^\tr{T}$ that will be used later in the proof.
\begin{itemize}
\item Since all pairwise differences $x_i-x_j=\sum_{k=0}^{m-1}(q_ {i,k}-q_{j,k})v_k$ are even numbers, and since $q_{i,k}-q_{j,k} \in \set{0,\pm 2}$, we conclude that $\mat{V} \in \ms{Z}^m$.
\item Because of \eqref{xi-pam}, the sum $\pm v_0 \pm v_1 \pm \cdots \pm v_{m-1}$, with all combinations of signs, generates all the elements in $\XX_\tr{PAM}$. Since $\XX_\tr{PAM}$ is formed by $M$ distinct elements, $\pm v_0 \pm v_1 \pm \cdots \pm v_{m-1}$ must yield $M$ different values, and therefore, $|v_k|$ for $k=0,\ldots,m-1$ must all be distinct.
\end{itemize}

Consider a given bit position $l \in \{0,\ldots,m-1\}$ and define, for $i=0,\ldots,M-1$,
\eqlab{si}{ s_i &\triangleq x_i \mod 2v_l \nonumber\\
     &= \sum_{\substack{k=0\\k\ne l}}^{m-1} q_{i,k} v_k \pm v_l        
\mod 2v_l	\nonumber\\
   &= \sum_{\substack{k=0\\k\ne l}}^{m-1} q_{i,k}v_k + v_l \mod 2v_l,
   }
where in the last step we have used the identity $(a \pm b) \mod 2b = (a + b) \mod 2b$. Because $x_i$ is an odd integer, $s_i \in \{1,3,\ldots,2v_l-1\}$ for all $i$. We will now study the vector $\mat{S}=[s_0,s_1,\ldots,s_{M-1}]^\tr{T}$ and in particular count how many times each odd integer occurs in this vector. We will do this in two ways, in order to determine which values $v_l$ can take on.

\begin{itemize}
\item It follows from \eqref{si} that $s_i$ is independent of $q_{i,l}$ for all $i$. Thus, if two codewords $\mb{c}_i$ and $\mb{c}_j$ differ only in bit $l$, then $s_i = s_j$. This proves that each value $1,3,\ldots,2v_l-1$ occurs an even number of times in $\mat{S}$.
\item Because $\mat{X}$ is a vector of odd integers in increasing order and $\mat{S}$ consists of the same elements counted modulo $2v_l$, $\mat{S}$ consists of identical segments $[1,3,\ldots,2v_l-1]^\tr{T}$ of length $v_l$. If $v_l$ divides $M$, then $\mat{S}$ contains a whole number of such segments and each value in $\{1,3,\ldots,2v_l-1\}$ occurs exactly $M/v_l$ times in $\mat{S}$. If on the other hand $v_l$ does not divide $M$, then the first and the last segment are truncated. In this case, $\mat{S}$ includes some values $\lfloor M/v_l \rfloor$ times and other values $\lfloor M/v_l \rfloor+1$ times, where  $\lfloor \cd \rfloor$ denotes the integer part.
\end{itemize}

Since either $\lfloor M/v_l \rfloor$ or $\lfloor M/v_l \rfloor+1$ is odd, and 
each value must occur in $\mat{S}$ an even number of times, we conclude from these two properties that $v_l$ must divide $M$. Furthermore, the number of occurrences $M/v_l$ must be even, so $v_l$ must divide $M/2$.

In conclusion, $v_0,\ldots,v_{m-1}$ must all divide $M/2$, and their absolute values must be all distinct. Since $M/2$ has only $m$ divisors, which are $1,2,4,\ldots,2^{m-1}$, they must all appear in $\mat{V}$, but they can do so in any order and with any sign. If $\mat{V} = [-1,-2,-4,\ldots,-2^{m-1}]^\tr{T}$, then \eqref{xi-pam} is fulfilled by the NBC $\mat{N}_m$ (see Example \ref{ex:mpam}). Negating $v_k$ for any $k$ corresponds to inverting bit $k$ of the NBC, whereas reordering the rows of $\mat{V}$ corresponds to permuting columns in $\mat{N}_m$.

\section{Proof of Theorem~\ref{Sec.PAM.QAM.PSK.QAMTheorem}}\label{Appendix.PAM.QAM.PSK.QAMTheorem}

Let $M' = 2^{m'}$, $M'' = 2^{m''}$, and $M = 2^m = 2^{m'+m''}$.
To prove the ``if'' part, we assume that there exist two FOO constellations $[\XX',\mat{L}',\mat{U}_{M'}]$ and $[\XX'',\mat{L}'',\mat{U}_{M''}]$. Then, from Theorem \ref{thD},
\eqlab{D1}{
x'_i &= \sum_{k=0}^{m'-1} q'_{i,k} v'_k, \qquad i=0,\ldots,M'-1 ,\\
x''_i &= \sum_{k=0}^{m''-1} q''_{i,k} v''_k, \qquad i=0,\ldots,M''-1 \label{D2}
.}
We analyze the two-dimensional constellation $\Omega$ constructed as $\Omega=[\XX,\mat{L},\mat{U}_M] = [\XX'\otimes\XX'',\mat{L}'\otimes\mat{L}'',\mat{U}_{M'M''}]$. It follows from the definition of the operator $\otimes$ that for all $l=0,\ldots,M'-1$, $j=0,\ldots,M''-1$, and $k=0,\ldots,m-1$,
\eq{
\mb{x}_{M''l+j} &= [x'_l,x''_j] \\
q_{M''l+j,k} &= 
\begin{cases}
q'_{l,k}, & k=0,\ldots,m'-1 \\
q''_{j,k-m'}, & k=m',\ldots,m-1 
\end{cases}
.}

We will now show that $\Omega$ is FOO by explicitly constructing a matrix $\mat{V}$ that satisfies Theorem \ref{thD}. To this end, we define the vectors
\eq{
\bv_k = \begin{cases}
[v'_k,0], & k = 0,\ldots,m'-1 \\
[0,v''_{k-m'}], & k = m',\ldots,m-1
\end{cases}
,}
with $v'_k$ and $v''_k$ that satisfy \eqref{D1}--\eqref{D2}. The vectors $\bv_k$ constructed in this manner have the property that for all $l=0,\ldots,M'-1$ and $j=0,\ldots,M''-1$,
\begin{align}
\nonumber
\sum_{k=0}^{m-1} q_{M''l+j,k}\bv_k
  &= \sum_{k=0}^{m'-1} q_{M''l+j,k}\bv_k + \sum_{k=0}^{m''-1} q_{M''l+j,k+m'}\bv_{k+m'} \\
\nonumber
  &= \sum_{k=0}^{m'-1} q'_{l,k} [v'_k,0] + \sum_{k=0}^{m''-1} q''_{j,k} [0,v''_k] \\
\nonumber
  &= \left[ \sum_{k=0}^{m'-1} q'_{l,k} v'_k , \sum_{k=0}^{m''-1} q''_{j,k} v''_k \right] \\
\nonumber
  &= [x'_l,x''_j] \\
\nonumber
  &= \mb{x}_{M''l+j}
.
\end{align}
Substituting $M''l+j = i$ yields \eqref{e7}, which shows that $\Omega$ is FOO. Finally, to show that the constellation $[\XX'\otimes\XX'',\Pi_\tr{C}(\mat{L}'\otimes\mat{L}''),\mat{U}_{M'M''}]$ is also FOO, it suffices to observe that synchronously permuting the columns of $\mat{Q}(\LL)$ and $\mat{V}^\tr{T}$ does not change the right-hand side of \eqref{eq:thD}, which completes the proof of the ``if'' part\footnote{An intuitive explanation for this is that reordering the bits of all codewords does not change the constellation's performance.}.

For the ``only if'' part, consider any two-dimensional FOO constellation $[\mat{X},\mat{L},\mat{U}_M]$. By Theorem \ref{thD}, the elements of $\mat{X}$ fulfill \eqref{e7}, which can be decomposed into scalar equalities as
\eqlab{xqv1}{
x_{i,n} &= \sum_{k=0}^{m-1} q_{i,k} v_{k,n}, \qquad i=0,\ldots,M-1,\quad n=0,1
}
where $\mb{x}_i = [x_{i,0},x_{i,1}]$ for $i=0,\ldots,M-1$ and $\bv_k = [v_{k,0},v_{k,1}]$ for $k=0,\ldots,m-1$. We will use this decomposition to characterize the points with the largest coordinate value in one of the dimensions. Because $q_{i,k} \in \set{-1,1}$, $x_{i,n}$ takes values in the set $\pm v_{0,n} \pm \cdots \pm v_{m-1,n}$. The largest of these values is
\eq{
  \hat{x}_n \triangleq \max_{i=0,\ldots,M-1}x_{i,n} = |v_{0,n}| + \cdots + |v_{m-1,n}|
.}
If $v_{k,n} \ne 0$ for all $k=0,\ldots,m-1$, then the symbol $\mb{x}_i$ for which $x_{i,n} = \hat{x}_n$ is unique. If $v_{k,n} = 0$ for one value of $k$, then $x_{i,n} = \hat{x}_n$ for two values of $i$, and so on.  Generalizing, there exist $2^a$ symbols for which $x_{i,n} = \hat{x}_n$ if and only if there are $a$ zeros among $v_{0,n},\ldots,v_{m-1,n}$. Analogous relations hold for the minimum of $x_{i,n}$.

For the special case when $\XX$ is obtained from two one-dimensional input alphabets $\XX'$ and $\XX''$ as $\XX = \XX' \otimes \XX''$, the two-dimensional symbols are $\mb{x}_{M''l+j} = [x'_l,x''_j]$ for $l=0,\ldots,M'-1$ and $j=0,\ldots,M''-1$. We will prove that there exist labelings $\LL'$ and $\LL''$ such that $[\XX',\LL',\mat{U}_{M'}]$ and $[\XX'',\LL'',\mat{U}_{M''}]$ are both FOO, and we will identify the set of all such labelings. We do this by analyzing $v_{k,n}$ for $n=0$ and 1 separately, beginning with $n=0$. There are $M''$ symbols $\mb{x}_i$ having $x_{i,0}=x'_l$ for each $l=0,\ldots,M'-1$. This holds in particular for $x'_l = \hat{x}_0$. From the result in the previous paragraph, there are therefore $m''$ zeros among $v_{0,0},\ldots,v_{m-1,0}$. We will first consider the special case when the zeros are $v_{m',0},\ldots,v_{m-1,0}$, i.e., when
\begin{align}\label{vzeros}
[v_{0,0}, \ld , v_{m'-1,0}, v_{m',0}, \ld , v_{m-1,0}] 
=
[\underbrace{v_{0,0}, \ld v_{m'-1,0} }_{\text{$m'$ nonzero elements}} , \underbrace{0, \ld ,0 }_{\text{$m''$ zeros}}]
,\end{align}
and will later generalize the obtained results to an arbitrary location of the $m''$ zeros.

Assuming that \eqref{vzeros} holds, $x'_l$ can, for all $l=0,\ldots,M'-1$, be written as
\eqlab{xj}{
  x'_l &= x_{M''l+j,0} \nonumber\\
   &= \sum_{k=0}^{m-1} q_{M''l+j,k} v_{k,0} \nonumber\\
   &= \sum_{k=0}^{m'-1} q_{M''l+j,k} v_{k,0}
}
where the second line follows from \eqref{xqv1} and the third from \eqref{vzeros}. The relation holds for all $j=0,\ldots,M''-1$.

We will now conclude from \eqref{xj} that
\eqlab{qp1}{
q_{M''l+j,k} = q_{M''l,k},\qquad l=0,\ldots,M'-1, \quad j=0,\ldots,M''-1, \quad k=0,\ldots,m'-1
.}
This can be seen as follows. The sequence $q_{M''l+j,0},\ldots,q_{M''l+j,m'-1}$ can take on $2^{m'} = M'$ different values, because each element is $\pm 1$. For given values of $v_{k,0}$, these sequences all yield different values of $x'_l$ in \eqref{xj}, because these values are, by assumption, all distinct. Thus the sequence $q_{M''l+j,0},\ldots,q_{M''l+j,m'-1}$ is uniquely determined by $x'_l$ and $v_{0,0},\ldots,v_{m'-1,0}$. Since both $x'_l$ and $v_{k,0}$ are independent of $j$, so is $q_{M''l+j,k}$. Therefore $q_{M''l+j,k} = q_{M''l,k}$.

From this conclusion, \eqref{xj} simplifies into
\eq{
  x'_l = \sum_{k=0}^{m'-1} q_{M''l,k} v_{k,0}, \qquad l=0,\ldots,M'-1
,}
which is a one-dimensional version of \eqref{e7}. It is satisfied only if $[\XX',\LL',\mat{U}_{M'}]$ is FOO, where the elements $q'_{l,k}$ of $\mat{Q}(\LL')$ are
\eqlab{qp2}{
  q'_{l,k} = q_{M''l,k},\qquad l=0,\ldots,M'-1, \quad k=0,\ldots,m'-1
.}

A similar analysis for $n=1$ shows that $[\XX'',\LL'',\mat{U}_{M''}]$ is also FOO and, furthermore, yields analogous expressions to \eqref{qp1} and \eqref{qp2} as
\eqlab{qpp1}{
  q_{M''l+j,k} = q_{j,k},\qquad &l=0,\ldots,M'-1, \quad j=0,\ldots,M''-1, \quad k=m',\ldots,m-1 \\
  q''_{j,k} = q_{j,k+m'},\qquad &j=0,\ldots,M''-1, \quad k=0,\ldots,m''-1 \label{qpp2}
}
where $q''_{j,k}$ are the elements of $\mat{Q}(\LL'')$.

Together, \eqref{qp1}, \eqref{qp2}, \eqref{qpp1}, and \eqref{qpp2} show that for $l=0,\ldots,M'-1$ and $j=0,\ldots,M''-1$,
\eq{
q_{M''l+j,k} = \begin{cases}
  q'_{l,k},     & k=0,\ldots,m'-1 \\
  q''_{j,k-m'}, & k=m',\ldots,m-1
\end{cases}
,}
or, equivalently, that $\mat{Q}(\LL) = \mat{Q}(\LL') \otimes \mat{Q}(\LL'')$. To convert this relation into a relation between (unmodified) labeling matrices $\LL$, $\LL'$, and $\LL''$, we can apply \eqref{qki_def} to conclude that $\LL$ is a column-permuted version of $\LL' \otimes \LL''$.

To complete the proof, we need to consider the case when the $m''$ zeros among $v_{0,0},\ldots,v_{m-1,0}$ are not the last $m''$ elements as in \eqref{vzeros}. To this end, we apply an arbitrary row permutation to the $\mat{V}$ matrix, whose first column is given by \eqref{vzeros}. Permuting the $m$ rows of $\mat{V}$ means permuting the $m$ elements of \eqref{vzeros}, which in turn means that the $m''$ zeros are shifted into arbitrary locations. Furthermore, as was observed in the first part of this proof, a row permutation of $\mat{V}$ corresponds to a column permutation of $\mat{Q}(\LL)$, or, equivalently, a column permutation of $\LL$. We can therefore conclude that regardless of where the $m''$ zeros are located, the labeling $\LL$ must be a column-permuted version of $\LL' \otimes \LL''$.

\section{Proof of Theorem~\ref{Sec.PAM.QAM.PSK.constant-energy}}\label{Appendix.PAM.QAM.PSK.constant-enery}

If $\mb{v}_0,\ldots,\mb{v}_{m-1}$ are orthogonal, then $\mb{v}_k\mb{v}_l^{\tr{T}} = 0$ for $k\ne l$. The symbol energies $\|\mb{x}_i\|^2$, for $i=0,\ldots,M-1$, can be calculated from \eqref{e7} as
\eq{
  \|\mb{x}_i\|^2 &= \sum_{k=0}^{m-1} \sum_{l=0}^{m-1}
    q_{i,k} q_{i,l} \mb{v}_k \mb{v}_l^{\tr{T}} \\
  &= \sum_{k=0}^{m-1} q_{i,k}^2 \|\mb{v}_k\|^2 \\
  &= \sum_{k=0}^{m-1} \|\mb{v}_k\|^2
,}
which is independent of $i$. This completes the ``if'' part of the theorem.

For the ``only if'' part, we make use of the identity
\eqlab{abc}{
  8\mb{b}\mb{c}^{\tr{T}} &=
    \|\mb{a}+\mb{b}+\mb{c}\|^2-\|\mb{a}+\mb{b}-\mb{c}\|^2  -\|\mb{a}-\mb{b}+\mb{c}\|^2+\|\mb{a}-\mb{b}-\mb{c}\|^2
,}
which holds for any vectors $\mb{a}$, $\mb{b}$, and $\mb{c}$. 
Let $\mat{X}$ be any FOO constant-energy input alphabet and let $k$ and $l$ be any pair of distinct integers $0\le k,l\le m-1$. Define
\eq{
\mb{a} \triangleq \sum_{\substack{j=0\\j\notin \{k,l\}}}^{m-1} \mb{v}_j
,}
$\mb{b} \triangleq \mb{v}_k$, and $\mb{c} \triangleq \mb{v}_l$. From \eqref{e7}, the four vectors $\mb{a}\pm \mb{b} \pm \mb{c}$ all belong to $\mat{X}$. Thus, all four have the same energy and the right-hand side of \eqref{abc} is zero. Thus $\mb{v}_k \mb{v}_l^{\tr{T}} = \mb{b} \mb{c}^{\tr{T}} = 0$. This holds for all pairs of distinct $k$ and $l$, which completes the proof.

\section{Proof of Theorem~\ref{Theorem.alpha.PAM}}\label{Appendix.Theorem.alpha.PAM}

For $\mat{P}=\mat{U}_M$, the average symbol energy is given by $\Es=\frac{M^2-1}{3}$,
and that the constellation is zero mean, \ie $\ms{E}_{{X}}[{X}]^2=0$. Therefore, the coefficient $\alpha_\Omega^\tr{BI}$ in \eqref{a1.BICM.general} is
\begin{align}\label{a1.bicm.zeromean}
\alpha_ \Omega^\tr{BI}=\frac{\log_2{\tr{e}}}{\Es}\sum_{k=0}^{m-1}\sum_{u\in\set{0,1}}\frac{1}{2}\ms{E}_{{X}|C_k=u}[{X}]^2.
\end{align}

For the BRGC, $\ms{E}_{{X}|C_k=u}[{X}]=0$ for $k=1,\ld,m-1$ and $u\in\set{0,1}$. For $k=0$ we find that 
\begin{align*}
\ms{E}_{{X}|C_0=u}[{X}]^2=\left(\sum_{i\in\mc{I}_{0,u}}\frac{2}{M}x_i\right)^2=\frac{M^2}{4},
\end{align*}
which used in \eqref{a1.bicm.zeromean} gives the desired result.

For the NBC, we note that
\begin{align*}
\ms{E}_{{X}|C_k=u}[{X}]^2=\left((-1)^{u+1}2^{m-k-1}\right)^2=\frac{M^2}{4}\left(\frac{1}{2}\right)^{2k}.
\end{align*}
Using the fact that 
\begin{align*}
\sum_{k=0}^{m-1}\left(\frac{1}{2}\right)^{2k}=\frac{4}{3}\left(1-\frac{1}{M^2}\right),
\end{align*}
the result $\alpha_ \Omega^\tr{BI}=\log_2{\tr{e}}$ is obtained.\footnote{A similar argument for the proof of the NBC has been used in \cite{Stierstorfer09a}.}

That $\alpha_\Omega^\tr{BI}=0$ if $\mat{L}_m=\mat{S}_m$ follows trivially because of the construction of the BSGC, \ie $\ms{E}_{{X}|C_k=u}[{X}]=0$ for $k=0,\ld,m-1$ and $u\in\set{0,1}$.

For the FBC, finally, its symmetry results in the same condition as for the BRGC, \ie $\ms{E}_{{X}|C_k=u}[{X}]=0$ for $k=1,\ld,m-1$ and $u\in\set{0,1}$. Moreover, since for $k=0$ the BRGC and the FBC are the same, the coefficient $\alpha_ \Omega^\tr{BI}$ is also the same. 

\section{Proof of Theorem~\ref{Theorem.alpha.PSK}}\label{Appendix.Theorem.alpha.PSK}

For PSK and any $k$, $P_{C_k}(0)\ms{E}_{\mb{X}|C_k=0}[\mb{X}]+P_{C_k}(1)\ms{E}_{\mb{X}|C_k=1}[\mb{X}]=\ms{E}_{\mb{X}}[\mb{X}]=\mb{0}$. Furthermore, since $P_{C_k}(0)=P_{C_k}(1)=1/2$, $\|\ms{E}_{\mb{X}|C_k=0}[\mb{X}]\|^2=\|\ms{E}_{\mb{X}|C_k=1}[\mb{X}]\|^2$.
From these equalities, \eqref{a1.BICM.general} reduces to
\begin{align}\label{a1.BICM.PSK.App}
\alpha_ \Omega^\tr{BI}=\log_2{\tr{e}} \sum_{k=0}^{m-1}\|\ms{E}_{\mb{X}|C_k=0}[\mb{X}]\|^2=\frac{4\log_2{\tr{e}}}{M^2} \sum_{k=0}^{m-1}\biggl\|\sum_{i\in\mc{I}_{k,0}}\mb{x}_i\biggr\|^2.
\end{align}

\subsection{Proof for the BRGC}\label{App.PSK.BRGC}

Because of the symmetry of PSK input alphabets and the BRGC, $\|\sum_{i\in\mc{I}_{k,0}}\mb{x}_i\|^2$ in \eqref{a1.BICM.PSK.App} is zero for $k=2,\ld, m-1$. Moreover, by symmetry, $\|\sum_{i\in\mc{I}_{0,0}}\mb{x}_i\|^2=\|\sum_{i\in\mc{I}_{1,0}}\mb{x}_i\|^2$. Since $\mc{I}_{0,0}=\set{0,\ld,M/2-1}$, the coefficient in \eqref{a1.BICM.PSK.App} is given by
\begin{align}
\label{alpha_BI-PSK.proof.2}
\alpha_ \Omega^\tr{BI}		& = \frac{8\log_2{\tr{e}}}{M^2}\biggl\|\sum_{i=0}^{M/2-1}\mb{x}_i\biggr\|^2 = \frac{8\log_2{\tr{e}}}{M^2}\left[\Biggl(\sum_{i=0}^{M/2-1}\cos\frac{(2i+1)\pi}{M} \Biggr)^2+\Biggl(\sum_{i=0}^{M/2-1}\sin\frac{(2i+1)\pi}{M}\Biggr)^2\right].
\end{align}
Using \cite[eq.~(1.341.3)]{Gradshteyn80_Book} we find that the first sum in \eqref{alpha_BI-PSK.proof.2} is zero, and from \cite[eq.~(1.341.1)]{Gradshteyn80_Book} the second sum in \eqref{alpha_BI-PSK.proof.2} is equal to $1/\sin(\pi/M)$. This completes the first part of the proof.

\subsection{Proof for the NBC}\label{App.PSK.NBC}

For the NBC, $\|\sum_{i\in\mc{I}_{k,0}}\mb{x}_i\|^2$ in \eqref{a1.BICM.PSK.App} is zero for $k=1,\ld, m-1$. Moreover, since the first column of $\mat{N}_m$ is always equal to the first column of $\mat{G}_m$, it is clear that the coefficient for the NBC is half of the one for the BRGC.

\subsection{Proof for the BSGC}\label{App.PSK.BSGC}

By construction, $\mat{S}_m=\mat{G}_m$ for all the columns except the first one, and therefore, only two bit positions contribute in the outer sum in \eqref{a1.BICM.PSK.App}, \ie $k=0$ and $k=1$. From the proof for the BRGC, the contribution for $k=1$ is known to be
\begin{align} \label{bsgc.sum.i01}
\biggl\|\sum_{i\in\mc{I}_{1,0}}\mb{x}_i\biggr\|^2 = \frac{1}{\sin^2(\pi/M)}
.\end{align}
For $k=0$, we need the index set (cf.~Example \ref{ExampleG3N3S3})
\begin{align} \label{bsgc.i00}
\mc{I}_{0,0} &= \{0,4,\ldots,M/2-4\} \cup \{3,7,\ldots,M/2-1\} \nonumber\\
  &\qquad \cup\{M/2+1,M/2+5,\ldots,M-3\} \cup \{M/2+2,M/2+6,\ldots,M-2\} \nonumber\\
  &= \bigcup_{k=0}^{M/8-1} \{4k, M/2-1-4k, M/2+1+4k, M-2-4k\}.
\end{align}
This partitioning of $\mc{I}_{0,0}$ into four subsets will now be used to calculate
\begin{align} \label{bsgc.sum.i00}
\biggl\|\sum_{i\in\mc{I}_{0,0}}\mb{x}_i\biggr\|^2
=\Biggl(\sum_{i\in\mc{I}_{0,0}}\cos\frac{(2i+1)\pi}{M} \Biggl)^2+
\Biggl(\sum_{i\in\mc{I}_{0,0}}\sin\frac{(2i+1)\pi}{M} \Biggl)^2.
\end{align}
We split the sum over $\mc{I}_{0,0}$ in the second term of \eqref{bsgc.sum.i00} into four sums, one for each subset in \eqref{bsgc.i00}, which yields
\begin{align} \label{bsgc.sum.sin1}
\sum_{i\in\mc{I}_{0,0}}\sin\frac{(2i+1)\pi}{M}
  &= \sum_{k=0}^{M/8-1} \bigg( \sin\frac{(1+8k)\pi}{M}
    + \sin\frac{(M-1-8k)\pi}{M} \nonumber\\
    &\qquad + \sin\frac{(M+3+8k)\pi}{M}
    + \sin\frac{(2M-3-8k)\pi}{M} \bigg) \nonumber\\
  &= \sum_{k=0}^{M/8-1} \bigg( 2\sin\frac{(1+8k)\pi}{M}
    - 2\sin\frac{(3+8k)\pi}{M} \bigg).
\end{align}
Applying \cite[eq.~(1.341.3)]{Gradshteyn80_Book} twice yields
\begin{align} \label{bsgc.sum.sin2}
\sum_{i\in\mc{I}_{0,0}}\sin\frac{(2i+1)\pi}{M}
  &= \left( 2\cos\frac{3\pi}{M} - 2\cos\frac{\pi}{M} \right)
    \csc\frac{4\pi}{M} \nonumber\\
  &= -2 \sin\frac{\pi}{M} \sec\frac{2\pi}{M}
,
\end{align}
where $\csc x = 1/\sin x$ is the cosecant function and $\sec x = 1/\sin x$ is the secant.

Expanding the first term of \eqref{bsgc.sum.i00} by the same method as in \eqref{bsgc.sum.sin1} reveals that this term is zero. Now the result follows from \eqref{a1.BICM.PSK.App}, \eqref{bsgc.sum.i01}, \eqref{bsgc.sum.i00}, and \eqref{bsgc.sum.sin2}.

\subsection{Proof for the FBC}

By construction, the first bit of the FBC is the same as for the BRGC and the other bits are symmetric around $M/2$. Therefore, the components in the second dimension of $\|\sum_{i\in\mc{I}_{k,0}}\mb{x}_i\|^2$ are zero for $k=1,\ldots, M-1$ and \eqref{a1.BICM.PSK.App} can be expressed as
\begin{align}
\alpha_\Omega^\tr{BI} &= \frac{4\log_2{\tr{e}}}{M^2} \left[\frac{1}{\sin^2(\pi/M)}+\sum_{k=1}^{m-1}\Biggl(\sum_{i\in\mc{I}_{k,0}}\cos\frac{(2i+1)\pi}{M}\Biggr)^2\right] \nonumber\\
  &=\frac{4\log_2{\tr{e}}}{M^2} \left[\frac{1}{\sin^2(\pi/M)}+4\sum_{k=1}^{m-1}\Biggl(\sum_{i\in\mc{I}_{k,0}^\tr{U}}\cos\frac{(2i+1)\pi}{M}\Biggr)^2\right], \label{App.PSK.FBC.1}
\end{align}
where $\mc{I}_{k,0}^\tr{U}\triangleq \{i\in\mc{I}_{k,0}: i<M/2\}$.

The indexes $\mc{I}_{k,0}^\tr{U}$ of the FBC for $k=1,\ldots,m-1$ are obtained as the indexes of the NBC of order $m-1$.
For example, for $M=32$, we obtain $\mc{I}_{1,0}^\tr{U}=\set{0,1,2,3,4,5,6,7}$, $\mc{I}_{2,0}^\tr{U}=\set{0,1,2,3,8,9,10,11}$, $\mc{I}_{3,0}^\tr{U}=\set{0,1,4,5,8,9,12,13}$, and $\mc{I}_{4,0}^\tr{U}=\set{0,2,4,6,8,10,12,14}$. This regularity results in a simplified expression of the inner sum in \eqref{App.PSK.FBC.1}, \ie
\begin{align}\label{App.PSK.FBC.2}
\sum_{i\in\mc{I}_{k,0}^\tr{U}}\cos\frac{(2i+1)\pi}{M} &=\sum_{j=0}^{2^{k-1}-1}\sum_{l=0}^{2^{m-k-1}-1}\cos{\left(\frac{\pi}{M}[2^{m-k+1}j+2l+1]\right)} \\
& =\frac{\tan(\pi/2^{k+1})}{2\sin(\pi/M)} \label{App.PSK.FBC.3},
\end{align}
where the final result was obtained by using
\cite[eq.~(1.341.3)]{Gradshteyn80_Book} twice in \eqref{App.PSK.FBC.2}, after some algebraic manipulation.
Using \eqref{App.PSK.FBC.3} in \eqref{App.PSK.FBC.1} gives the desired result.

\bibliography{IEEEabrv,references_all}

\begin{thebibliography}{10}
\providecommand{\url}[1]{#1}
\csname url@samestyle\endcsname
\providecommand{\newblock}{\relax}
\providecommand{\bibinfo}[2]{#2}
\providecommand{\BIBentrySTDinterwordspacing}{\spaceskip=0pt\relax}
\providecommand{\BIBentryALTinterwordstretchfactor}{4}
\providecommand{\BIBentryALTinterwordspacing}{\spaceskip=\fontdimen2\font plus
\BIBentryALTinterwordstretchfactor\fontdimen3\font minus
  \fontdimen4\font\relax}
\providecommand{\BIBforeignlanguage}[2]{{%
\expandafter\ifx\csname l@#1\endcsname\relax
\typeout{** WARNING: IEEEtran.bst: No hyphenation pattern has been}%
\typeout{** loaded for the language `#1'. Using the pattern for}%
\typeout{** the default language instead.}%
\else
\language=\csname l@#1\endcsname
\fi
#2}}
\providecommand{\BIBdecl}{\relax}
\BIBdecl

\bibitem{Nyquist24}
H.~Nyquist, ``Certain factors affecting telegraph speed,'' \emph{Bell System
  Technical Journal}, pp. 324--346, Apr. 1924.

\bibitem{Nyquist28}
------, ``Certain topics in telegraph transmission theory,'' \emph{Trans. of
  the A.I.E.E.}, vol.~47, no.~2, pp. 617--644, Apr. 1928.

\bibitem{Hartley28}
R.~V.~L. Hartley, ``Transmission of information,'' \emph{Bell System Technical
  Journal}, pp. 535--563, July 1928.

\bibitem{Shannon48}
C.~E. Shannon, ``A mathematical theory of communications,'' \emph{Bell System
  Technical Journal}, vol.~27, pp. 379--423 and 623--656, July and Oct. 1948.

\bibitem{Shannon49}
------, ``Communication in the presence of noise,'' \emph{Proceedings of the
  IRE}, vol.~37, no.~1, pp. 10--21, Jan. 1949.

\bibitem{Pierce73}
J.~R. Pierce, ``The early days of information theory,'' \emph{{IEEE} Trans.
  Inf. Theory}, vol. IT-19, no.~1, pp. 3--8, Jan. 1973, (Invited Paper).

\bibitem{Massey74}
J.~L. Massey, ``Coding and modulation in digital communications,'' in
  \emph{International Zurich Seminar on Digital Communications}, Zurich,
  Switzerland, Mar. 1974.

\bibitem{Ungerboeck76}
G.~Ungerboeck and I.~Csajka, ``On improving data-link performance by increasing
  channel alphabet and introducing sequence decoding,'' in \emph{International
  Symposium on Information Theory (ISIT)}, Ronneby, Sweden, June 1976.

\bibitem{Ungerboeck82}
G.~Ungerboeck, ``Channel coding with multilevel/phase signals,'' \emph{{IEEE}
  Trans. Inf. Theory}, vol.~28, no.~1, pp. 55--67, Jan. 1982.

\bibitem{Imai77}
H.~Imai and S.~Hirakawa, ``A new multilevel coding method using
  error-correcting codes,'' \emph{{IEEE} Trans. Inf. Theory}, vol. IT-23,
  no.~3, pp. 371--377, May 1977.

\bibitem{Imai77b}
------, ``Correction to `{A} new multilevel coding method using
  error-correcting codes','' \emph{{IEEE} Trans. Inf. Theory}, vol. IT-23,
  no.~6, p. 784, Nov. 1977.

\bibitem{Zehavi92}
E.~Zehavi, ``8-{PSK} trellis codes for a {Rayleigh} channel,'' \emph{{IEEE}
  Trans. Commun.}, vol.~40, no.~3, pp. 873--884, May 1992.

\bibitem{Caire98}
G.~Caire, G.~Taricco, and E.~Biglieri, ``Bit-interleaved coded modulation,''
  \emph{{IEEE} Trans. Inf. Theory}, vol.~44, no.~3, pp. 927--946, May 1998.

\bibitem{3gpp25212_2009}
3GPP, ``Universal mobile telecommunications system {(UMTS)}; multiplexing and
  channel coding {(FDD)},'' 3GPP, Tech. Rep. TS 25.212, V8.5.0 Release 8, Mar.
  2009.

\bibitem{Dahlman08_Book}
E.~Dahlman, S.~Parkvall, J.~Sk\"old, and P.~Beming, \emph{{3G} Evolution:
  {HSPA} and {LTE} for Mobile Broadband}, 2nd~ed.\hskip 1em plus 0.5em minus
  0.4em\relax Academic Press, 2008.

\bibitem{IEEE80211_July99}
{IEEE 802.11}, ``Part 11: Wireless {LAN} medium access control ({MAC}) and
  physical layer ({PHY}) specifications: High-speed physical layer in the
  5{GH}z band,'' IEEE Std 802.11a-1999(R2003), Tech. Rep., July 1999.

\bibitem{IEEE80211n_October09}
{IEEE 802.11n}, ``Part 11: Wireless {LAN} medium access control ({MAC}) and
  physical layer ({PHY}) specifications. {A}mendment 5: Enhancements for higher
  throughout,'' IEEE Std 802.11n-2009, Tech. Rep., Oct. 2009.

\bibitem{ETSI_EN_302_755_v111}
ETSI, ``Digital video broadcasting ({DVB}); {F}rame structure channel coding
  and modulation for a second generation digital terrestrial television
  broadcasting system ({DVB-T2}),'' ETSI, Tech. Rep. ETSI EN 302 755 V1.1.1
  (2009-09), Sep. 2009.

\bibitem{ETSI_EN_302_307_v121}
------, ``Digital video broadcasting ({DVB}); {S}econd generation framing
  structure, channel coding and modulation systems for broadcasting,
  interactive services, news gathering and other broadband satellite
  applications ({DVB-S2}),'' ETSI, Tech. Rep. ETSI EN 302 307 V1.2.1 (2009-08),
  Aug. 2009.

\bibitem{ETSI_EN_302_769_v111}
------, ``Digital video broadcasting ({DVB}); {F}rame structure channel coding
  and modulation for a second generation digital transmission system for cable
  system ({DVB-C2}),'' ETSI, Tech. Rep. ETSI EN 302 769 V1.1.1 (2010-04), Apr.
  2010.

\bibitem{Martinez08b}
A.~Martinez, A.~{Guill\'en i F\`abregas}, and G.~Caire, ``Bit-interleaved coded
  modulation in the wideband regime,'' \emph{{IEEE} Trans. Inf. Theory},
  vol.~54, no.~12, pp. 5447--5455, Dec. 2008.

\bibitem{Alvarado10c}
A.~Alvarado, E.~Agrell, A.~{Guill\'en i F\`abregas}, and A.~Martinez,
  ``Corrections to `{Bit}-interleaved coded modulation in the wideband
  regime','' \emph{{IEEE} Trans. Inf. Theory}, vol.~56, no.~12, p. 6513, Dec.
  2010.

\bibitem{Stierstorfer09a}
C.~Stierstorfer and R.~F.~H. Fischer, ``Asymptotically optimal mappings for
  {BICM} with {$M$-PAM} and {$M^2$-QAM},'' \emph{IET Electronics Letters},
  vol.~45, no.~3, pp. 173--174, Jan. 2009.

\bibitem{Stierstorfer07a}
------, ``({G}ray) {M}appings for bit-interleaved coded modulation,'' in
  \emph{IEEE Vehicular Technology Conference (VTC-Spring)}, Dublin, Ireland,
  Apr. 2007.

\bibitem{MacWilliams77}
F.~J. MacWilliams and N.~J.~A. Sloane, \emph{The Theory of Error-Correcting
  Codes}.\hskip 1em plus 0.5em minus 0.4em\relax Amsterdam, The Netherlands:
  North-Holland, 1977.

\bibitem{Barsoum07}
M.~Barsoum, C.~Jones, and M.~Fitz, ``Constellation design via capacity
  maximization,'' in \emph{IEEE International Symposium on Information Theory
  (ISIT)}, Nice, France, June 2007.

\bibitem{Agrell04}
E.~Agrell, J.~Lassing, E.~G. Str\"{o}m, and T.~Ottosson, ``On the optimality of
  the binary reflected {G}ray code,'' \emph{{IEEE} Trans. Inf. Theory},
  vol.~50, no.~12, pp. 3170--3182, Dec. 2004.

\bibitem{Gray53}
F.~Gray, ``Pulse code communications,'' U.~S. Patent 2\,632\,058, Mar. 1953.

\bibitem{Agrell07}
E.~Agrell, J.~Lassing, E.~G. Str\"om, and T.~Ottosson, ``Gray coding for
  multilevel constellations in {G}aussian noise,'' \emph{{IEEE} Trans. Inf.
  Theory}, vol.~53, no.~1, pp. 224--235, Jan. 2007.

\bibitem{Lassing03b}
J.~Lassing, E.~G. Str{\"o}m, E.~Agrell, and T.~Ottosson, ``Unequal bit-error
  protection in coherent {$M$}-ary {PSK},'' in \emph{IEEE Vehicular Technology
  Conference (VTC-Fall)}, Orlando, FL, USA, Oct. 2003.

\bibitem{Viterbi98}
A.~J. Viterbi, ``An intuitive justification and a simplified implementation of
  the {MAP} decoder for convolutional codes,'' \emph{{IEEE} J. Sel. Areas
  Commun.}, vol.~16, no.~2, pp. 260--264, Feb. 1998.

\bibitem{Tsgr1_15_1093}
Ericsson, Motorola, and Nokia, ``Link evaluation methods for high speed
  downlink packet access ({HSDPA}),'' TSG-RAN Working Group 1 Meeting \#15,
  TSGR1\#15(00)1093, Tech. Rep., Aug. 2000.

\bibitem{Classon02}
B.~Classon, K.~Blankenship, and V.~Desai, ``Channel coding for {4G} systems
  with adaptive modulation and coding,'' \emph{{IEEE} Wireless Commun. Mag.},
  vol.~9, no.~2, pp. 8--13, Apr. 2002.

\bibitem{Alvarado06c}
A.~Alvarado, H.~Carrasco, and R.~Feick, ``On adaptive {BICM} with finite
  block-length and simplified metrics calculation,'' in \emph{IEEE Vehicular
  Technology Conference (VTC-Fall)}, Montreal, QC, Canada, Sep. 2006.

\bibitem{Szczecinski08b}
L.~Szczecinski, R.~Bettancourt, and R.~Feick, ``Probability density function of
  reliability metrics in {BICM} with arbitrary modulation: Closed-form through
  algorithmic approach,'' \emph{{IEEE} Trans. Commun.}, vol.~56, no.~5, pp.
  736--742, May 2008.

\bibitem{Benjillali06b}
M.~Benjillali, L.~Szczecinski, and S.~Aissa, ``Probability density functions of
  logarithmic likelihood ratios in rectangular {QAM},'' in \emph{Twenty-Third
  Biennial Symposium on Communications}, Kingston, ON, Canada, May 2006.

\bibitem{Alvarado07d}
A.~Alvarado, L.~Szczecinski, R.~Feick, and L.~Ahumada, ``Distribution of
  {L}-values in {G}ray-mapped ${M}^2$-{QAM}: Closed-form approximations and
  applications,'' \emph{{IEEE} Trans. Commun.}, vol.~57, no.~7, pp. 2071--2079,
  July 2009.

\bibitem{Szczecinski07f}
L.~Szczecinski, A.~Alvarado, and R.~Feick, ``Distribution of max-log metrics
  for {QAM}-based {BICM} in fading channels,'' \emph{{IEEE} Trans. Commun.},
  vol.~57, no.~9, pp. 2558--2563, Sep. 2009.

\bibitem{Kenarsari10}
A.~Kenarsari-Anhari and L.~Lampe, ``An analytical approach for performance
  evaluation of {BICM} over {N}akagami-$m$ fading channels,'' \emph{{IEEE}
  Trans. Commun.}, vol.~58, no.~4, pp. 1090--1101, Apr. 2010.

\bibitem{Knagenhjelm96}
P.~Knagenhjelm and E.~Agrell, ``The {H}adamard transform---a tool for index
  assignment,'' \emph{{IEEE} Trans. Inf. Theory}, vol.~42, no.~4, pp.
  1139--1151, July 1996.

\bibitem{Khan10}
F.~A. Khan, E.~Agrell, and M.~Karlsson, ``Electronic dispersion compensation by
  {H}adamard transformation,'' in \emph{Optical Fiber Conference (OFC)}, San
  Diego, CA, USA, Mar. 2010.

\bibitem{Sutter91}
E.~E. Sutter, ``The fast $m$-transform: a fast computation of
  cross-correlations with binary $m$-sequences,'' \emph{SIAM J. Comput.},
  vol.~20, no.~4, pp. 686--694, Aug. 1991.

\bibitem{Pratt69}
W.~K. Pratt, J.~Kane, and H.~C. Andrews, ``Hadamard transform image coding,''
  \emph{Proceedings of the IEEE}, vol.~57, no.~1, pp. 58--72, Jan. 1969.

\bibitem{Cover06_Book}
T.~Cover and J.~Thomas, \emph{Elements of Information Theory}, 2nd~ed.\hskip
  1em plus 0.5em minus 0.4em\relax New York, USA: John Wiley \& Sons, 2006.

\bibitem{Cover02}
T.~Cover and M.~Chiang, ``Duality between channel capacity and rate distortion
  with two-sided state information,'' \emph{{IEEE} Trans. Inf. Theory},
  vol.~48, no.~6, pp. 1629--1638, June 2002.

\bibitem{Schreckenbach07_Thesis}
F.~Schreckenbach, ``Iterative decoding of bit-interleaved coded modulation,''
  Ph.D. dissertation, Technische Universit\"at M\"unchen, Munich, Germany,
  2007, available at http://mediatum2.ub.tum.de/doc/644160/644160.pdf.

\bibitem{Brannstrom09}
F.~Br\"annstr\"om and L.~K. Rasmussen, ``Classification of unique mappings for
  8{PSK} based on bit-wise distance spectra,'' \emph{{IEEE} Trans. Inf.
  Theory}, vol.~55, no.~3, pp. 1131--1145, Mar. 2009.

\bibitem{Martinez09}
A.~Martinez, A.~{Guill\'en i F\`abregas}, and G.~Caire, ``Bit-interleaved coded
  modulation revisited: A mismatched decoding perspective,'' \emph{{IEEE}
  Trans. Inf. Theory}, vol.~55, no.~6, pp. 2756--2765, June 2009.

\bibitem{Stierstorfer09_Thesis}
C.~Stierstorfer, ``A bit-level-based approach to coded multicarrier
  transmission,'' Ph.D. dissertation, Friedrich-Alexander-Universit\"at
  Erlangen-N\"{u}rnberg, Erlangen, Germany, 2009, available at
  http://www.opus.ub.uni-erlangen.de/opus/volltexte/2009/1395/.

\bibitem{Fischer02_Book}
R.~F.~H. Fischer, \emph{Precoding and Signal Shaping for Digital
  Transmission}.\hskip 1em plus 0.5em minus 0.4em\relax John Wiley \& Sons,
  2002.

\bibitem{Wachsmann99}
U.~Wachsmann, R.~F.~H. Fischer, and J.~B. Huber, ``Multilevel codes:
  Theoretical concepts and practical design rules,'' \emph{{IEEE} Trans. Inf.
  Theory}, vol.~45, no.~5, pp. 1361--1391, July 1999.

\bibitem{Alvarado09c}
A.~Alvarado, E.~Agrell, L.~Szczecinski, and A.~Svensson, ``Exploiting {UEP} in
  {QAM}-based {BICM}: Interleaver and code design,'' \emph{{IEEE} Trans.
  Commun.}, vol.~58, no.~2, pp. 500--510, Feb. 2010.

\bibitem{Legoff05}
S.~{Le Goff}, B.~S. Sharif, and S.~A. Jimaa, ``Bit-interleaved turbo-coded
  modulation using shaping coding,'' \emph{{IEEE} Commun. Lett.}, vol.~9,
  no.~3, pp. 246--248, Mar. 2005.

\bibitem{LeGoff07}
S.~{Le Goff}, B.~K. Khoo, C.~C. Tsimenidis, and B.~S. Sharif, ``Constellation
  shaping for bandwidth-efficient turbo-coded modulation with iterative
  receiver,'' \emph{{IEEE} Trans. Wireless Commun.}, vol.~6, no.~6, pp.
  2223--2233, June 2007.

\bibitem{Raphaeli04}
D.~Raphaeli and A.~Gurevitz, ``Constellation shaping for pragmatic turbo-coded
  modulation with high spectral efficiency,'' \emph{{IEEE} Trans. Commun.},
  vol.~52, no.~3, pp. 341--345, Mar. 2004.

\bibitem{Martinez06}
A.~Martinez, A.~{Guill\'en i F\`abregas}, and G.~Caire, ``Error probability
  analysis of bit-interleaved coded modulation,'' \emph{{IEEE} Trans. Inf.
  Theory}, vol.~52, no.~1, pp. 262--271, Jan. 2006.

\bibitem{Fabregas08_Book}
A.~{Guill\'en i F\`abregas}, A.~Martinez, and G.~Caire, ``Bit-interleaved coded
  modulation,'' \emph{Foundations and Trends in Communications and Information
  Theory}, vol.~5, no. 1--2, pp. 1--153, 2008.

\bibitem{Verdu02}
S.~Verd\'{u}, ``Spectral efficiency in the wideband regime,'' \emph{{IEEE}
  Trans. Inf. Theory}, vol.~48, no.~6, pp. 1319--1343, June 2002.

\bibitem{Guo05}
D.~Guo, S.~Shamai, and S.~Verd\'{u}, ``Mutual information and minimum
  mean-square error in {G}aussian channels,'' \emph{{IEEE} Trans. Inf. Theory},
  vol.~51, no.~4, pp. 1261--1282, Apr. 2005.

\bibitem{Simoens08}
F.~Simoens, H.~Wymeersch, and M.~Moeneclaey, ``Linear precoders for
  bit-interleaved coded modulation on {AWGN} channels: Analysis and design
  criteria,'' \emph{{IEEE} Trans. Inf. Theory}, vol.~54, no.~1, pp. 87--99,
  Jan. 2008.

\bibitem{Brannstrom09b}
F.~Br\"annstr\"om, L.~K. Rasmussen, and A.~J. Grant, ``Optimal puncturing
  ratios and energy allocation for multiple parallel concatenated codes,''
  \emph{{IEEE} Trans. Inf. Theory}, vol.~55, no.~5, pp. 2062--2077, May 2009.

\bibitem{Peleg98}
M.~Peleg and S.~Shamai, ``On the capacity of the blockwise incoherent {MPSK}
  channel,'' \emph{{IEEE} Trans. Commun.}, vol.~46, no.~5, pp. 603--609, May
  1998.

\bibitem{Stark85}
W.~E. Stark, ``Capacity and cutoff rate of noncoherent {FSK} with nonselective
  {R}ician fading,'' \emph{{IEEE} Trans. Commun.}, vol. COM-33, no.~11, pp.
  1153--1159, Nov. 1985.

\bibitem{Fabregas10a}
A.~{Guill\'en i F\`abregas} and A.~Martinez, ``Bit-interleaved coded modulation
  with shaping,'' in \emph{IEEE Information Theory Workshop (ITW)}, Dublin,
  Ireland, Aug.--Sep. 2010.

\bibitem{Guo08}
D.~Guo, S.~Shamai, and S.~Verd\'{u}, ``Estimation of non-{G}aussian random
  variables in {G}aussian noise: Properties of the {MMSE},'' in \emph{IEEE
  International Symposium on Information Theory (ISIT)}, Toronto, ON, Canada,
  July 2008.

\bibitem{Guo11}
------, ``Estimation in {G}aussian noise: Properties of the minimum mean-square
  error,'' \emph{{IEEE} Trans. Inf. Theory}, vol.~57, no.~1, Jan. 2011,
  available at http://arxiv.org/abs/1004.3332.

\bibitem{Zwillinger03_Book}
D.~Zwillinger, \emph{Standard Mathematical Tables and Formulae}, 31st~ed.\hskip
  1em plus 0.5em minus 0.4em\relax Boca Raton, FL: CRC Press, 2003.

\bibitem{Shannon59b}
C.~E. Shannon, ``Coding theorems for a discrete source with a fidelity
  criterion,'' in \emph{IRE National Convention Record}, New York, NY, Mar.
  1959, pp. 142--163, vol. 7, pt. 4.

\bibitem{Vitthaladevuni03}
P.~K. Vitthaladevuni and M.-S. Alouini, ``A recursive algorithm for the exact
  {BER} computation of generalized hierarchical {QAM} constellations,''
  \emph{{IEEE} Trans. Inf. Theory}, vol.~49, no.~1, pp. 297--307, Jan. 2003.

\bibitem{Morimoto95}
M.~Morimoto, M.~Okada, and S.~Komaki, ``A hierarchical image transmission
  system in fading channel,'' in \emph{IEEE International Conference on
  Universal Personal Communications (ICUPC)}, Tokyo, Japan, Oct. 1995, pp.
  769--772.

\bibitem{Hossain06b}
{Md. J. Hossain}, M.-S. Alouini, and V.~K. Bhargava, ``Hierarchical
  constellation for multi-resolution data transmission over block fading
  channels,'' \emph{{IEEE} Trans. Wireless Commun.}, vol.~5, no.~4, pp.
  849--857, Apr. 2006.

\bibitem{Divsalar95b}
D.~Divsalar and F.~Pollara, ``On the design of turbo codes,'' Jet Propulsion
  Laboratory, Pasadena, CA, TDA Progr. Rep. 42-123, pp.~99--121, Nov. 1995,
  available at http://tmo.jpl.nasa.gov/progress\_report/42-123/123D.pdf.

\bibitem{Prelov04}
V.~V. Prelov and S.~Verd\'{u}, ``Second-order asymptotics of mutual
  information,'' \emph{{IEEE} Trans. Inf. Theory}, vol.~50, no.~8, pp.
  1567--1580, Aug. 2004.

\bibitem{Gradshteyn80_Book}
I.~S. Gradshteyn and I.~M. Ryzhik, \emph{Tables of Integrals, Series and
  Products}, 6th~ed.\hskip 1em plus 0.5em minus 0.4em\relax New York, NY:
  Academic Press, 1980.

\end{thebibliography}
\bibliographystyle{IEEEtran}

\end{document}